\documentclass[11pt,british]{article}
\usepackage[T1]{fontenc}
\usepackage[latin9]{inputenc}
\usepackage{geometry}
\usepackage{amsmath}
\usepackage{amssymb}
\usepackage{stackrel}
\usepackage{graphicx}
\usepackage{wasysym}
\usepackage{esint}
\usepackage{cite}
\usepackage[caption=false]{subfig}
\providecommand{\tabularnewline}{\\}
\usepackage{babel}
\usepackage{graphicx}
\usepackage[colorlinks,bookmarks=false,citecolor=blue,linkcolor=blue,urlcolor=blue]{hyperref}
\usepackage{color}

\makeatletter
\numberwithin{equation}{section}
\numberwithin{figure}{section}
\numberwithin{table}{section}

\makeatother

\def\ket#1{\mathinner{|{#1}\rangle}}

\begin{document}

\frenchspacing

\title{Hamiltonian truncation approach to quenches in the \\Ising field theory }

\author{T. Rakovszky$\;{}^{1}$, M. Mestyán$\;{}^{2}$
M. Collura$\;{}^{2}$, M. Kormos$\;{}^{3,4}$, and G. Takács$\;{}^{3,4}$\\
 ~\\
 $^{1}${\small{}Institute of Physics, Eötvös University}\\
 {\small{}1117 Budapest, Pázmány Péter sétány 1/A, Hungary}\\
~\\
 $^{2}${\small{}SISSA and INFN, via Bonomea 265, 34136 Trieste, Italy.}\\
~\\
 $^{3}${\small{}MTA-BME ``Momentum'' Statistical
Field Theory Research Group}\\
 {\small{}1111 Budapest, Budafoki út 8., Hungary}\\
 ~\\
 $^{4}${\small{}Department of Theoretical Physics, }\\
 {\small{}Budapest University of Technology and Economics}\\
 {\small{}1111 Budapest, Budafoki út 8., Hungary}}

\date{\today}
\maketitle
\begin{abstract}

In contrast to lattice systems where powerful numerical techniques such as 
matrix product state based methods are available to study the non-equilibrium 
dynamics, the non-equilibrium behaviour of continuum systems is much harder to 
simulate. We demonstrate here that Hamiltonian truncation methods can be 
efficiently applied to this problem, by studying the quantum quench dynamics of the 1+1 
dimensional Ising field theory using a truncated free fermionic space approach. 
After benchmarking the method with integrable quenches corresponding to changing 
the mass in a free Majorana fermion field theory, we study the effect of an 
integrability breaking perturbation by the longitudinal magnetic field. In both 
the ferromagnetic and paramagnetic phases of the model we find persistent 
oscillations with frequencies set by the low-lying particle excitations not only 
for small, but even for moderate size quenches. In the ferromagnetic phase 
these particles are the 
various non-perturbative confined bound states of the domain wall excitations, 
while in the paramagnetic phase the single magnon excitation governs the 
dynamics, allowing us to capture the time evolution of the magnetisation 
using a combination of known results from perturbation theory and 
form factor based methods. We point out that the dominance of low lying 
excitations allows for the numerical or experimental determination of the
mass spectra through the study of the quench dynamics.

\end{abstract}

\newpage

\tableofcontents{}

\section{Introduction}

In this work we investigate global quantum quenches in a quantum field
theory in one spatial dimension. A global quantum quench is the non-equilibrium
time evolution in isolated quantum systems which is initiated by
a sudden change of one or more parameters in the Hamiltonian \cite{2006PhRvL..96m6801C,2007JSMTE..06....8C}.
This protocol is routinely engineered in current cold-atom experiments
\cite{2006Natur.440..900K,2007Natur.449..324H,2012NatPh...8..325T,2012Sci...337.1318G,2013NatPh...9..640L,2013PhRvL.111e3003M,2013Natur.502...76F,2015Sci...348..207L,2016arXiv160304409K},
and has become a theoretical and experimental paradigm in non-equilibrium
dynamics of quantum systems. 

Quantum dynamics in one-dimensional integrable systems is especially
interesting due to the lack of thermalisation as a consequence of the infinite 
number of conserved quantities present in these systems. 
This has been observed experimentally
\cite{2006Natur.440..900K,2012Sci...337.1318G} and
led to the theoretical proposal of the Generalised Gibbs Ensemble
(GGE) \cite{2007PhRvL..98e0405R} which has also been 
observed experimentally\cite{2015Sci...348..207L}. In a surprising development,
recent studies \cite{2014PhRvL.113k7202W,2014PhRvL.113k7203P,
%2014JSMTE..12..009B,2015JSMTE..04..001M,
2015PhRvL.115o7201I}
led to the conclusion that the completeness of the GGE requires the inclusion
of novel quasi-local charges \cite{
%2013PhRvL.111e7203P,2014JSMTE..09..037P,2015PhRvL.115l0601I,2015arXiv151204454I,
2016arXiv160300440I}
which are sufficient to uniquely determine all post-quench quasi-particle
distributions; indeed this latter condition had previously been shown
to be necessary for the GGE to work at all \cite{2014PhRvA..90d3625G,2014JSMTE..09..026P
%,2014JSMTE..10..045P
}. 

In many cases, non-integrable systems can be considered as integrability
breaking perturbations of an otherwise integrable system. Generally,
non-integrable systems are expected to relax to a thermal (Gibbs)
state, at least for a suitable class of (local, or few-body) observables;
the principle underlying thermalisation in closed quantum systems
is the Eigenstate Thermalisation Hypothesis (ETH) \cite{1991PhRvA..43.2046D,1994PhRvE..50..888S}.
This leads to some important questions, chief among them is what aspects
of integrability are still retained after such a perturbation.

A natural scenario is dubbed prethermalisation \cite{2004PhRvL..93n2002B,2011PhRvB..84e4304K,2015PhRvL.115r0601B}
which posits that for a weak breaking of integrability the system
first relaxes to a deformed version of the GGE and is trapped there
for some (possibly long) time before eventual thermalisation. Prethermalisation,
in the sense of the system being trapped in a non-thermal metastable
state, have also been observed experimentally \cite{2012Sci...337.1318G}.
Vestiges of integrability may also be reflected in a quantum version
of the KAM (Kolmogorov--Arnold--Moser) theorem \cite{2015PhRvX...5d1043B}.

However, recent work \cite{2016arXiv160403571K} on the quantum Ising
spin chain has drawn attention to the possibility that in some systems
the integrability breaking interaction is never perturbative for any
nonzero value of its coupling, leading to a dramatic change in the dynamics
due to non-perturbative effects such as confinement.

The system considered in this work is the massive scaling
Ising field theory, the scaling limit of the quantum Ising spin chain.
Due to the scaling limit, quenches in the field theory correspond
to very small (infinitesimal) quenches on the spin chain close to
its quantum critical point. While these quenches form a somewhat special class of
processes from the lattice point of view, the dynamics
are still very nontrivial due to the proximity of the critical point.
An important motivation for considering these quenches is that in this
limit one can study aspects of quenches that only depend on the universality
class of the system. Apart from this, field theoretic quenches are
also interesting in their own right, since quantum field
theories are valid descriptions of physical phenomena, in particular
in high energy physics \cite{2004AIPC..739....3B}. 

In the context of quantum field theory in one spatial dimension, quenches to integrable
models have been widely considered \cite{2010NJPh...12e5015F,2011PhRvL.106s1601B,2012JSMTE..02..017S,2012JSMTE..04..017S,mussardoPRL111,2014PhLB..734...52S,2014JSMTE..10..035B,2014JPhA...47N2001D,2015PhRvA..91e1602E,2015JSMTE..11..004S,2016NuPhB.902..508H,2016JSMTE..03.3115C,2016JSMTE..06.3102B,2016arXiv160403879C}.
However, much less is known about the dynamics of quenches to non-integrable
quantum field theories; for integrable pre-quench dynamics, a perturbative 
approach was proposed in \cite{2014JPhA...47N2001D}. While perturbation theory
has limited applicability in quantum quenches, there are some situations for 
which it leads to interesting predictions that can be tested within our framework.

As a result of the above situation, in this work we are particularly interested in the effects of 
integrability breaking in quantum field theory quenches. To address this problem, we adopt a 
nonperturbative Hamiltonian truncation approach that has been successfully applied to study 
equilibrium properties of both integrable and non-integrable two-dimensional quantum field theories. 

The particular version of Hamiltonian truncation used in this work is based on a truncated fermionic space 
approach developed for the Ising field theory \cite{1991IJMPA...6.4557Y,2001hep.th...12167F} abbreviated 
as TFSA. Note, however, that similar Hamiltonian truncation methods apply to a much
wider range of field theory models such as perturbed minimal conformal field theories
\cite{1990IJMPA...5.3221Y,1991NuPhB.348..591L}, sine--Gordon \cite{1998PhLB..430..264F},
$\Phi^{4}$ and Landau--Ginzburg \cite{2014JSMTE..12..010C,2015arXiv151206901B,2015PhRvD..91h5011R,2016PhRvD..93f5014R},
and Wess--Zumino models \cite{2013NuPhB.877..457B,2015NuPhB.899..547K,2016arXiv160102979A},
and can even be extended to more than one spatial dimension \cite{2015PhRvD..91b5005H}.
As a result, there are many potential directions to extend the studies
in the present work. 

There have also been some other recent approaches to quantum quenches
incorporating Hamiltonian truncation, either combined with
algebraic Bethe Ansatz methods to study one-dimensional Bose gases
\cite{2015PhRvX...5d1043B}, or the truncated Hamiltonian approach
forming part of a chain array matrix product state algorithm \cite{2015PhRvB..92p1111J}
to study the non-equlibrium time evolution of two-dimensional spin
systems. 
However, our approach is essentially different from the above examples 
in that the simulation of the time evolution is performed
within the truncated Hamiltonian setting. Since the method works with 
microscopic data, many quantities relevant to quench problems are accessible, 
such as the overlaps between states, the full statistics of work performed in the 
quench, the time dependent return probability (Loschmidt echo) as well as 
the time evolution of expectation values.

To establish the validity of this approach, first we consider the integrable case where
it is possible to compare the numerical results with theoretical predictions,
and then extend our studies to the non-integrable case. To have an
independent verification in the latter case, the TFSA numerics are compared 
to infinite-volume time evolved block decimation (iTEBD) simulations 
on the lattice near the critical point. 

The outline of the paper is the following. Section \ref{sec:Quenches-in-Ising}
contains a review of the necessary facts about the Ising field theory,
and discusses the setup and general properties of quantum quenches.
The Hamiltonian truncation method is introduced and described 
in Section \ref{sec:The-truncated-fermionic}.
In Section \ref{sec:Integrable-quenches}, the results of
the numerical time evolution applied to integrable quenches
are compared to theoretical predictions. 
We demonstrate that it gives a correct description for integrable
quenches, at least for quenches of moderate size;
in particular, we avoid quenching across the phase transition. In
Section \ref{sec:Non-integrable-quenches-in-FM} considers non-integrable
quenches in the ferromagnetic phase. Non-integrable
quenches in the paramagnetic phase are discussed in 
Section \ref{sec:Non-integrable-quenches-in-PM}.
In Section \ref{sec:TFSA-versus-iTEBD} we present a direct, 
parameter-free comparison with a lattice simulation performed 
for an infinite spin chain near the scaling limit. 
Section \ref{sec:Conclusions} contains our conclusions. 
Some of the more technical details 
regarding the cut-off extrapolation and finite size effects, 
and the description of the meson spectrum in the ferromagnetic phase are
relegated to appendices. 

\section{Ising field theory and quench protocols \label{sec:Quenches-in-Ising}}

In this section we introduce the Ising field theory and discuss its relation to the quantum Ising spin chain. We also define the integrable and non-integrable quantum quenches that are studied in the rest of the paper.

\subsection{From the quantum Ising spin chain to the scaling Ising field theory}\label{subsec:scalingIFT}

The dynamics of the quantum Ising chain (QIC) is governed by the Hamiltonian
\begin{equation}
H_\text{QIC}=-J\left(\sum_{i=1}^{L}\sigma_{i}^{x}\sigma_{i+1}^{x}+h_{x}\sum_{i=1}^{L}\sigma_{i}^{x}+h_{z}\sum_{i=1}^{L}\sigma_{i}^{z}\right)\,,
\label{eq:QIM_Hamiltonian}
\end{equation}
where $\sigma_{i}^{\alpha}$ are Pauli matrices acting at site $i$,
$J>0$ and periodic boundary conditions $\sigma_{L+1}^{\alpha}\equiv\sigma_{1}^{\alpha}$ 
are imposed.
The parameters $h_{z}$ and $h_x$ are called transverse and longitudinal magnetic fields, respectively.

For $h_{x}=0$ the model (\ref{eq:QIM_Hamiltonian}) simplifies to 
the transverse field quantum Ising spin chain
that can be mapped exactly to free spinless Majorana fermions with the dispersion relation
\cite{1961AnPhy..16..407L,1970AnPhy..57...79P}
\begin{equation}
\epsilon(k)=2J\sqrt{1+h_{z}^{2}-2h_{z}\cos k}
\end{equation}
having a gap $\Delta=2J|1-h_{z}|$. At $h_z=1$ the system has a quantum critical point separating the ordered or ferromagnetic phase ($h_z<1$) and the disordered or paramagnetic phase ($h_z>1$). 
The order parameter is the longitudinal magnetisation $\langle \sigma_i^x\rangle$ that has 
a vanishing expectation value in the paramagnetic phase, while in the ferromagnetic phase 
\begin{equation}
\langle\sigma_i^x\rangle = (1-h_z^2)^{1/8}\,.
\end{equation}

In the scaling limit, $J\rightarrow\infty$ and $h_{z}\rightarrow1$
with 
\begin{equation}
\label{eq:M-hz}
M=2J\left|1-h_{z}\right|
\end{equation}
fixed, the model scales to a free Majorana fermion field theory
\begin{align}
H_\text{FF} &=\int_{-\infty}^{\infty}dx\,\frac{1}{2\pi}\left[\frac{i}{2}\left(\psi(x)\partial_{x}\psi(x)-\bar{\psi}(x)\partial_{x}\bar{\psi}(x)\right)-iM\bar{\psi(}x)\psi(x)\right]\,, \nonumber \\
 & \left\{ \psi(x,t),\bar{\psi}(y,t)\right\} =2\pi\delta(x-y)\:.
\label{eq:HamiltonianMajorana}
\end{align}
Here we used units in which the lattice spacing is $a=2/J$
and consequently the resulting speed of light is $c=1.$
The dispersion relation of the excitations becomes (upon the substitution $k=pa$) that of free relativistic particles
\begin{equation}
\epsilon(k)\to \sqrt{M^2+p^2}\,.
\end{equation}
We note that the distinction between the two phases is lost at the level of the field theory Hamiltonian
\eqref{eq:HamiltonianMajorana}. In fact, the two phases correspond
to realisations of the same Hamiltonian on different Hilbert spaces,
as described in Appendix \ref{sub:The-ferro/paramagnetic-phases}.

The longitudinal magnetic field $h_z$ in the field theory couples to the spin field $\sigma(x)$, 
which is the continuum limit of the longitudinal magnetization operator $\sigma^x_i$. 
Choosing the conformal field theory normalisation
\begin{equation}
\lim_{x\rightarrow\infty}|x|^{1/4}\langle0|\sigma(x)\sigma(0)|0\rangle=1
\end{equation}
fixes the relation between the lattice and the continuum field as 
\begin{equation}
\sigma(n a) =2^{1/12}e^{-1/8}\mathcal{A}^{3/2}J^{1/8} \sigma_n^x\equiv\bar s \,J^{1/8} \sigma_n^x\,,
\label{eq:sigmanorm}
\end{equation}
where $\mathcal{A}=1.2824271291\dots$ is Glaisher's constant.
The expectation value of the spin field in the ferromagnetic
phase is given by
\begin{equation}
\langle0|\sigma(x)|0\rangle\equiv\bar{\sigma}=\bar s M^{1/8}\,.
\label{eq:sigmabar}
\end{equation}

If the longitudinal field $h_x$ is non-zero, the lattice model  \eqref{eq:QIM_Hamiltonian} is non-integrable. The spectrum and the dynamics of the Hamiltonian depend strongly on the phase of the system set by $h_z.$ In the ferromagnetic phase $h_{z}<1$,
the fermions correspond to domain walls, and a longitudinal magnetic
fields leads to their confinement \cite{1978PhRvD..18.1259M}. This
is a non-perturbative effect which is drastic even for small values
of $h_{x}$, as discussed in more detail in Section \ref{sec:Non-integrable-quenches-in-FM}.
In the paramagnetic phase $h_{z}>1$, the elementary fermions correspond
to spin waves (magnons), and a small $h_{x}$ only introduces some
perturbative corrections to their masses and other properties.

The full scaling Ising field theory accounting for the nonzero longitudinal field has the Hamiltonian
\begin{equation}
H_\text{IFT}=\int_{-\infty}^{\infty}dx\,\left\{ \frac{1}{2\pi}\left[\frac{i}{2}\left(\psi(x)\partial_{x}\psi(x)-\bar{\psi}(x)\partial_{x}\bar{\psi}(x)\right)-iM\bar{\psi(}x)\psi(x)\right]+h\sigma(x)\right\} \,.
\label{eq:IFT_Hamiltonian}
\end{equation}
This Hamiltonian describes the universal behaviour of the spin chain \eqref{eq:QIM_Hamiltonian}
in the vicinity of the lattice quantum critical point $h_{z}=1$, $h_{x}=0$
on scales much larger than the lattice spacing $a$. The scaling relation between $h$ 
and $h_{x}$ is given by
\begin{equation}
\label{h-hx}
h = \frac2{\bar s}J^{15/8} \,h_x\,,
\end{equation}
where $\bar s$ is defined in Eq. \eqref{eq:sigmanorm}.
For later convenience we also introduce the dimensionless combination
\begin{equation}
\bar{h}=hM^{-15/8}=\frac{2^{-7/8}}{\bar s}(1-h_z)^{-15/8}\,h_x\,.
\label{eq:hbardef}
\end{equation}

In the ferromagnetic phase it is important to keep track of the relative orientation of 
the magnetic field $h$ with respect to the spontaneous magnetisation Since in 
our conventions the spontaneous magnetisation is positive: 
$\langle0|\sigma(x)|0\rangle>0$,
a field with $h>0$ is in the ``wrong'' direction in the sense that it counteracts 
the spontaneous magnetisation by increasing its energy cost.

\subsection{Quantum quenches in the spin chain and the field theory}

In this work we consider the setting of an instantaneous global quench \cite{2006PhRvL..96m6801C,2007JSMTE..06....8C} which consists of an abrupt change of a parameter in the
Hamiltonian. Here we study global quantum
quenches in which case both the pre-quench ($t<0$) Hamiltonian and
post-quench ($t>0$) Hamiltonian are translation invariant. 
As usual, the initial state at $t=0$ is taken to be 
the ground state of the pre-quench Hamiltonian which is then 
evolved by the post-quench Hamiltonian, although
more general choices for the initial state such as excited \cite{2014JPhA...47q5002B}
or thermal states \cite{2009EL.....8720002S} are also possible.

\subsubsection{Integrable and non-integrable quenches}\label{subsec:integrable_quenches}

\begin{figure}[t]
\centering{}\includegraphics[width=0.5\textwidth]{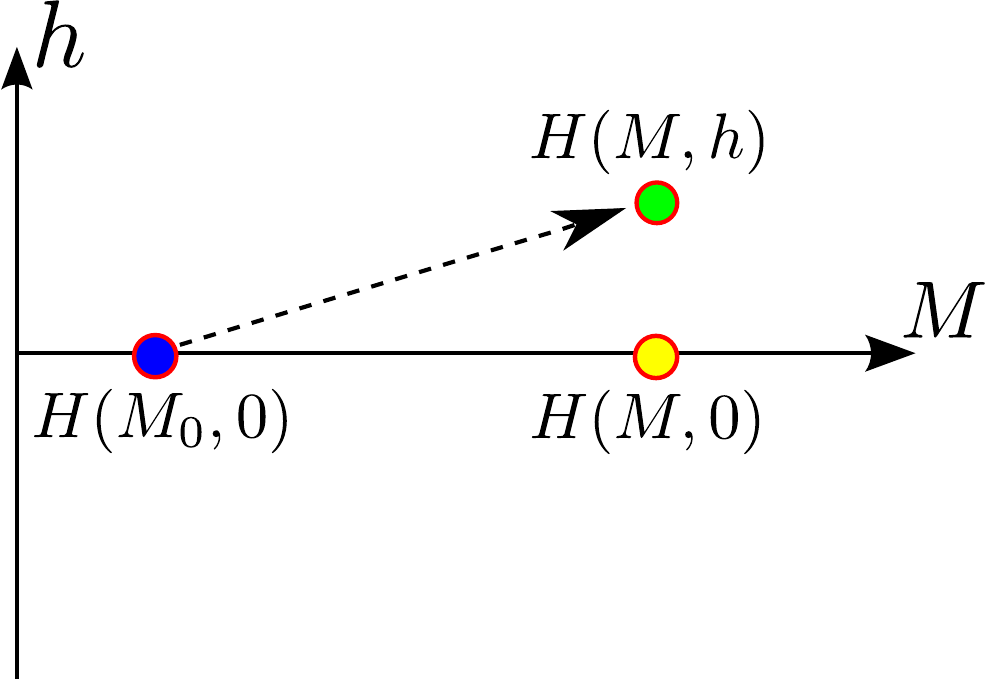}
\caption{Schematic illustration of the quenches in the parameter 2D space spanned by the
transverse and longitudinal couplings. To describe the quench from the Hamiltonian 
$H(M_{0},0)$ (blue dot) to the Hamiltonian $H(M,h)$ (green dot), one can 
use either the basis defined by the eigenvectors of pre-quench Hamiltonian $H(M_0,0)$
(blue dot) or the ``intermediate'' Hamiltonian $H(M,0)$ (yellow dot), both of which 
correspond to free fermions.}
\label{fig:Quench protocol basis} 
\end{figure}

For the quantum Ising chain (\ref{eq:QIM_Hamiltonian}), a special class 
of quenches corresponds to abruptly
changing $h_{z}$ while keeping $h_{x}=0$. The dynamics of these
{\it integrable quenches} can be studied analytically using free fermion techniques
both on the chain \cite{2011PhRvL.106v7203C,2012JSMTE..07..016C,2012JSMTE..07..022C}
and in the field theory \cite{2012JSMTE..04..017S}. The time evolution of several
quantities have been computed; we quote these results when needed
in the sequel.

Another scenario that we consider is {\it breaking integrability} during
the quench, in our case starting with $h_{x}=0$ and switching on
a nonzero $h_{x}$ while also changing $h_{z}$; these quenches are called 
{\it non-integrable}. In the ferromagnetic phase, integrability breaking
leads to an interesting dynamics due to the onset of confinement
\cite{2016arXiv160403571K}; some of the quenches considered here correspond
to this case. However, the dynamics of confinement is most explicitly
manifested in the evolution of the two-point function and entanglement
entropy, which is not considered in this work due to the limitations
of the truncated Hamiltonian method. On the other hand, an aspect
of dynamical confinement is the domination of the power spectrum by
the mesonic excitations which can be studied in detail with our method
thereby complementing the results of \cite{2016arXiv160403571K}.

In the context of the Ising field theory Hamiltonian \eqref{eq:IFT_Hamiltonian},
the quenches we consider correspond to the parameter change 
(cf. Fig. \ref{fig:Quench protocol basis})
\begin{equation}
(M_{0},h_{0}=0)\longrightarrow(M,h)\,.
\end{equation}
These quenches are integrable when $h=0$ and break integrability
otherwise. The initial state $|\Psi_{0}\rangle$ is taken to be the ground
state of the pre-quench system governed by the Hamiltonian $H_{0}$
which is $H_\text{IFT}$ in \eqref{eq:IFT_Hamiltonian} with parameters
$M=M_{0}$ and $h=0$, and the time evolution is governed by the post-quench
Hamiltonian $H$ identical to $H_\text{IFT}$ in \eqref{eq:IFT_Hamiltonian}, thus
\begin{equation}
|\Psi(t)\rangle=e^{-iHt}|\Psi_{0}\rangle\,.
\label{eq:time_evolution_psi}
\end{equation}
The above time evolution is numerically simulated using a truncated Hamiltonian
scheme which is described in Sec. \ref{sec:The-truncated-fermionic}. Before turning 
to the numerical simulation, let us first recall the general expectations regarding 
the stationary state.

\subsubsection{Stationary state and diagonal ensemble \label{sub:Stationary-state-and-diagonal-ensemble}}

After a sudden quantum quench, the system is expected to evolve towards a stationary state which is described by the so-called diagonal ensemble. Writing the initial state in terms of eigenstates $|\alpha\rangle$ of the post-quench 
Hamiltonian as
\begin{equation}
|\Psi_{0}\rangle  =\sum_{\alpha}C_{\alpha}|\alpha\rangle\, ,
\end{equation}
the time evolution of the density matrix is given by
\begin{equation}
\rho(t)=|\Psi(t)\rangle\langle\Psi(t)|=\sum_{\alpha,\beta}C_{\alpha}C_{\beta}^{*}e^{-i(E_{\alpha}-E_{\beta})t}|\alpha\rangle\langle\beta|\,.
\end{equation}
The amplitudes $C_{\alpha}$ are called overlaps and together with the post-quench energies determine the 
post-quench time-evolution of the system.

The post-quench time evolution of an observable $\mathcal{O}$ is given by
$\left\langle \mathcal{O}(t)\right\rangle =\mbox{Tr}\left(\rho(t)\mathcal{O}\right)$,
while the stationary value coincides with the infinite time average 
\begin{equation}
\left\langle \mathcal{O}\right\rangle _\text{DE}=\sum_{\alpha}\left|C_{\alpha}\right|^{2}\langle\alpha|\mathcal{O}|\alpha\rangle\,.
\label{eq:DE}
\end{equation}
The sum \eqref{eq:DE} can be viewed as an ensemble average over the diagonal ensemble density matrix
\begin{equation}
\rho_\text{DE}=\sum_{\alpha}\left|C_{\alpha}\right|^{2}|\alpha\rangle\langle\alpha|\,.
\label{eq:DE_rho}
\end{equation}
It is expected that a general quench results in a stationary state
which is equivalent to a Gibbs ensemble for local observables $\mathcal{O}$, so
\begin{equation}
\mbox{Tr}\rho_\text{DE}\mathcal{O}=\mbox{Tr}e^{-\beta H}\mathcal{O}\,.
\end{equation}
This is guaranteed if the so-called eigenstate thermalisation hypothesis
\cite{1991PhRvA..43.2046D,1994PhRvE..50..888S,2008Natur.452..854R}
is valid. For integrable models, the stationary point cannot be described
by a Gibbs ensemble \cite{2007PhRvL..98e0405R}; the upshot of recent
investigations \cite{2014PhRvL.113k7202W,2014PhRvL.113k7203P,2014JSMTE..12..009B,2015JSMTE..04..001M,2015PhRvL.115o7201I}
is that a suitably defined generalisation must be introduced. 

For integrable quenches in the Ising model, exact results for
the time evolution and stationary properties have been obtained by
direct calculation \cite{2011PhRvL.106v7203C,2012JSMTE..04..017S,2012JSMTE..07..016C,2012JSMTE..07..022C};
these are used to check the validity of the TFSA in Section \ref{sec:Integrable-quenches}.

\section{The truncated fermionic space approach \label{sec:The-truncated-fermionic}}

The main idea underlying all the variants of the Hamiltonian truncation is to split 
the Hamiltonian into a solvable part $H_0$ (usually a free theory or a conformal 
field theory) and a perturbation $H_1$, which is assumed to be relevant in the 
renormalisation group sense. In order to have a discrete spectrum, the system 
is put in a finite volume. Using the basis spanned by eigenstates of $H_0$,  
the matrix elements of $H_1$ and consequently the full Hamiltonian matrix can be computed 
exactly. As the spectrum of a field theory is unbounded from above, the basis 
must be truncated at some cut-off energy value, yielding a finite dimensional matrix. 
Numerical diagonalisation of this matrix gives an approximate spectrum, and expectation 
values or matrix elements can be computed in a straightforward manner. 

The effectiveness of this method is explained by  renormalisation group considerations.
For a relevant perturbation the effective coupling goes to zero at high energies,  
and the high energy 
states affect only weakly the low energy properties. In any practical simulation, 
this can, and indeed should, be checked 
numerically by gradually raising the energy cut-off. Moreover, 
the cut-off dependence is well-understood by now through a
renormalisation group in terms of the energy cut-off
\cite{2006hep.th...12203F,2007PhRvL..98n7205K,2015PhRvD..91b5005H}. 
We emphasise that truncated Hamiltonian approaches do not rely on the 
integrability of the full Hamiltonian, and can be applied equally 
well to gapped (massive) and gapless (massless) systems. 

\subsection{Truncating the Hilbert space \label{sub:Truncating-the-Hilbert}}

In setting up the truncated Hamiltonian approach we follow \cite{2001hep.th...12167F}
and use the eigenstates of the free fermion theory \eqref{eq:HamiltonianMajorana}
in finite volume as a basis, truncated at a certain value of energy;
this is called the truncated fermionic space approach (TFSA). Since the
initial state has zero momentum and the Hamiltonian \eqref{eq:IFT_Hamiltonian} 
conserves momentum, it is sufficient to include states
with zero total momentum only.

Due to the finite size of the system it is necessary to specify boundary
conditions. The periodic boundary conditions on the chain allow for
periodic or anti-periodic boundary conditions for the Majorana fermion
field, splitting the Hilbert space into the Ramond (periodic) and
the Neveu--Schwarz (anti-periodic) sectors (see Appendix \ref{sub:The-ferro/paramagnetic-phases}).

The eigenstates of the free fermion Hamiltonian \eqref{eq:HamiltonianMajorana} 
can be constructed using the mode expansion 
\begin{align}
\psi(x,t) & =  \sqrt{\frac{\pi}{L}}\sum_{n}\frac{e^{\theta_{n}/2}}{\sqrt{\cosh\theta_{n}}}\left(\omega a(\theta_{n})e^{ip_{n}x-iE_{n}t}+\bar{\omega}a^{\dagger}(\theta_{n})e^{-ip_{n}x+iE_{n}t}\right),\nonumber \\
\bar{\psi}(x,t) & =  -\sqrt{\frac{\pi}{L}}\sum_{n}\frac{e^{-\theta_{n}/2}}{\sqrt{\cosh\theta_{n}}}\left(\bar{\omega}a(\theta_{n})e^{ip_{n}x-iE_{n}t}+\omega a^{\dagger}(\theta_{n})e^{-ip_{n}x+iE_{n}t}\right),
\label{eq:modeexp}
\end{align}
where 
\begin{equation}
\left\{ a(\theta_{n}),a^{\dagger}(\theta_{n'})\right\} =\delta_{n,n'}\:.
\end{equation}
The momentum and energy of the quasi-particles are quantised as
\begin{subequations}
\label{eq:PE}
\begin{align}
p_{n} & =  M\sinh\theta_{n}=\frac{2\pi n}{L}\:,\\
E_{n} & =  M\cosh\theta_{n}\:,\label{eq:finvol_rapidity}
\end{align}
\end{subequations}
where $n$ is the momentum quantum number and we also introduced the
rapidity parameter $\theta$ for later convenience. 
The  quantum number $n$ is integer in the Ramond sector and half-integer
in the Neveu--Schwarz sector, and we use the phase factors $\omega=e^{i\pi/4}$
and $\bar{\omega}=e^{-i\pi/4}$ following the conventions of \cite{2001hep.th...12167F}.
All
physical quantities are measured in units of the post-quench fermion
mass $M$: energies are simply given in units of $M$ while volume
is parameterised by the dimensionless variable $ML$.

Using the mode expansion \eqref{eq:modeexp} and the rapidity variables in \eqref{eq:PE}, the free fermion Hamiltonian \eqref{eq:HamiltonianMajorana} can be
rewritten as 
\begin{equation}
H_{\textrm{FF}}=M\sum_{n}\cosh(\theta_{n})a^{\dagger}(\theta_{n})a(\theta_{n})\,.
\end{equation}
The eigenvectors of this Hamiltonian are given as
\begin{equation}
\label{eq:Fock}
|\theta_{1},\theta_{2},...,\theta_{n}\rangle  =  a^{\dagger}(\theta_{1})a^{\dagger}(\theta_{2})...a^{\dagger}(\theta_{n})|0\rangle\,,
\end{equation}
with energy eigenvalues $M\sum_{n}\cosh(\theta_{n})$, and the vacuum defined by $a(\theta)|0\rangle=0$. 

The truncation of the Hilbert space is performed by keeping 
states with energies less than an imposed cut-off $\Lambda$:
\begin{equation}
\label{eq:TFSAcut-off}
M\sum_{n}\cosh(\theta_{n}) \le \Lambda\,.
\end{equation}
The number of states retained is denoted by $N_\text{cut}.$

For the construction of the matrix elements of the Hamiltonian (\ref{eq:IFT_Hamiltonian}) and the calculation of expectation values we need the finite volume matrix elements of the energy operator 
\begin{equation}
\varepsilon(x)=i:\bar{\psi}(x)\psi(x):
\label{eq:energy_operator}
\end{equation}
and the magnetisation operator $\sigma(x)$. 
The matrix elements of $\varepsilon(x)$ can be computed in a straightforward way from the mode expansion, 
while those of $\sigma(x)$ were given in \cite{2001hep.th...12167F}; for the sake of 
completeness and in order to fix our conventions they are summarised in Appendix \ref{sub:sigmaFF}.

\subsection{Choice of the TFSA basis}
\label{sec:TFSAbasis}

In order to improve the accuracy of the TFSA, one can optimise the
choice of basis when implementing the quench from $H(M_{0},h=0)$ to
$H(M,h)$ so that the initial state $|\Psi_{0}\rangle$ and the post-quench
Hamiltonian $H$ are both represented with the highest possible precision.
This is best achieved by using the eigenbasis of the Hamiltonian $H(M,0)$,
i.e. the free fermion theory closest to the actual post-quench Hamiltonian
(see Fig. \ref{fig:Quench protocol basis}). 

For integrable quenches from $H(M_0,0)$ to $H(M,0)$ the above choice of basis 
implies that the time evolution operator $e^{-iH(M,0)t}$ is diagonal and exact apart from the truncation.
The overlaps between the ground state of $H(M_0,0)$ with the eigenstates 
of $H(M,0)$ are exactly known (cf. Eq. \eqref{eq:initialStateExpansion}). 
The accuracy of the TFSA can be verified by numerically computing the ground 
state of the pre-quench Hamiltonian written as
\begin{equation}
H(M_0,0)=H(M,0)+(M_0-M)\int dx\, \epsilon(x)
\label{eq:prequenchHam}\end{equation}
in the post-quench basis and comparing its overlaps with the basis states 
to the exact predictions. As discussed below, we found that the TFSA approach 
reproduces these overlaps to a good accuracy for quenches of moderate size. 

For non-integrable quenches from $H(M_{0},h=0)$ to $H(M,h)$ we use these exact overlaps when expanding the initial state in the eigenbasis of $H(M,0).$ The matrix of the post-quench Hamiltonian,
\begin{equation}
H(M,h)=H(M,0)+h\int dx\,\sigma(x)\,,
\end{equation}
is also computed in the eigenbasis of $H(M,0).$

\subsection{TFSA implementation of the time evolution}

For integrable quenches to $H(M,0)$, the time evolution operator $e^{-iH(M,0)t}$ is diagonal 
in the eigenbasis of $H(M,0)$ we use and can be trivially calculated. This is not the case for 
non-integrable quenches to $H(M,h\neq0).$ For these quenches the action of $e^{-iH(M,h)t}$ on the initial state can be evaluated efficiently using an expansion in terms of Chebyshev polynomials, which is known to give 
the best finite order polynomial approximation of continuous functions on a 
finite interval. More details are given in Appendix \ref{sub:Implementing-the-quench}.  

As we are are principally interested in the infinite volume behaviour, the 
simulation time is bounded by $t=L/2,$ i.e. by the time needed for the fastest 
excitations moving along the light cone to travel half-way around the circle 
and meet again. For $t>L/2$ the behaviour of observables is affected by partial revivals absent in the infinite volume case. Note that numerical errors accumulated during the TFSA 
simulation may in principle invalidate the results before this time is reached.

\section{Integrable quenches \label{sec:Integrable-quenches}}

Before applying the TFSA to non-integrable quenches, we first study 
integrable quenches from $H(M_0,0)$ to $H(M,0)$. 
As discussed in Section \ref{subsec:integrable_quenches}, there is a large number of 
analytic results for these quenches, therefore they can serve as a benchmark of the 
TFSA simulation. Following the  discussion in Section \ref{sec:TFSAbasis}, we work 
in the basis of eigenstates 
of the post-quench free fermion Hamiltonian $H(M,0)$ and find the initial state as the ground 
state of the pre-quench Hamiltonian $H(M_0,0)$ that is numerically computed in 
this basis. The time evolution is trivial to implement since $e^{-iH(M,0)t}$ is diagonal
and its eigenvalues are exactly known. We test the TFSA by comparing its results to analytic expressions for the overlaps and for the post-quench dynamics of the energy operator, the magnetisation, and the Loschmidt echo. 

\subsection{Time evolution of the energy operator}

\begin{figure}[t!]
\subfloat[Quench from $M_{0}=1.5M$ to $M$.]{\includegraphics[width=.5\textwidth]{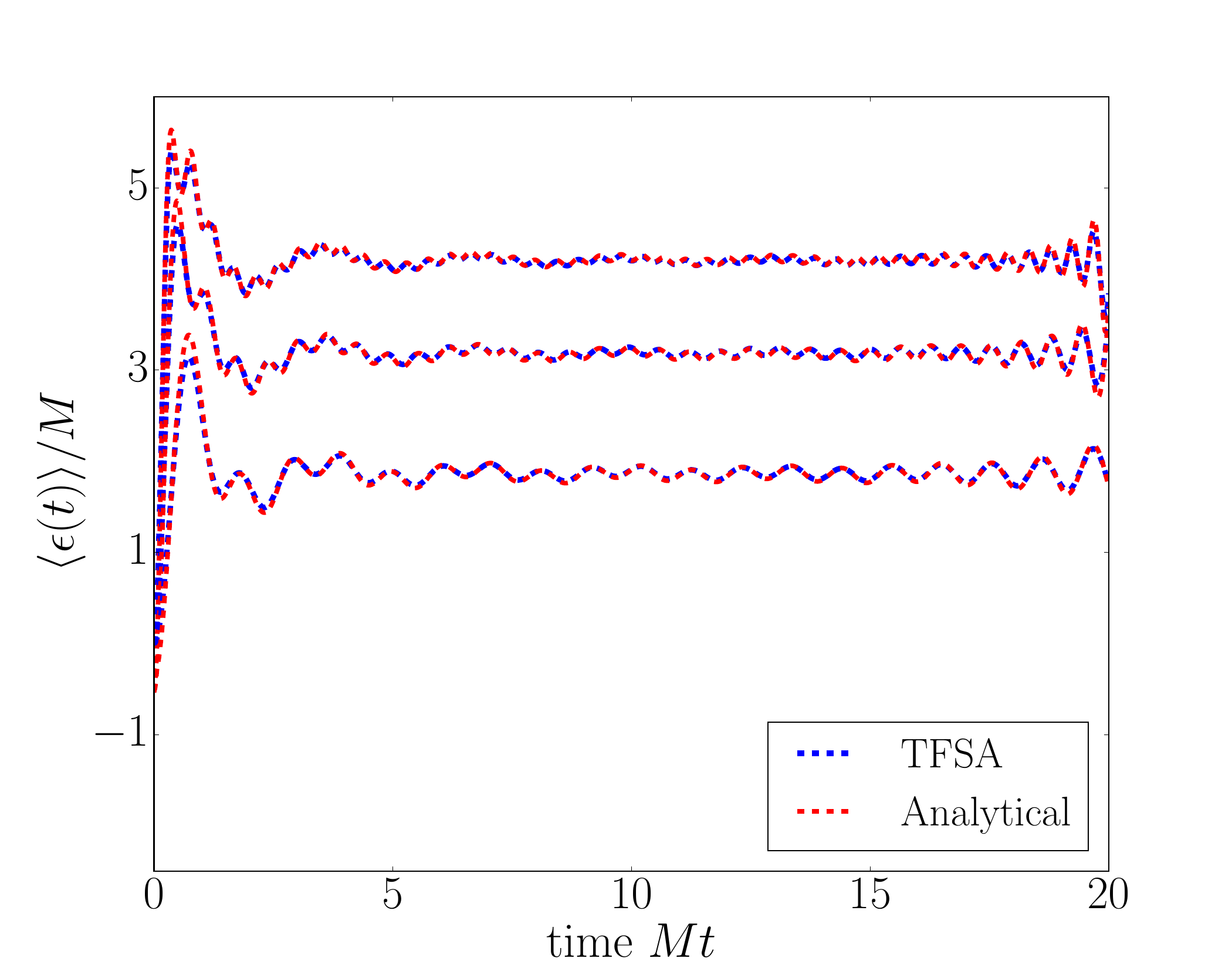}\label{fig:epsilonOK}}
\subfloat[Quench from $M_{0}=3M$ to $M$.]{\includegraphics[width=.5\textwidth]{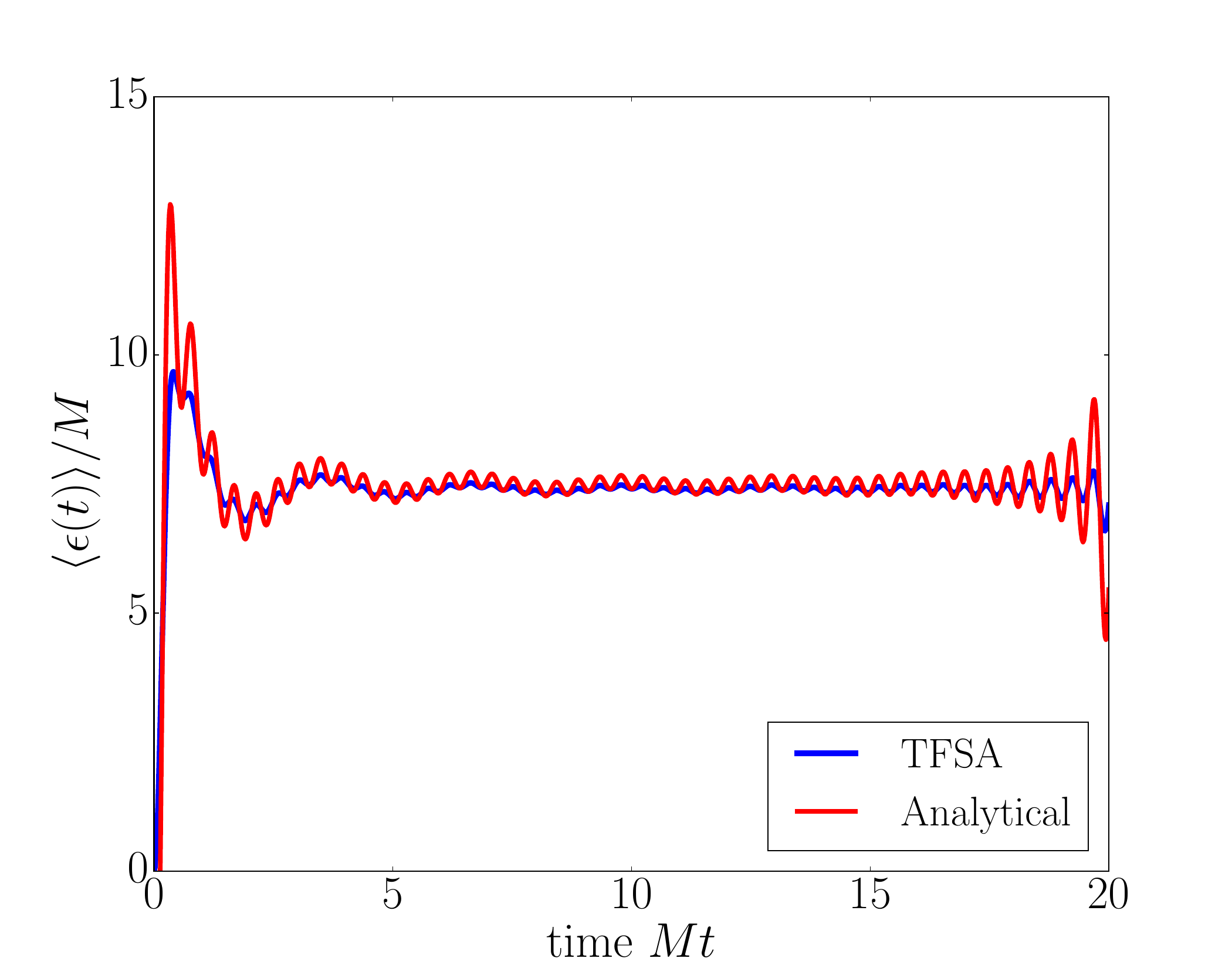}\label{fig:epsilonfail}}
\caption{\label{fig:epsilon} Time-dependent expectation value of the energy
operator $\varepsilon=i\!:\!\bar{\psi}\psi\!:$
after integrable mass quenches in the ferromagnetic phase at system size $ML=40.$ Left panel: $M_0=1.5M$ with cut-offs (from bottom to top) $\Lambda=6M,10M,14M.$ 
The TFSA results are shown in blue while the analytical result \eqref{eq:epsilontheory} is plotted in red. Right panel: $M_0=3M$ with cut-off $\Lambda=14M.$ 
}
\end{figure}

We start our investigation of the out of equilibrium dynamics of various observables by studying the so-called energy operator, $\varepsilon(x)=i:\bar{\psi}(x)\psi(x):$, which is the continuum limit of the transverse magnetisation
on the lattice. The known analytical 
result on the lattice \cite{2012JSMTE..07..022C}
has the scaling limit 
\begin{equation}
\left\langle \varepsilon(t)\right\rangle =\langle\Psi(t)|\varepsilon|\Psi(t)\rangle=-\sum_{n}\left(\cos\beta_{n}\cos\Delta_{n}+\sin\beta_{n}\sin\Delta_{n}\cos(2E_{n}t)\right),
\label{eq:epsilontheory}
\end{equation}
where 
\begin{align}
\beta_{n} & =  \eta\arctan\left(\sinh\theta_{n}\right)\,, \nonumber\\
\Delta_{n} & =  \beta_{n}-\eta\arctan\left(\frac{M}{M_{0}}\sinh\theta_{n}\right)\,.
\end{align}
with $\eta=\pm1$ in the paramagnetic and ferromagnetic phase, respectively.
In \eqref{eq:epsilontheory}, the summation over momentum quantum numbers is logarithmically divergent but
the expression can be evaluated by imposing a momentum cut-off. In order to compare the TFSA simulations with this analytical formula, we regularise the expression by introducing a cut-off $n_\text{cut}$ for the momentum quantum numbers such that
\begin{equation}
2E(p=2\pi n_\text{cut}/L)=2\sqrt{M^2+(2\pi n_\text{cut}/L)^2}<\Lambda\,.
\end{equation}
Here the factor of $2$ reflects the fact that the initial state consists of particle pairs. This is a sort of a ``single particle'' cut-off and as 
such it does not exactly correspond to the ``many-body'' TFSA truncation specified in \eqref{eq:TFSAcut-off}. 
In spite of this mismatch, for quenches of moderate size a 
very good agreement is observed, as shown in Fig. \ref{fig:epsilonOK}. 
For large quenches, as shown in Fig. \ref{fig:epsilonfail}, the analytical 
formula yields high frequency oscillations 
that are present in the result of the TFSA simulation, but with a 
much smaller amplitude. However, apart from this 
mismatch the agreement with the exact result is still quite satisfactory.

It is immediately apparent from Fig. \ref{fig:epsilon} that the
frequency of the high-frequency oscillations is related to the cut-off. 
Numerical checks reveal that their frequencies are indeed very close to $\Lambda$. 
These oscillations are present
in the numerical TFSA results obtained for $\langle\varepsilon(t)\rangle$ and the Loschmidt
echo in the integrable quenches; for the non-integrable quenches some
remnants of them survive depending on the dynamics as discussed later. 
The cut-off related oscillations can be suppressed
by the Gaussian smoothening procedure described in Appendix \ref{sub:Cut-off-extrapolation-schemes}.
For the energy operator, however, we do not apply smoothening as the theoretical prediction itself involves a cut-off due to the logarithmic ultraviolet divergence and consequently it also features these oscillations.
The logarithmic divergence is also manifest in the data shown in Fig.
\ref{fig:epsilon}: apart from the cut-off related oscillations, $\langle\varepsilon(t)\rangle$
attains a stationary value for $Mt\gtrsim5$, but this value steadily increases
with the cut-off. Due to this divergence it makes no
sense to perform a cut-off extrapolation in this case, unlike for other observables.

\subsection{Statistics of work and the Lochsmidt echo}

\begin{figure}[t!]
\subfloat[Quench from $M_{0}=0.5M$ to $M$.]{\includegraphics[width=.5\textwidth]{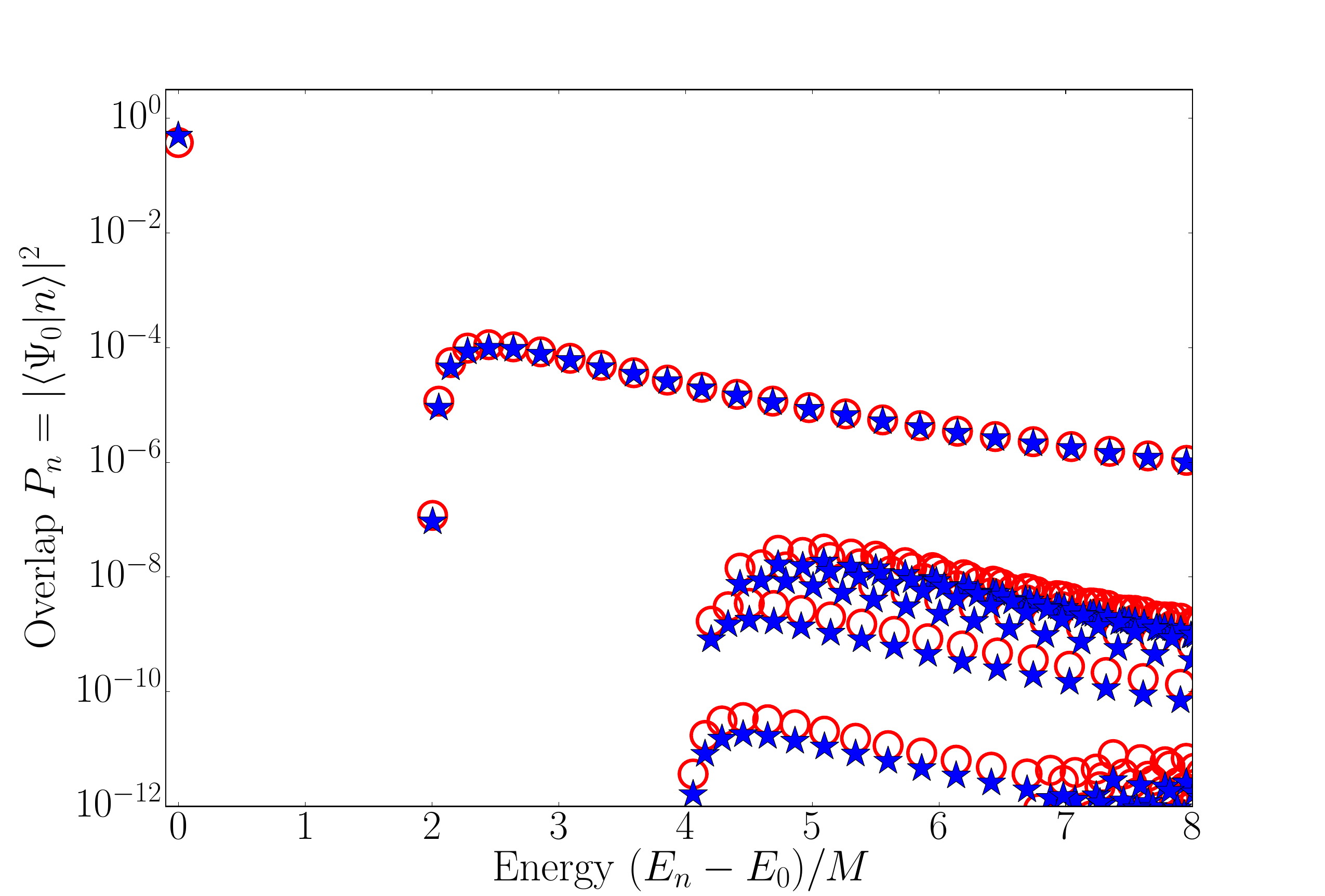}\label{fig:statistics_of_work_0.5}}
\subfloat[Quench from $M_{0}=1.5M$ to $M$.]{\includegraphics[width=.5\textwidth]{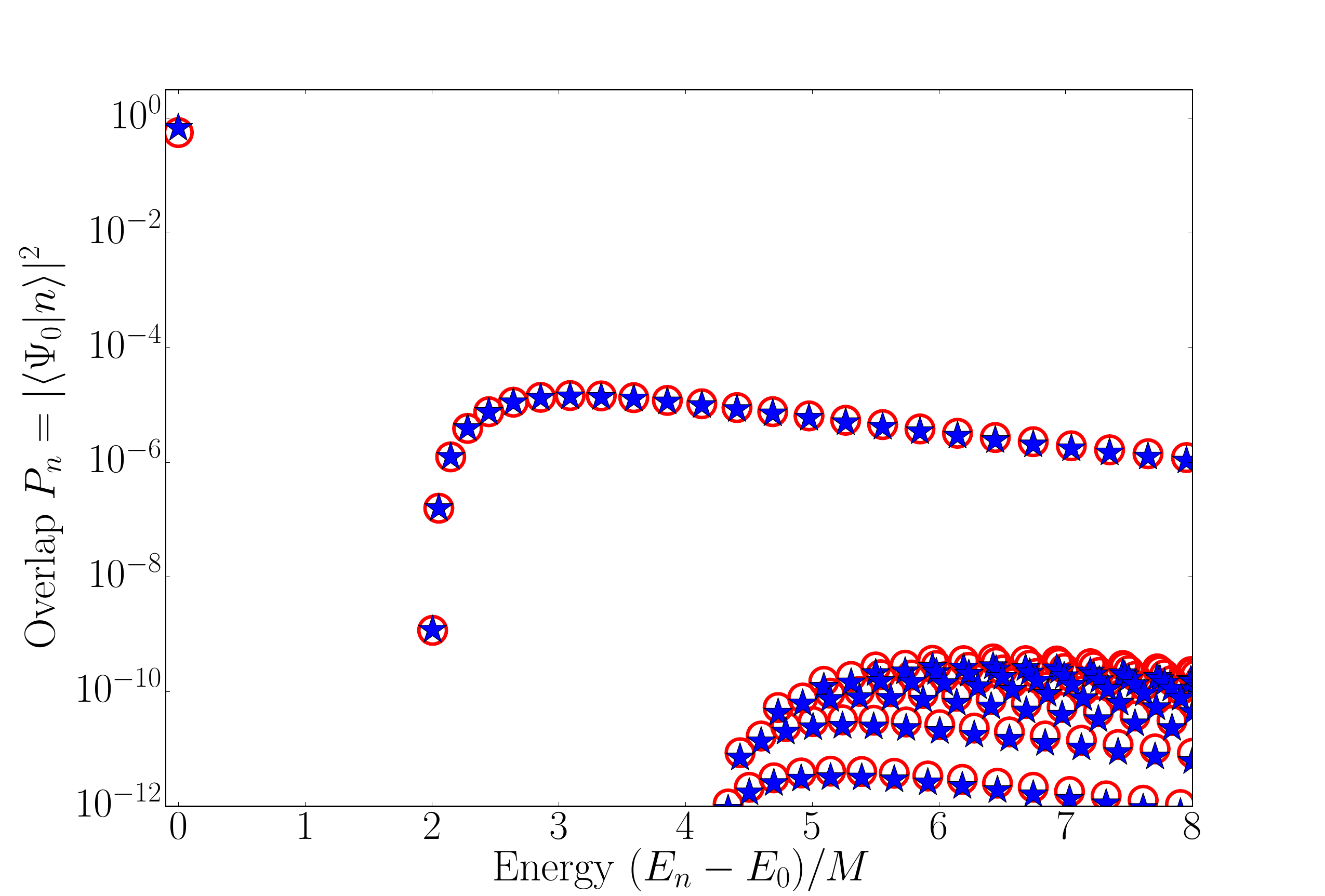}\label{fig:statistics_of_work_1.5}}
\caption{
The statistics of work $P(W)$ for integrable mass quenches within the ferromagnetic
phase. Here, for convenience, the work is defined relative to the post-quench ground state as opposed to Eq. \eqref{eq:PW}, which amounts to a horizontal shift by $E_0-\mathcal{E}_0.$ The dimensionless length
of the system is $ML=40$. The blue stars show the result of the TFSA
with cut-off $\Lambda=14M$ while the red open circles denote the
theoretical result corresponding to Eq. \eqref{eq:initialStateExpansion}.}
\label{fig:statistics_of_work}  
\end{figure}

The statistics of work is an important characteristic of quantum quenches
which gives the distribution of work performed during the 
quench \cite{2008PhRvL.101l0603S}. 
For a system with discrete energy levels, the probability of performing work $E_{\alpha}-\mathcal{E}_{0}$ is 
\begin{equation}
\label{eq:PW}
P(W) = \sum_n \delta[W-(E_n-\mathcal{E}_0)]\left|\langle\Psi_{0}|n\rangle\right|^{2}\,,
\end{equation}
where the initial state $|\Psi_{0}\rangle$ is the ground state of the pre-quench Hamiltonian,
\begin{equation}
H_{0}|\Psi_{0}\rangle=\mathcal{E}_{0}|\Psi_{0}\rangle\:,
\end{equation}
and $|\alpha\rangle$ are the eigenstates of the post-quench Hamiltonian with energies $E_n.$ 
In other words, the statistics of work is given as the set of pairs $\left(E_n,P_n\right)$ 
of possible energy values and their respective probabilites 
$P_n=\left|\langle\Psi_{0}|n\rangle\right|^{2}.$
 
The overlaps appearing in \eqref{eq:PW} are exactly known for integrable mass quenches $(M_0,0)\to (M,0)$, 
which are continuum limits of quenches of the transverse magnetic field in the Ising quantum spin chain. 
The components of the initial state in the Ramond and Neveu--Schwarz sectors can be expanded in terms of the Fock eigenvectors \eqref{eq:Fock} of the post-quench 
Hamiltonian as \cite{2012JSMTE..04..017S,2012JSMTE..07..016C,2012JSMTE..07..022C}
\begin{align}
|\Psi_{0}\rangle_\text{NS} &=\mathcal{N}_\text{NS}\,\exp\left\{ i\sum_{n\in \mathbb{Z} + \frac{1}{2}}K\left(\theta_{n},M,M_{0}\right)\,a^{\dagger}\left(\theta_{n}\right)a^{\dagger}\left(-\theta_{n}\right)\right\}|0\rangle_\text{NS} \,,\nonumber\\
|\Psi_{0}\rangle_\text{R} &=\mathcal{N}_\text{R}\,\exp\left\{ i\sum_{n\in \mathbb{Z}}K\left(\theta_{n},M,M_{0}\right)\,a^{\dagger}\left(\theta_{n}\right)a^{\dagger}\left(-\theta_{n}\right)\right\}|0\rangle_\text{R} \,,\nonumber\\
K\left(\theta,M,M_{0}\right)  &=\tan\left[\frac{1}{2}\arctan\left(\sinh\theta\right)-\frac{1}{2}\arctan\left(\frac{M}{M_0}\sinh\theta\right)\right]\:,
\label{eq:initialStateExpansion}
\end{align}
where the Ramond component only exists for quenches starting in the ferromagnetic  phase
and the normalisation factors are given by
\begin{equation}
\mathcal{N}_\text{R(NS)} = \prod_{n\in\mathbb{Z}\;(\mathbb{Z}+\frac12)}\left(1+|K(\theta_n,M,M_0)|^2\right)^{-1/2}\;.
\label{eq:normalisation_factor}\end{equation}
The initial state is thus a combination of zero-momentum particle pairs. In the paramagnetic phase the ground state only has components in the Neveu--Schwarz sector. In the ferromagnetic phase in infinite volume the ground state is doubly degenerate, but in finite volume this degeneracy is split and the eigenstates of the Hamiltonian have zero magnetisation. In order to study the time evolution starting from a state with finite magnetisation, the initial state must be chosen as a linear combination of the lowest energy states in the Ramond and Neveu--Schwarz sectors with equal weights: 
\begin{align}
|\Psi_{0}\rangle_\text{para} &= |\Psi_{0}\rangle_\text{NS} \nonumber\,,\\
|\Psi_{0}\rangle_\text{ferro} &= \frac{|\Psi_{0}\rangle_\text{NS} + |\Psi_{0}\rangle_\text{R}}{\sqrt{2}}\:.
\label{eq:InitialStateFerroPara}
\end{align}

The reliability of the TFSA can be checked by comparing the numerically obtained work statistics 
to the analytical result following from expansion \eqref{eq:initialStateExpansion} in 
Fig. \ref{fig:statistics_of_work}. Within the volume and cut-off range considered, the ground state always 
has the largest overlap with the initial state, while the arcs in the figure correspond to the 2-particle, 4-particle 
etc. states. 
The TFSA is found to reproduce the statistics of work well for moderate quenches. Moreover, raising the energy 
cut-off improves the accuracy of the overlaps given by the TFSA. Note that even though the ratios of 
the overlaps were perfectly reproduced by the TFSA, the truncation of the Hilbert space 
changes the overall normalisation of the overlaps compared to the exact prediction due to the omitted states. 

A further check is provided by the Loschmidt echo or fidelity defined as
\begin{equation}
\mathcal{L}\left(t\right)=|\langle\Psi_{0}|e^{-iHt}|\Psi_{0}\rangle|^{2}\,.
\end{equation}
Physically, $\mathcal{L}(t)$ is the probability
of observing the system in the initial state at time $t$ after the
quench. It is also the modulus square of the characteristic function of the work statistics
(\ref{eq:PW}) 
\begin{equation}
\mathcal{L}(t)=\left|\int dW P(W) e^{-iWt}\right|^2=
\left|\sum_{\alpha}|\langle\Psi_{0}|n\rangle|^{2}e^{-iE_nt}\right|^{2}=\left|\sum_nP_ne^{-iE_nt}\right|^{2}\,.
\label{eq:Loschmidt_exact}
\end{equation}

The Loschmidt echo is known analytically after an instantaneous
quench of the transverse field on the lattice \cite{2008PhRvL.101l0603S}
which in the scaling limit becomes
\begin{equation}
\mathcal{L}(t)=\left|\exp\left(\sum_{n}\log\left[\frac{1+|K(\theta_{n},M,M_0)|^{2}e^{-2iE_{n}t}}{1+|K(\theta_{n},M,M_0)|^{2}}\right]\right)\right|^{2},\label{eq:Lecho_prediction}
\end{equation}
where $K(\theta,M,M_0)$ is the kernel function in the expansion of the initial state (\ref{eq:initialStateExpansion}).

\begin{figure}
\subfloat[Quench from $M_{0}=0.5M$ to $M$.]{\includegraphics[width=.5\textwidth]{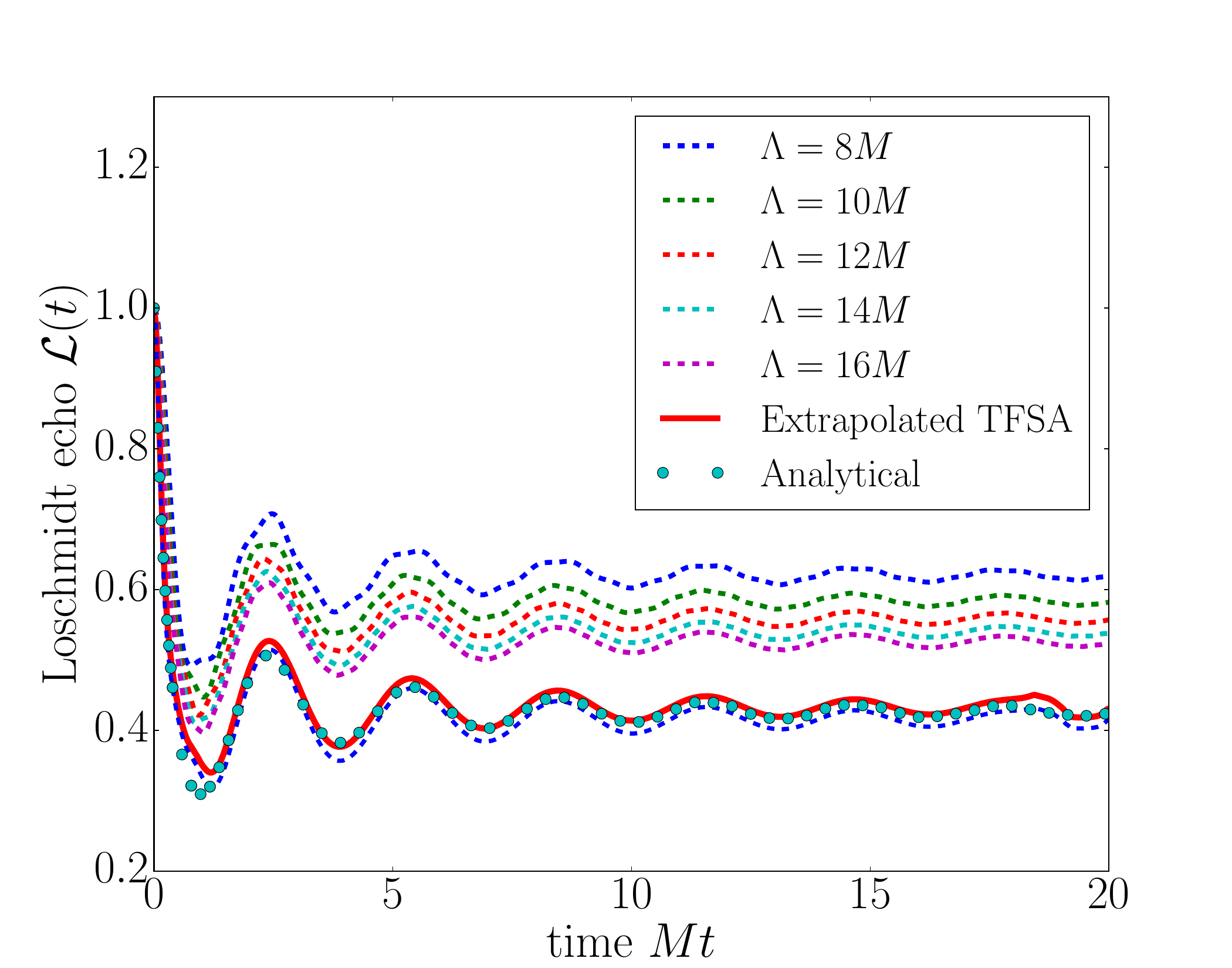}\label{fig:loschmidt_echo_m0_0,5}}
\subfloat[Quench from $M_{0}=1.5M$ to $M$.]{\includegraphics[width=.5\textwidth]{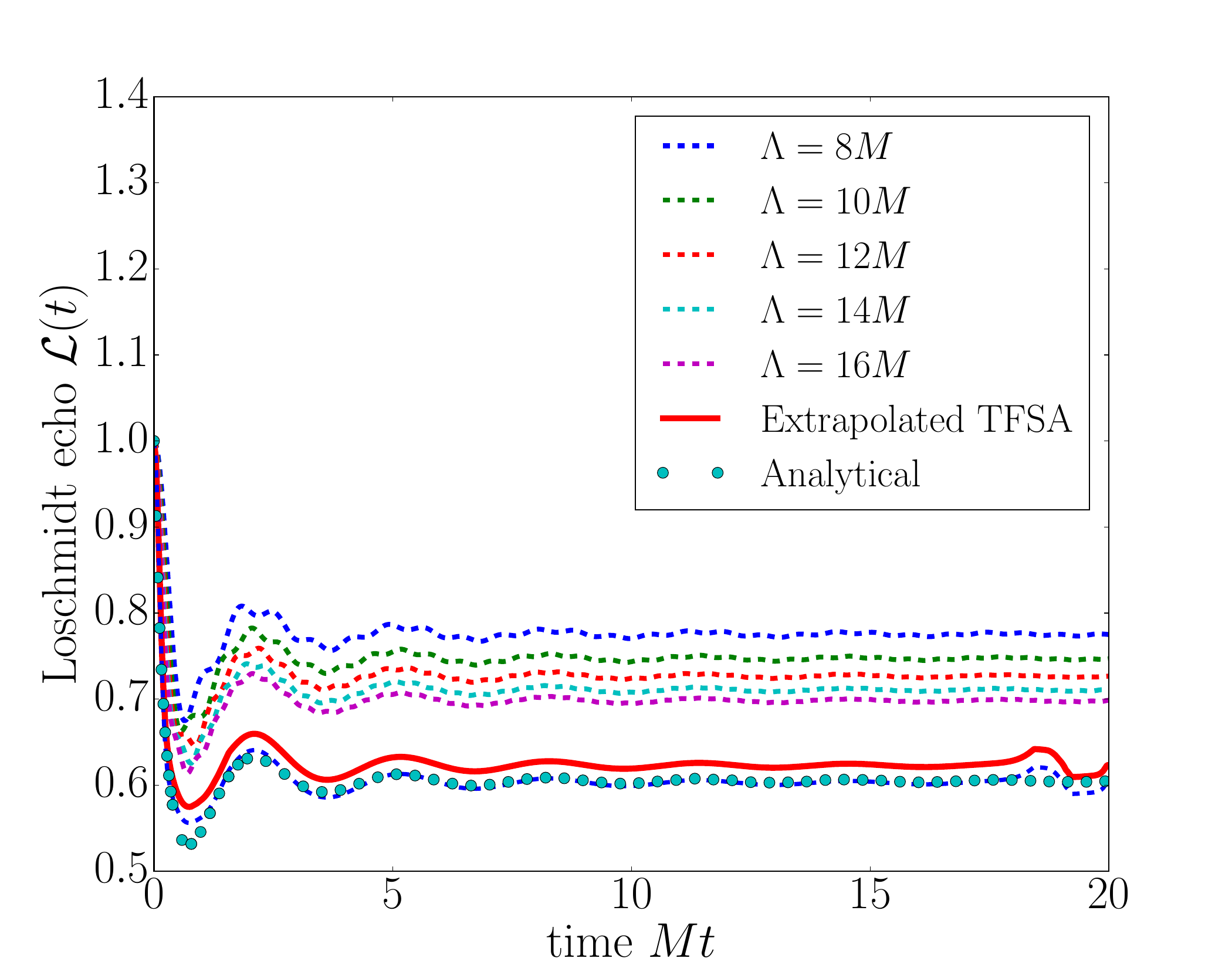}\label{fig:loschmidt_echo_m0_1,5}}
\caption{
The Loschmidt echo $\mathcal{L}(t)$ for integrable mass quenches in the ferromagnetic
phase 
%with zero magnetic field
for system size $ML=40$. The dashed lines are raw TFSA data, the solid red line
is the cut-off extrapolated TFSA result, while the exact analytical result
\eqref{eq:Lecho_prediction} is shown in green dots. The thin dashed line shows the TFSA result obtained by using the exact overlaps. 
The small deviations towards the end of the curve are artifacts of the smoothening procedure.}
\label{fig:loschmidt_echo_ferro}
\end{figure}

The TFSA results for $\mathcal{L}(t)$ are plotted against the analytical result \eqref{eq:Loschmidt_exact} in Fig. \ref{fig:loschmidt_echo_ferro}. The cut-off oscillations
for a given cut-off are smoothened using the procedure described in Appendix \ref{sub:Cut-off-extrapolation-schemes}
with a Gaussian smoothening kernel of width tuned to the frequency of the cut-off oscillations
$\omega=\Lambda$. The smoothened curves are then extrapolated
with a cut-off dependence $\Lambda^{-1}$, as illustrated in Fig.
\ref{fig:Loschmidt_extrap}. 
Details of extrapolation procedures are given in Appendix \ref{sub:Cut-off-extrapolation-schemes}.

After the cut-off extapolation, the TFSA result for the time-dependent Loschmidt 
echo agrees well with the theoretical prediction for moderate mass quenches 
within the same phase. In addition, the thin dashed line in Fig. 
\ref{fig:loschmidt_echo_m0_1,5} shows the TFSA result obtained by using the 
exact overlaps computed from Eq. \eqref{eq:initialStateExpansion}. This 
eliminates completely the numerical error from the overlaps, but still leaves 
the error coming from truncating the basis to a finite number of vectors. For 
the case of non-integrable quenches considered later this leads to the important 
observation that using the exact overlaps 
(as explained in Sec. \ref{sec:TFSAbasis}) improves the accuracy considerably, 
since
it takes into account the integrable part of the quench almost exactly.

\subsection{Time evolution of the order parameter}

\begin{figure}
\center

\begin{centering}
\subfloat[Quench from $M_{0}=0.5M$ to $M$.]{\includegraphics[width=.5\textwidth]{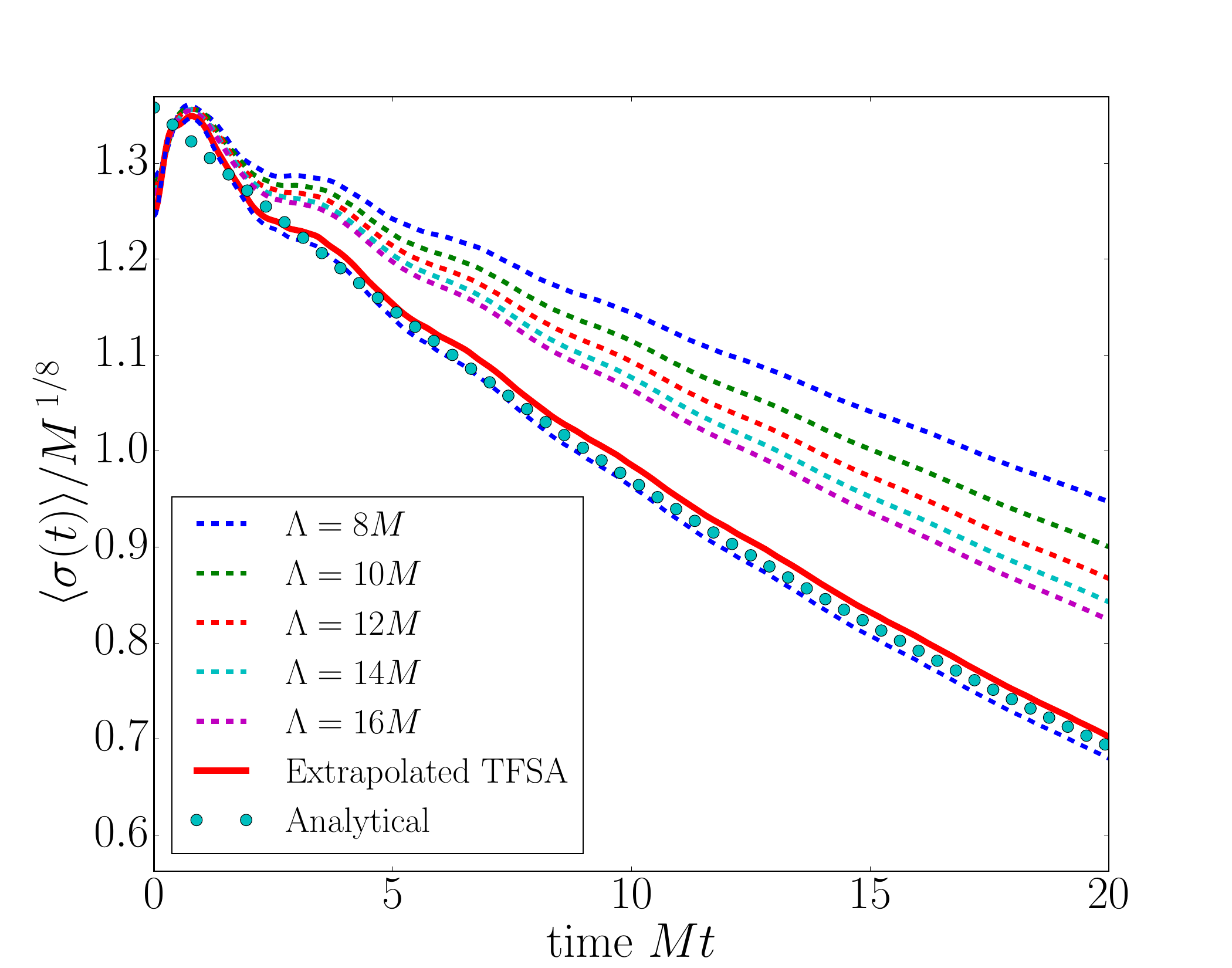}}
\subfloat[Quench from $M_{0}=1.5M$ to $M$.]{\includegraphics[width=.5\textwidth]{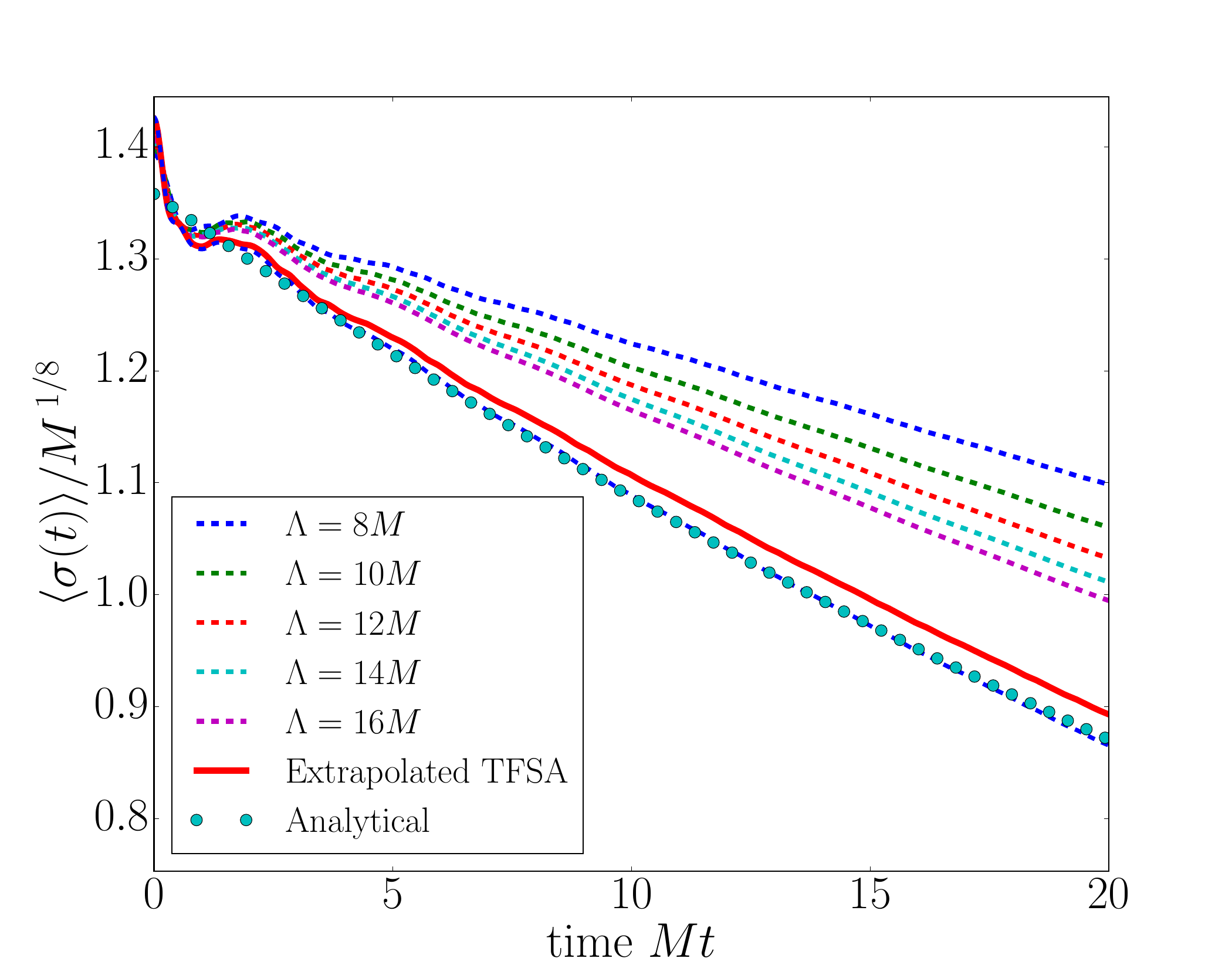}\label{fig:sigmaM0_1.5}}
\end{centering}
\caption{\label{fig:sigma_integrable} Time-dependent expectation value of
the order parameter $\langle\sigma(t)\rangle$ after integrable mass quenches in the ferromagnetic phase
for system size $ML=40$. The finite
cut-off energy results are shown in dashed lines, the cut-off
extrapolated value (solid red line) is 
compared to the theoretical result \eqref{eq:sigma_theor}
for the large time asymptotics (green dots). The TFSA results using the exact overlaps are shown in thin dashed line.}
\end{figure}

Fig. \ref{fig:sigma_integrable} summarises the TFSA results for $\langle\sigma(t)\rangle$ after mass quenches in the ferromagnetic phase at various fixed cut-offs together with the result after cut-off extrapolation. For short times we find a behaviour very similar to that on the lattice \cite{2012JSMTE..07..016C}. For quenches away from the critical point $M=0$ the order parameter initially increases, while for quenches towards the critical point it decreases. In both cases, after an initial oscillating transient an 
exponential decay sets in. This is in agreement with the analytical results both on the lattice
\cite{2012JSMTE..07..016C} and in the
Ising field theory \cite{2012JSMTE..04..017S}. In the latter, for small quenches a resummed form factor expansion in Ref. \cite{2012JSMTE..04..017S} yielded
\begin{align}
\left\langle \sigma(t)\right\rangle  & =  \bar{\sigma}\;[1+\lambda(t)]\;e^{-t/\tau}\qquad\mbox{for large }t\,,\label{eq:sigma_theor}\\
\tau & =  \left[\frac{2M}{\pi}\int_{0}^{\infty}d\xi|K(\xi)|^{2}\sinh\xi\right]^{-1}+O(K^{6})\:,\label{eq:tau_theor}\\
\lambda(t) & = \frac{\alpha}{Mt}-\frac{1-M/M_0}{8\sqrt\pi}\frac{\cos(2Mt-\pi/4)}{(Mt)^{3/2}}+\dots\,,\label{eq:lambda}
\end{align}
where $K(\theta)$ is the kernel function in the expansion of the
initial state (\ref{eq:initialStateExpansion}) and $\alpha$ was conjectured to be $\alpha=-(1-M/M_0)^2/(32\pi).$ This expression agrees with the continuum limit of the lattice result, and apart from the subleading prefactor $\lambda(t)$ it can also be obtained using semiclassical arguments \cite{2009PhRvL.102l7204R,2010PhRvB..82n4302R,2011PhRvB..84p5117R,2015arXiv150702708K}. 

The analytic result \eqref{eq:sigma_theor} (without $\lambda(t)$) is plotted using green dots in Fig. \ref{fig:sigma_integrable}. The agreement with the extrapolated TFSA data is quite good. 
The decay rates $\tau^{-1}_\text{TFSA}$ and the amplitudes $A_\text{TFSA}$ can be extracted fitting our results  for $8<Mt<20$ by an exponential $A_\text{TFSA}\exp(-t/\tau_\text{TFSA}),$ and 
they are in good agreement with the exact values $\tau^{-1}$ and $\bar\sigma$ for both quenches:
%
%\bigskip
\begin{center}
\begin{tabular}{l|c|c||c|c}
&$\tau^{-1}$&$\tau^{-1}_\text{TFSA}$&$\bar\sigma$&$A_\text{TFSA}$\\
\hline
$M_0=0.5M$&$0.0336M$&$0.0334M$&1.36&1.37\\
$M_0=1.5M$&$0.0222M$&$0.0214M$&1.36&1.37\\
\end{tabular}
\end{center}
\bigskip
Similarly to the Loschmidt echo, Fig. \ref{fig:sigmaM0_1.5} also shows in thin dashed line the TFSA result obtained using the exact overlaps that agrees almost perfectly with the analytic result.

\begin{figure}[t!]
\subfloat[$M_0=1.5M$]{\includegraphics[width=0.5\textwidth]{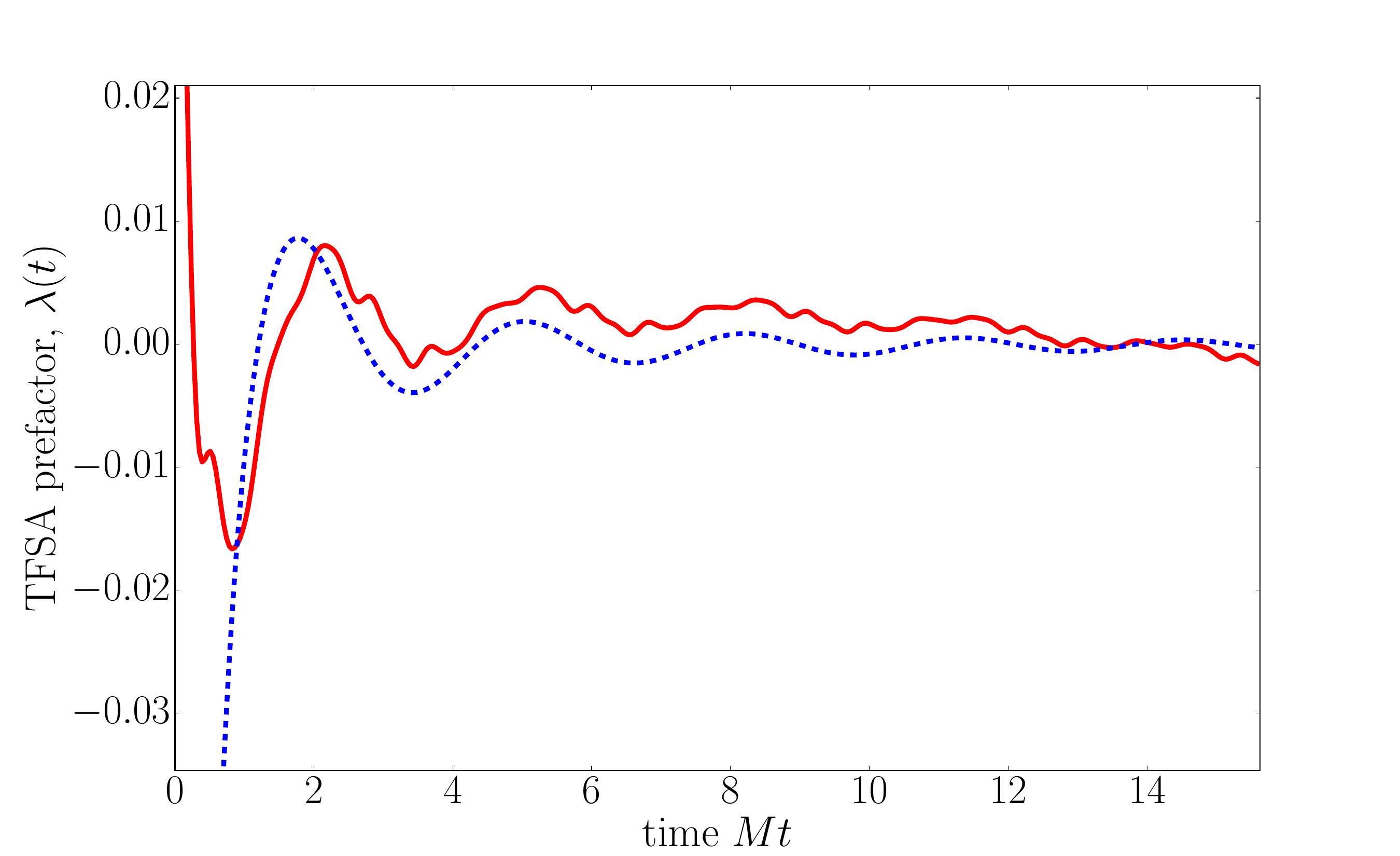}\label{fig:prefactor1,5}}
\subfloat[$M_0=0.5M$]{\includegraphics[width=0.5\textwidth]{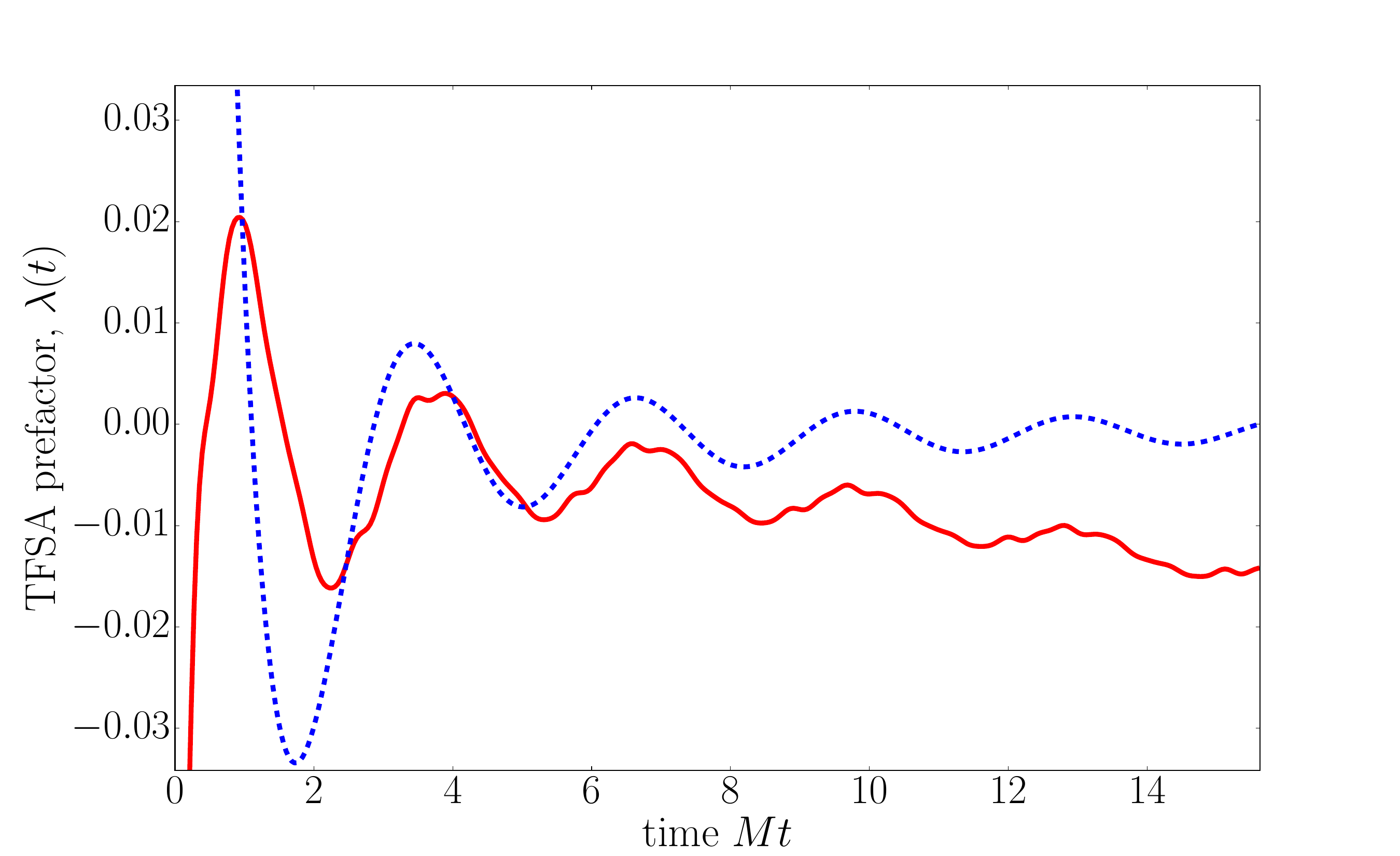}\label{fig:prefactor0,5}}
\caption{The prefactor extracted from the extrapolated TFSA data as $\langle\sigma(t)\rangle_\text{TFSA}/(\bar \sigma e^{-t/\tau})-1$ (solid red line) and the prediction $\lambda(t)$ in Eq. \eqref{eq:lambda} (blue dashed line). }
\label{fig:prefactor}
\end{figure}

A time dependent oscillatory prefactor can also be extracted from
the TFSA results by dividing it by the asymptotic expression $\bar\sigma e^{-t/\tau}.$
As can be seen in Fig. \ref{fig:prefactor}, this prefactor agrees in its main features with with the prediction $\lambda(t)$ in Eq. \eqref{eq:lambda}, 
in particular, its Fourier analysis reveals a single strong peak at frequency $2M.$ The high frequency oscillation visible in the TFSA results is a cut-off effect. 

%The effect of this oscillatory factor can be seen at short times in Fig. \ref{fig:sigma_integrable}.  }

\subsection{Summary of results for integrable quenches}

To sum up, we have demonstrated that for quenches of moderate size within the same phase of the Ising field theory, the TFSA method is able to reproduce the theoretical results for various quantities, including the statistics of work $P(W)$, the Loschmidt echo $\mathcal{L}(t)$, and the expectation values $\langle\varepsilon(t)\rangle$, $\langle\sigma(t)\rangle$ to a good accuracy. 

We emphasise that good agreement was reached between the theoretical predictions valid in infinite volume and our finite volume TFSA calculations performed at system size $M L=40$. 
In particular, this implies that the cut-off extrapolated numerical results have small finite size effects, which is expected for massive field theories that typically show exponentially suppressed finite size effects, as noted previously in \cite{2015PhRvB..92p1111J}.

Finally we note that the TFSA is less accurate for larger changes in the Hamiltonian 
parameters, in particular, reliable simulations for quenches across the critical point featuring non-analyticities in the Loschmidt echo would require much larger energy cut-offs.

\begin{figure}[t!]
\subfloat[Ferromagnetic spectrum]{\includegraphics[width=0.5\textwidth]{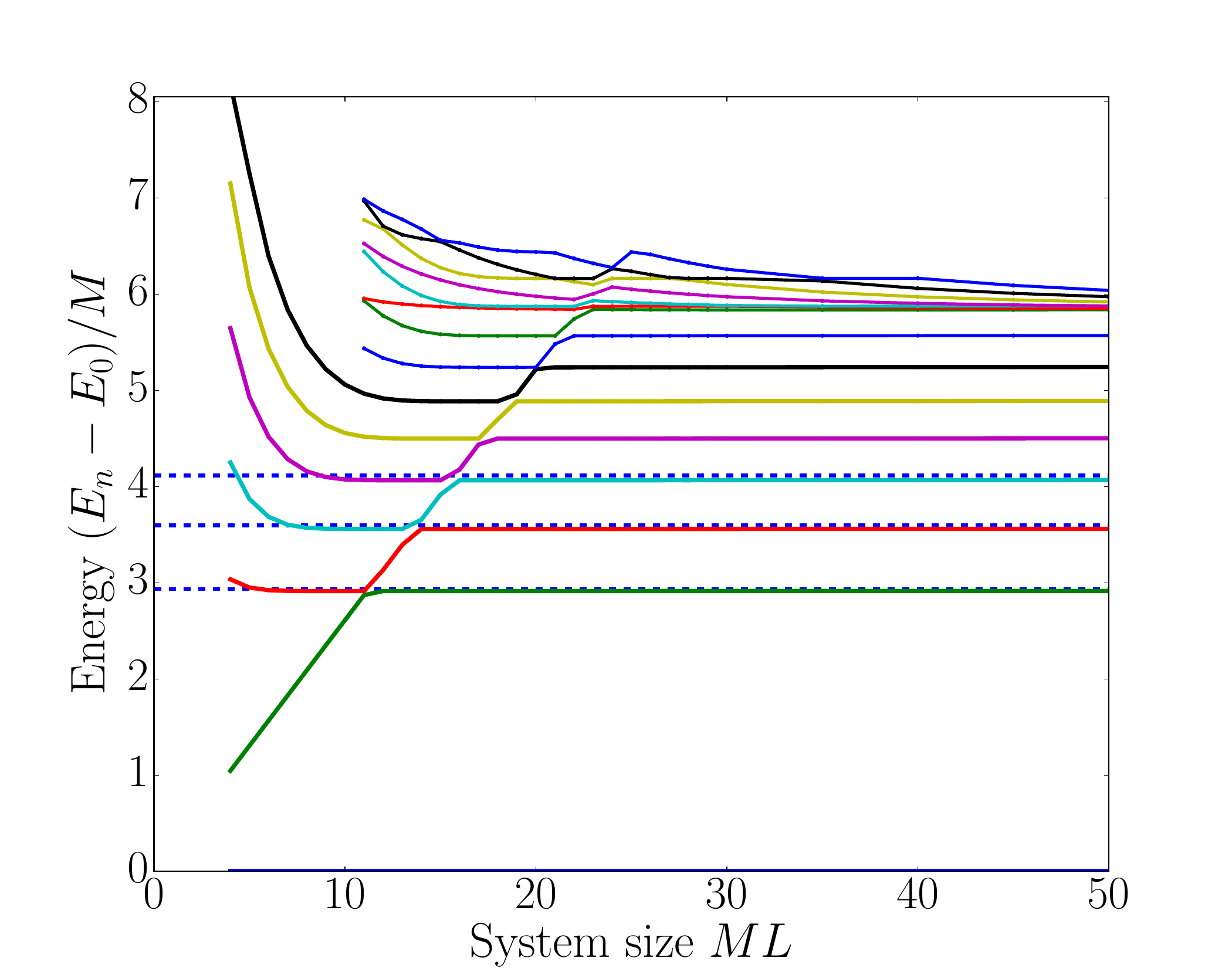}\label{fig:Spectrum_ferro} }
\subfloat[Paramagnetic spectrum]{\includegraphics[width=0.5\textwidth]{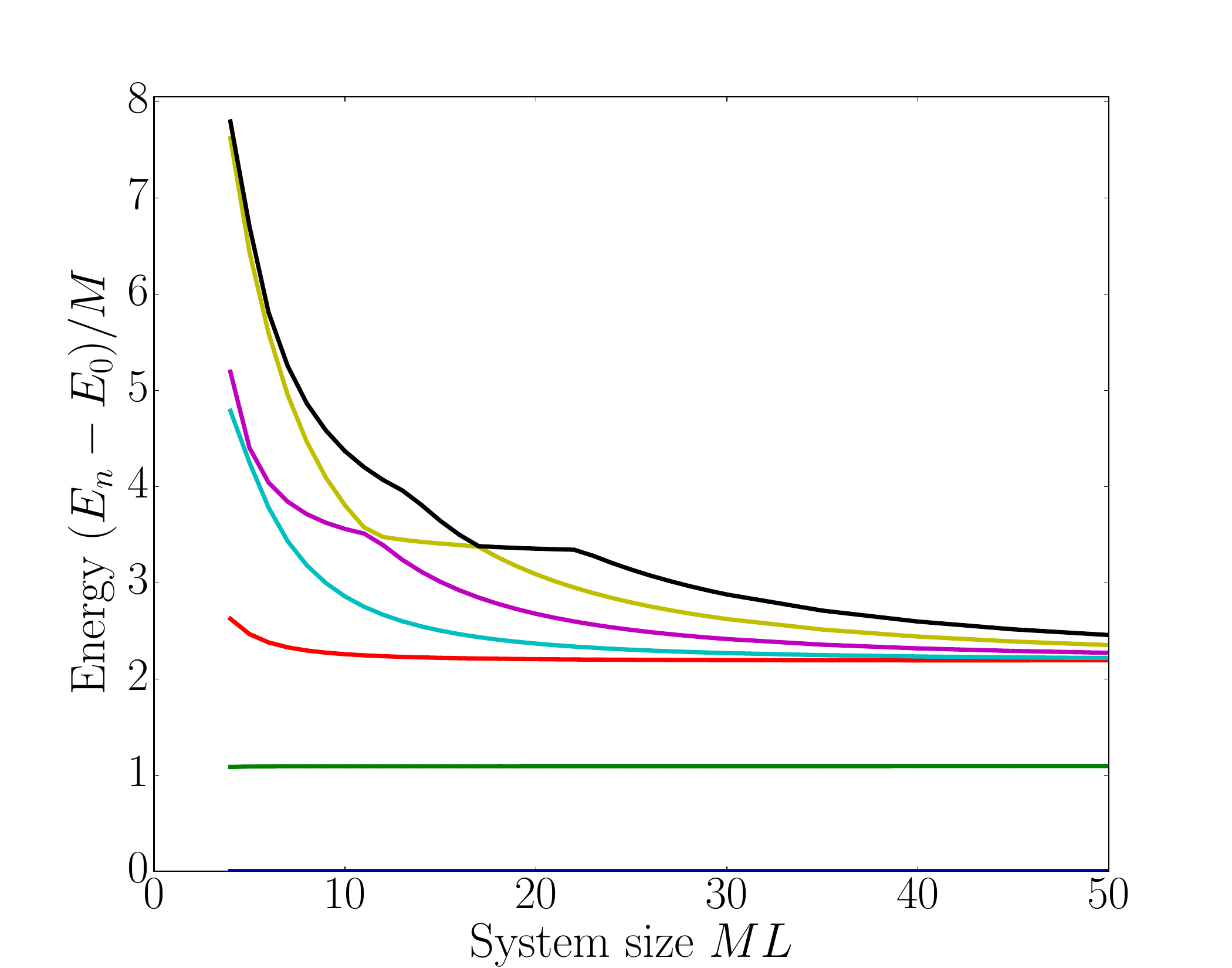}\label{fig:Spectrum_para} }
\caption{Spectra of zero momentum states of the post-quench Hamiltonian for $\bar{h}=-0.1$ at a cut-off
$\Lambda=8M$ in the (a) ferromagnetic and (b) paramagnetic case. The excitation energies quickly reach their infinite volume values. In the ferromagnetic case seven mesons are visible below the continuum, the blue dashed lines show the first three meson masses computed for an infinite system. The state with linearly increasing energy is the false vacuum corresponding to the ferromagnetic ground state with negative magnetisation. In the paramagnetic case there is a singe massive particle.}
\label{fig:Spectra}
\end{figure}

\section{Non-integrable quenches in the ferromagnetic phase \label{sec:Non-integrable-quenches-in-FM}}

Having tested the TFSA method let us turn now to the case of non-integrable quenches.
As discussed in Sec. \ref{sec:TFSAbasis}, for quenches from $H(M_0,h_0=0)$ to $H(M,h)$ it is possible to use the eigenbasis of the Hamiltonian $H(M,0)$ and construct the initial state 
truncating the expressions (\ref{eq:initialStateExpansion},\ref{eq:InitialStateFerroPara}). 

In this Section  we consider quenches within the ferromagnetic phase where 
the time evolution is found to be dominated by the meson bound states. 
They appear in the spectrum as a result of turning on the external magnetic
field in the longitudinal direction.

\subsection{Spectrum \label{sub:SpectrumFM}}

\begin{figure}[t!]
\label{fig:sigma_extrapolation}
\subfloat[$\bar{h}=-0.05$]{\includegraphics[width=.5\textwidth]{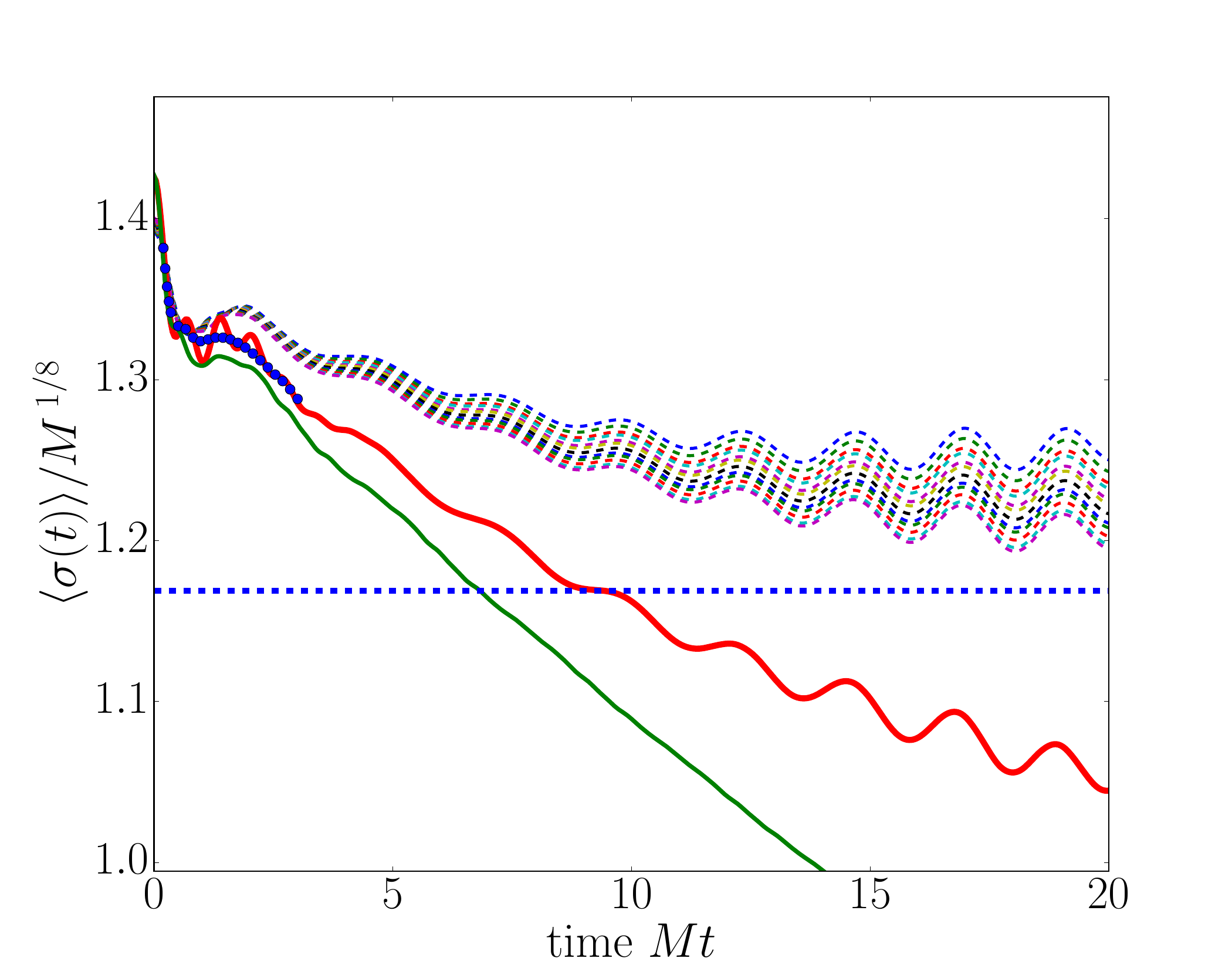}\label{fig:sigma_extrapolation_smallh}}
\subfloat[$\bar{h}=-0.1$]{\includegraphics[width=.5\textwidth]{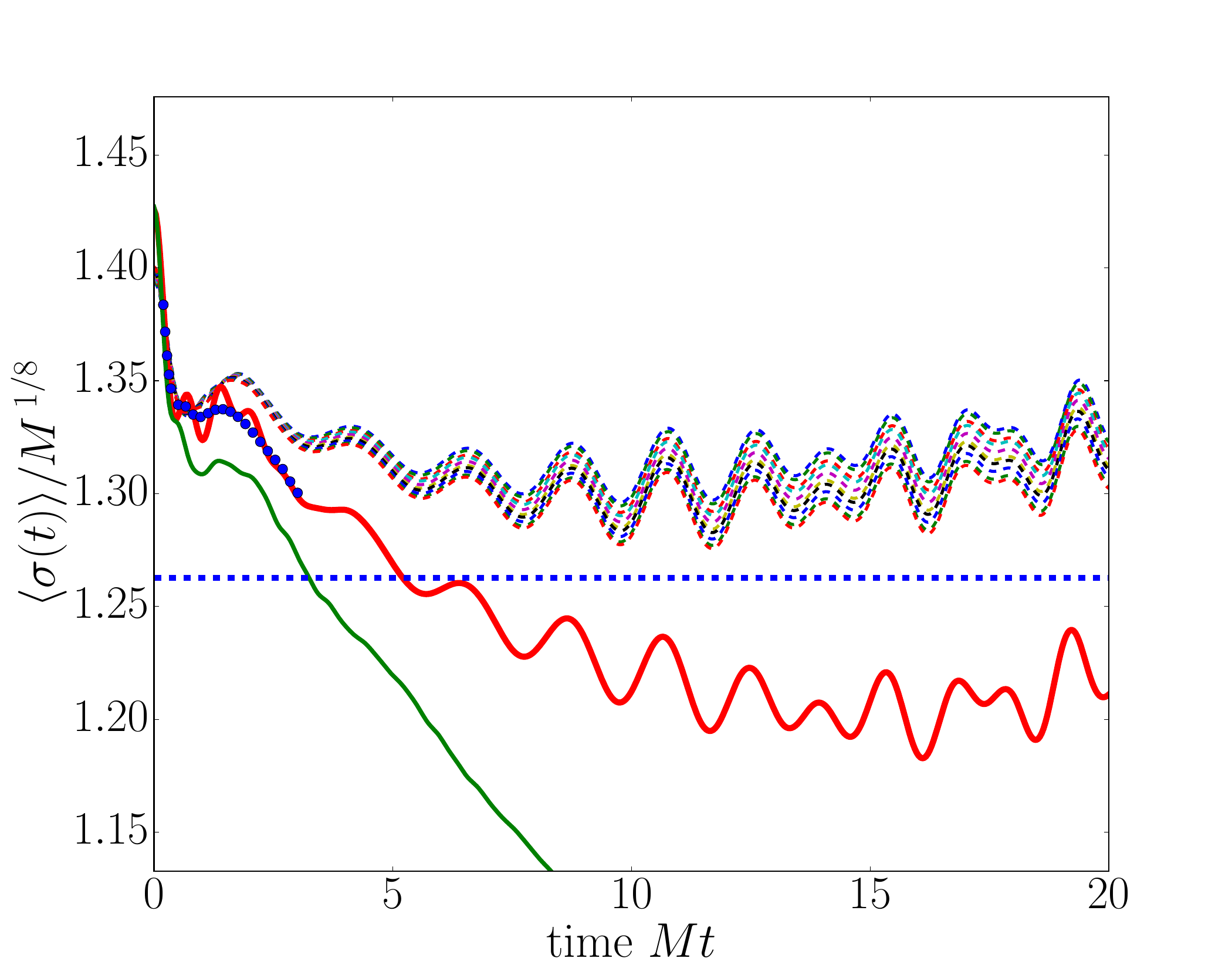}\label{fig:sigma_extrapolation_largeh}}\\
\subfloat[$\bar{h}=-0.05$]{\includegraphics[width=.5\textwidth]{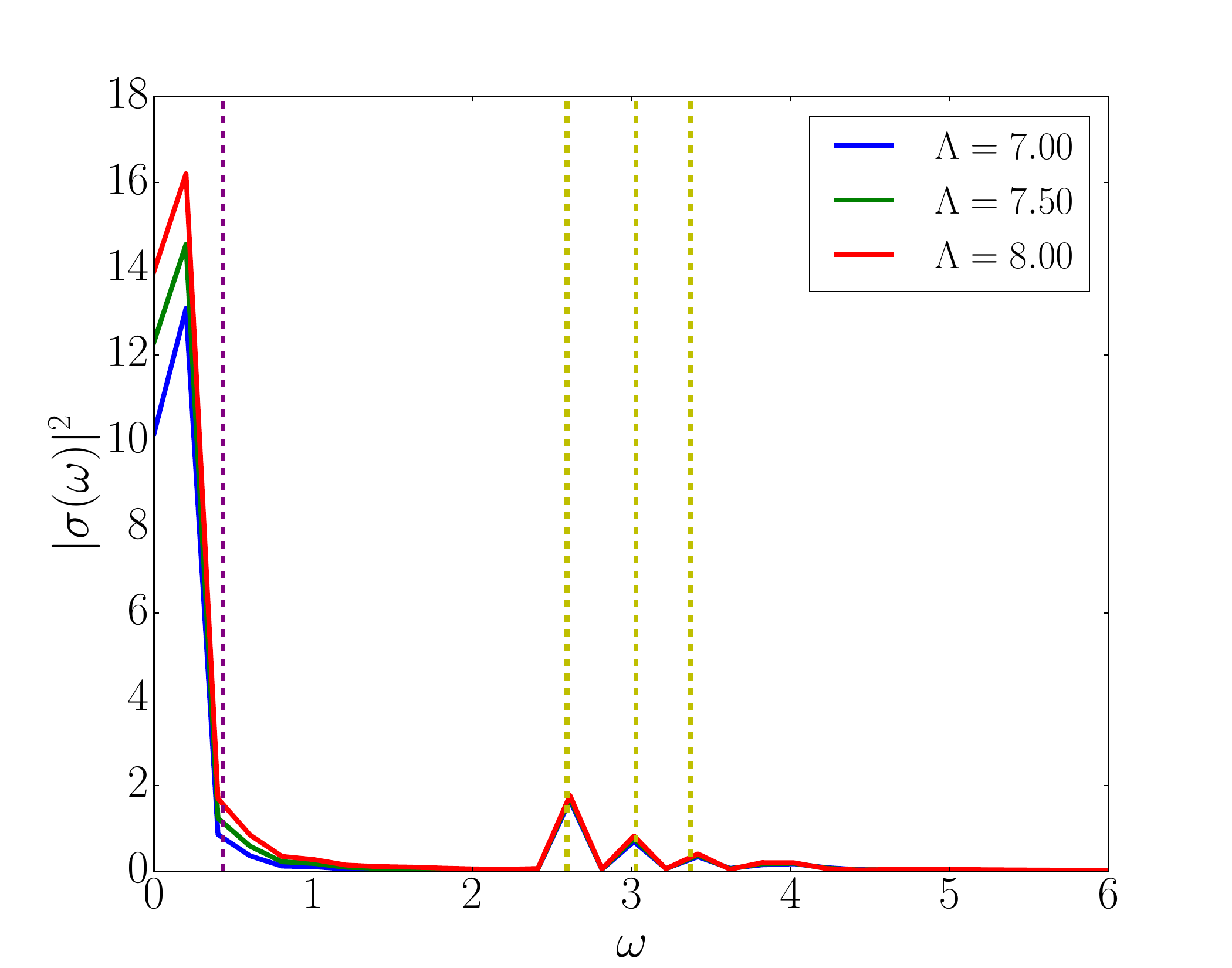}\label{fig:Fourier_sigma(t)_1}}
\subfloat[$\bar{h}=-0.1$]{\includegraphics[width=.5\textwidth]{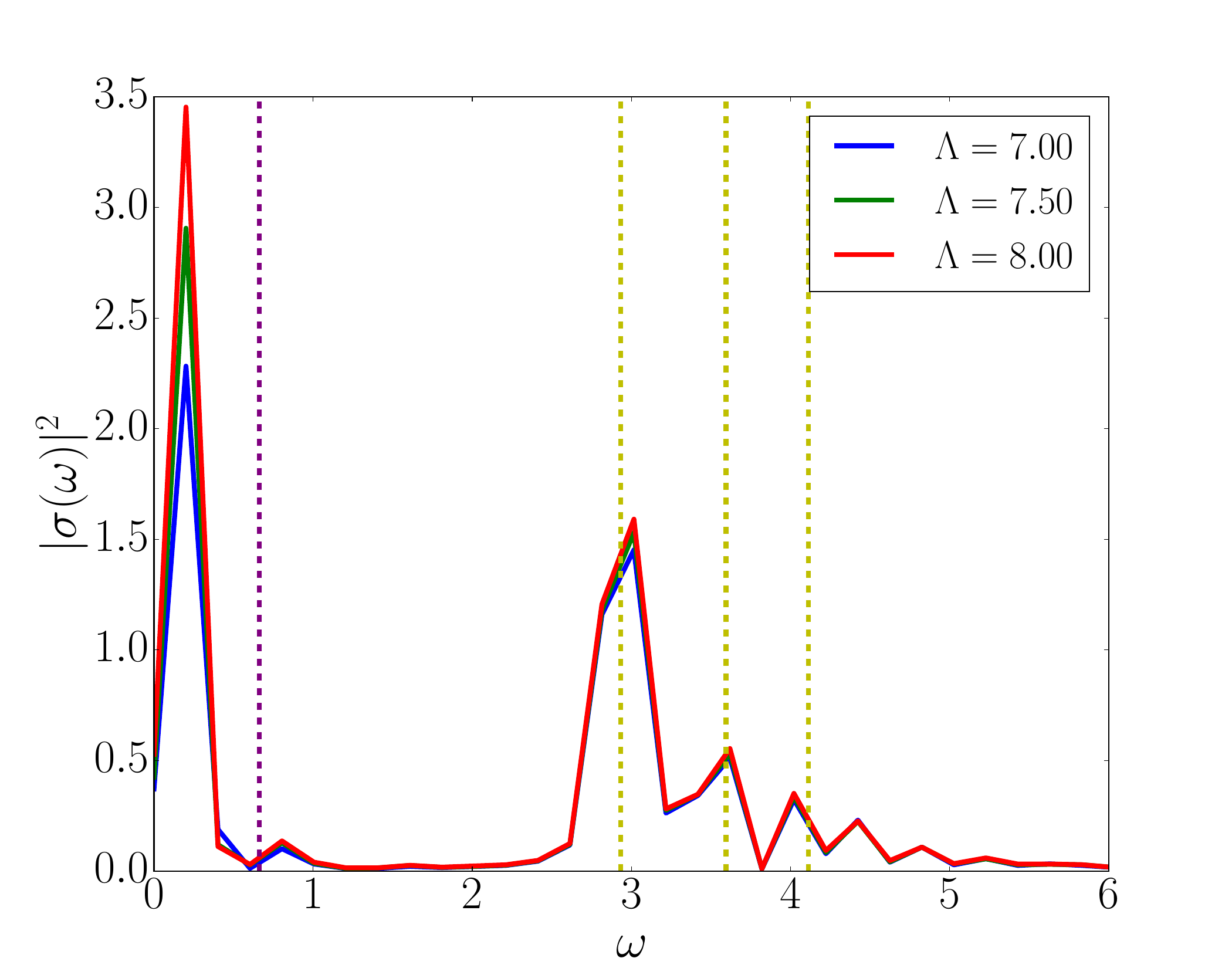}\label{fig:Fourier_sigma(t)_2}}
\caption{ {\it Top row:} Time evolution of the expectation value of the order parameter $\left\langle \sigma(t)\right\rangle$ after a quench from $M_{0}=1.5M$ to $M$ for system size $ML=40$ with (a) $\bar h = -0.05$, (b) $\bar h=-0.1$. Curves of $\left\langle \sigma(t)\right\rangle $ are shown for different values of the TFSA cut-off between $\Lambda=4.25M$
and $\Lambda=9.25M$ (dashed lines) along with the extrapolated curve (solid red line). Some details of the extrapolation can be found in Fig. \ref{fig:Cut-off-extrapolation-for-ferro-sigma-largeh} of Appendix \ref{sub:Cut-off-extrapolation-schemes}.  The horizontal blue dashed
line is the (cut-off extrapolated) diagonal ensemble average \eqref{eq:DE} computed using the
overlaps numerically determined from TFSA. For comparison, the cut-off extrapolated result for the integrable quench ($\bar h=0$) from the same $M_0$ is also shown in thin green line. {\it Bottom row:} 
Fourier spectra of $\left\langle \sigma(t)\right\rangle $
for the same parameters as in the top row but with system size $ML=80$
for different cut-offs $\Lambda/M=7,7.5,8.$  The yellow dashed vertical lines represent the
first three meson masses $m_{1,2,3}$ computed from the WKB approximation,
while the leftmost purple dashed vertical line is the difference of the first
two masses, $m_2-m_1.$ 
}
\end{figure}

Before turning to the quenches, we discuss the spectrum
of the post-quench Hamiltonian. In the ferromagnetic phase, switching
on a longitudinal magnetic field $\bar{h}$ causes a non-perturbative change in the spectrum which can be understood most easily on the lattice. In the ferromagnetic phase, the elementary excitations at $\bar h=0$ corresponding to free fermions are domain walls.  A longitudinal field induces a linear potential between pairs of domain walls that are separated by a distance $d$,
which for a small longitudinal magnetic field $h$ has the form
\begin{equation}
V(d)=h\bar{\sigma}\cdot d\label{eq:linearpot}
\end{equation}
leading to a confinement of the fermions into so-called mesons, a scenario first proposed by McCoy and
Wu \cite{1978PhRvD..18.1259M}. Based on this picture, it is possible to compute the spectrum
using various theoretical approaches, for details see Appendix \ref{sec:Meson-masses-in}.

The spectrum of the first few zero-momentum excited states as a function of the
dimensionless system size $ML$ is shown in Fig. \ref{fig:Spectrum_ferro}.
Note that the energy values quickly reach a stable value, already
around $ML=30$, which are close to the infinity volume values even
at this relatively small cut-off. The plot shows the values of the
first three meson masses for comparison as computed using the WKB
approximation (see Appendix \ref{sec:Meson-masses-in}). The excellent agreement demonstrates both the validity of the WKB approximation and the precision of the TFSA method.

\subsection{Time evolution of the order parameter}

\begin{figure}[t]
\centering{}\includegraphics[width=0.55\textwidth]{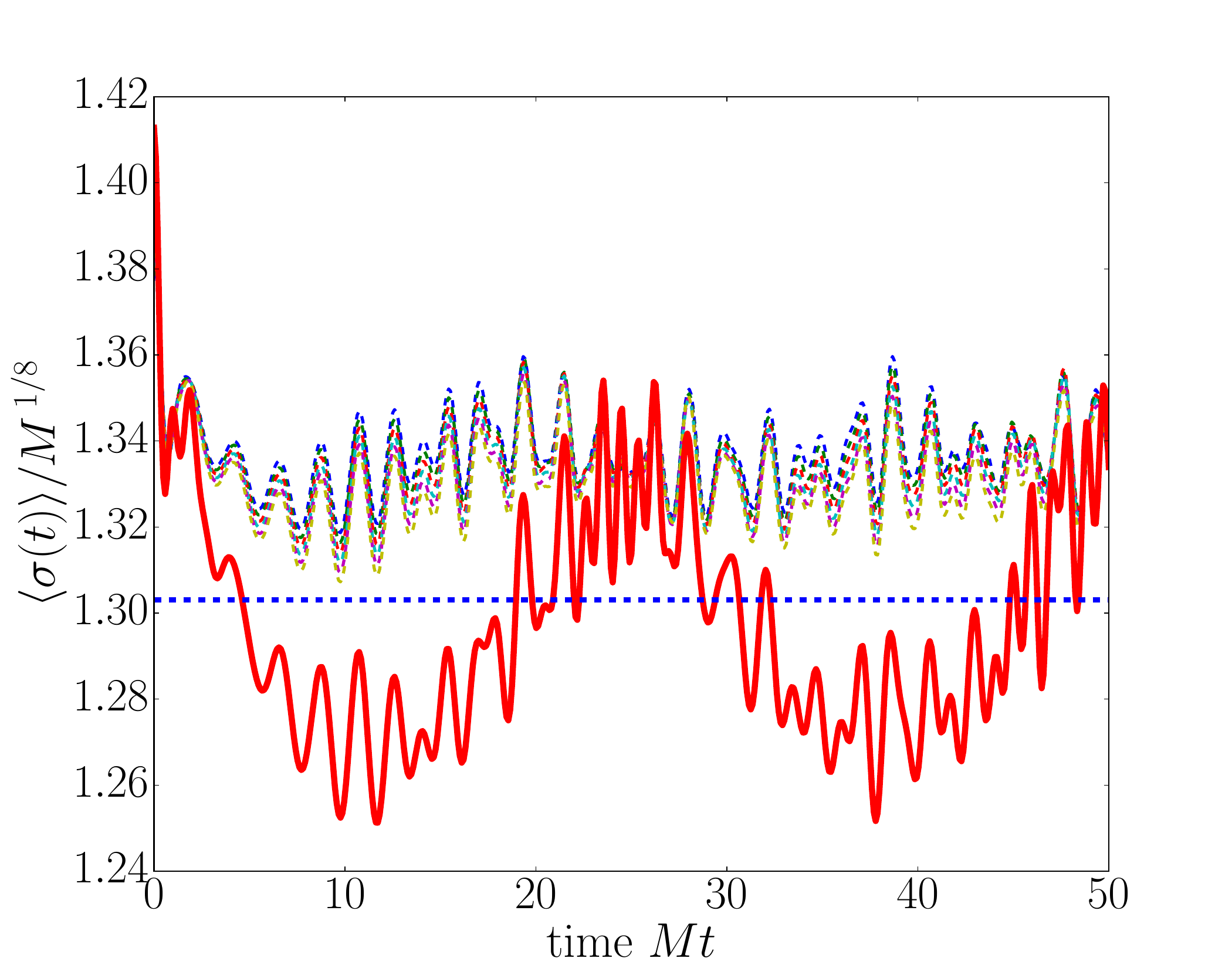}
\caption{ Time evolution of the expectation
value of the order parameter $\left\langle \sigma(t)\right\rangle$ after a quench from $M_{0}=1.5M$ to $M$ for system size $ML=100$ and $\bar h=-0.1$. Curves of $\left\langle \sigma(t)\right\rangle $ are shown
for different values of the TFSA cut-off between $\Lambda=4.25M$
and $\Lambda=8.25M$ (dashed lines) along with the extrapolated curve (solid red line). The blue dashed
line is the (cut-off extrapolated) diagonal ensemble average obtained by using the
overlaps numerically determined from TFSA.}
\label{fig:sigma_L100}
\end{figure}

Turning to the time evolution of the expectation value of the
magnetisation operator $\sigma$ after the quench, we present 
numerical data for quenches from $M_{0}=1.5M$
to $M$ while the magnetic field is changed from $\bar{h}_{0}=0$
to either $\bar{h}=-0.05$ (Fig. \ref{fig:sigma_extrapolation_smallh})
or $\bar{h}=-0.1$ (Fig. \ref{fig:sigma_extrapolation_largeh}).
A larger value of $\bar{h}$ causes the features of integrability
breaking to appear on a shorter time scale. 
\footnote{Quenching to a field in the ``wrong'' direction $\bar{h}>0$
(i.e. counteracting the magnetisation of the initial state) leads
to unreliable numerical results, so we do not treat this case here.}

The extrapolation introduces spurious oscillations at short times that are not present in the raw data. Smoothing the raw curves using a Gaussian kernel before extrapolation (see Appendix \ref{sec:smooth_extrapol}) eliminates these oscillations (see blue dots in Figs. \ref{fig:sigma_extrapolation}a,b) but spoils the large time behaviour.

Already a small magnetic field results in a drastic change in the time evolution 
of the magnetisation. Instead of the exponential decay seen for integrable 
quenches with $\bar h=0$, it does not decay to zero but features slow and fast 
oscillations.
The various oscillation frequencies can be extracted by taking the Fourier 
transform of the time series $\left\langle\sigma(t)\right\rangle$ in a time 
window $4<Mt<36$ in order to avoid the spurious oscillations at short times 
induced by the extrapolation procedure. As demonstrated in Figs. 
\ref{fig:Fourier_sigma(t)_1}, \ref{fig:Fourier_sigma(t)_2} the resulting power 
spectrum shows clear peaks related to the meson masses, suggesting that the 
confinement has drastic effects on the quench dynamics. Note that a non-zero 
asymptotic magnetisation is expected due to the presence of the external 
magnetic field. Both the overall behaviour of the magnetisation and the relation 
between the oscillation frequencies and the meson masses are in full agreement 
with the results found in the earlier work
\cite{2016arXiv160403571K} which considered the quench dynamics
on the spin chain governed by the Hamiltonian \eqref{eq:QIM_Hamiltonian}.

Fig. \ref{fig:sigma_L100} shows $\left\langle\sigma(t)\right\rangle$ computed in a larger volume $ML=100$ so that
the time evolution can be followed further, up to $Mt=ML/2=50.$ The magnetisation is seen to oscillate around its infinite time average given by the diagonal ensemble average. The oscillations persist up to the latest times we were able to study, even when the system size $L$ is much bigger than the equilibrium correlation length $M^{-1}$. Note that the studied times are already long in the natural unit $M^{-1}$, so any qualitative change will supposedly take place at some much larger, possibly astronomical time scales, if ever. The questions whether these oscillations eventually die out and if they do, what is the time scale of this damping, should be the subject of further study.

\subsection{Statistics of work and Loschmidt echo}

\begin{figure}[t!]
\subfloat[Statistics of work]{\includegraphics[width=0.48\textwidth]{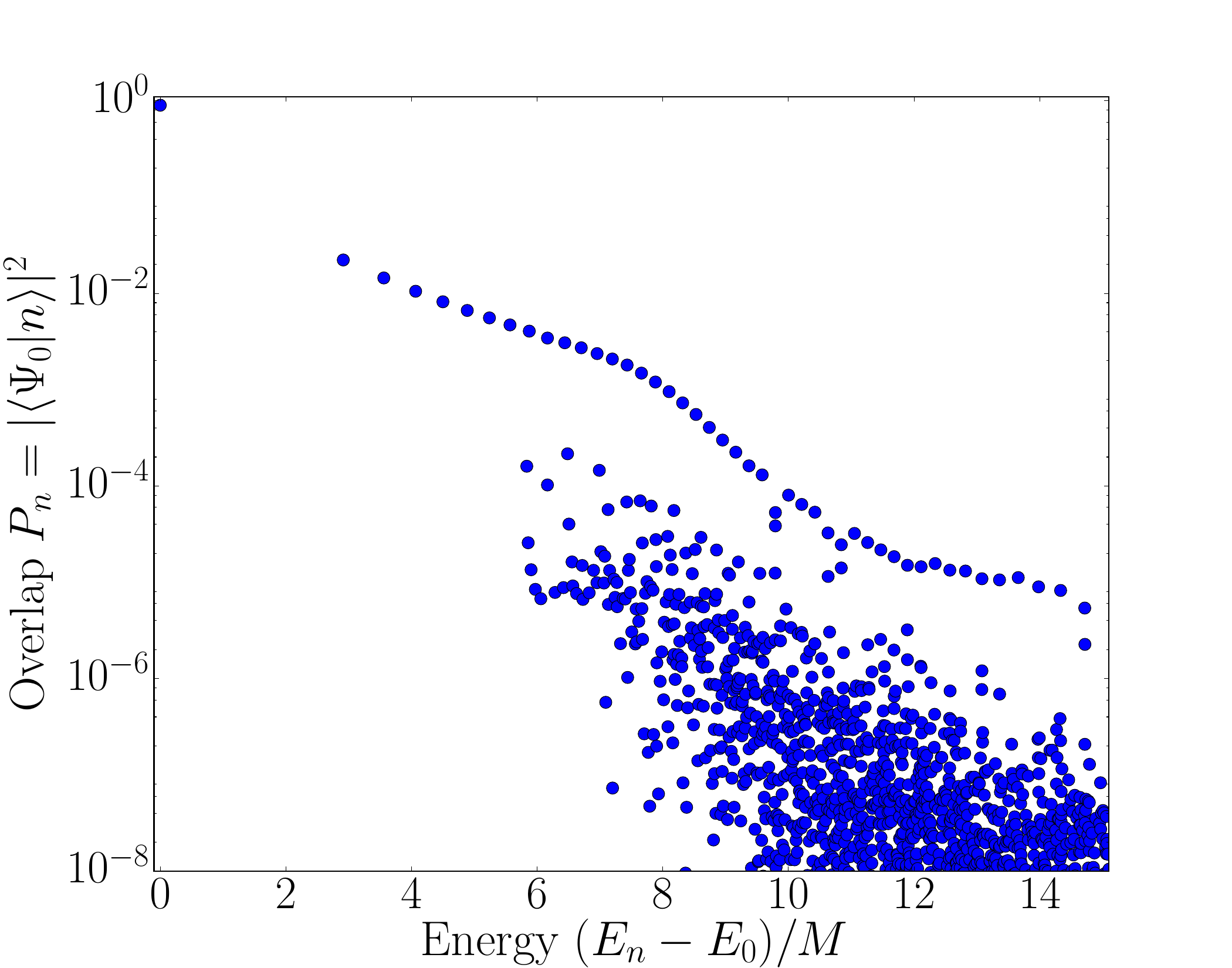}\label{fig:pw_ferro_nonint}}
\subfloat[Loschmidt echo]{\includegraphics[width=0.52\textwidth]{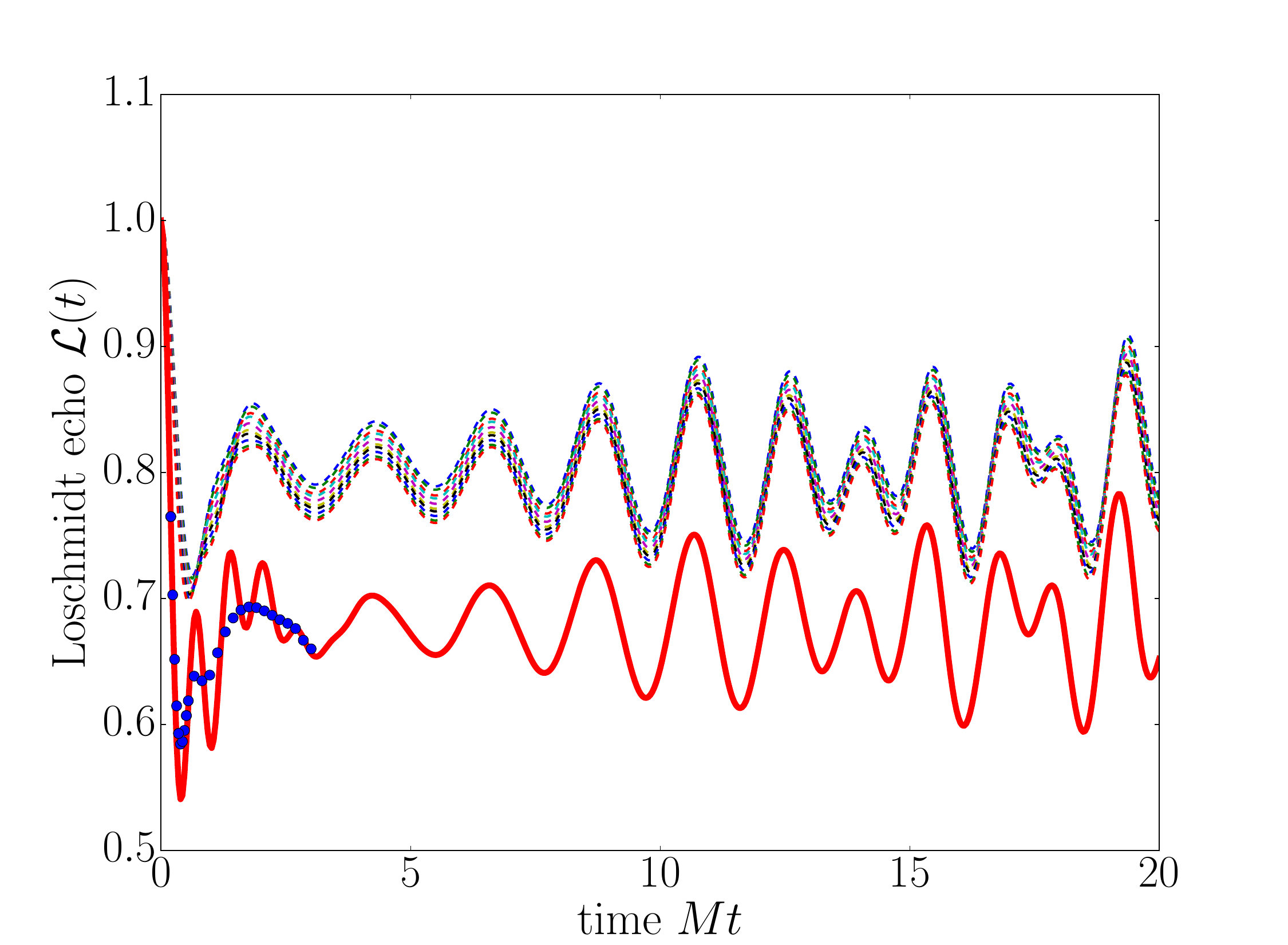}\label{fig:ferro_Loschmidt}}
\caption{(a) Statistics of work $P(W)$ for a quench within the ferromagnetic phase with $M_{0}=1.5M$,
$ML=40$, $\bar{h}=-0.1$, and cut-off $\Lambda=8M$.
(b) Loschmidt echo $\mathcal{L}(t)$ for the same quench. 
The family of dashed lines shows the raw data for different cut-offs
between $\Lambda=4.25M$ and $\Lambda=9.25M$ while the solid red line is
the extrapolated curve, with the 6 highest cut-offs used for the extrapolation. The blue dots at short times are obtained by smoothening the data before extrapolation.}
\end{figure}

The dominant role played by mesons is further supported by the statistics
of work which is shown in Fig. \ref{fig:pw_ferro_nonint} for a specific
but illustrative quench in the ferromagnetic phase. It is clear that
a few of the lowest lying states in the spectrum dominate and are responsible for the oscillatory behaviour, while all
other states have much smaller overlap with the initial state.

\begin{figure}[t]
\begin{centering}
\includegraphics[width=0.5\textwidth]{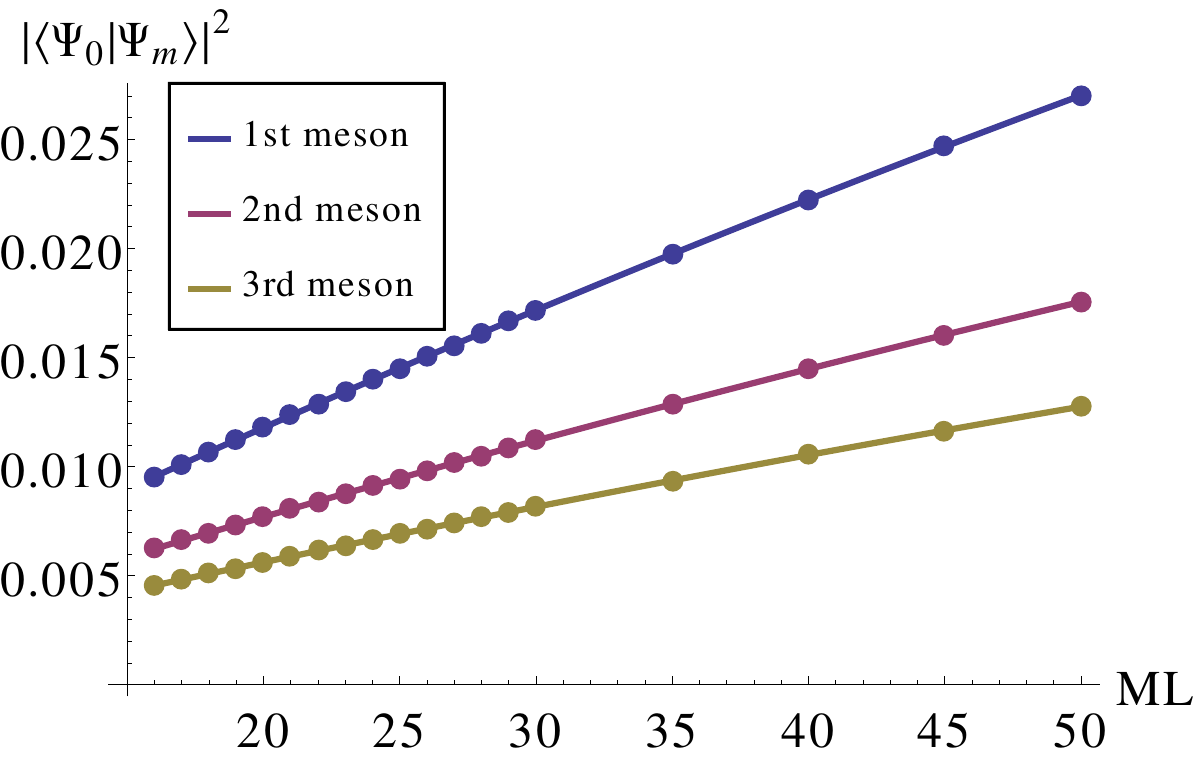}\\
\bigskip
\begin{tabular}{|c|c|c|}
\hline 
$m$ & $A_{m}/M$ & $B/M$\tabularnewline
\hline 
\hline 
$1$ & $6.2\cdot10^{-4}$ & $2.9\cdot10^{-3}$\tabularnewline
\hline 
$2$ & $4.1\cdot10^{-4}$ & $3.0\cdot10^{-3}$\tabularnewline
\hline 
$3$ & $3.0\cdot10^{-4}$ & $3.2\cdot10^{-3}$\tabularnewline
\hline 
\end{tabular}
\par\end{centering}
\caption{ The first three meson overlaps as functions
of the volume for a quench $M_{0}=1.5M$ and $\bar{h}=-0.1$ with
cut-off $\Lambda=8.0M$ in the ferromagnetic phase, as measured from
TFSA. The fits correspond to the theoretical curve (\ref{eq:overlaptheor})
with parameters given in the table.}
\label{fig:meson_overlap}
\end{figure}

The dominance of low lying states runs contrary to normal expectations:
in a quantum quench usually the highly excited states in the
middle of the post-quench spectrum are the relevant ones. On the one hand, 
the agreement with infinite volume lattice simulations indicates
that this feature is robust against changing the cut-off or the volume
of the system. On the other hand, as discussed in detail in Appendix 
\ref{sec:overlaps}, the overlaps of the 1-meson states $|\Psi_m\rangle_L$ are eventually expected to decay 
as 
\begin{equation}
\left|\langle\Psi_{0}|\Psi_{m}\rangle_{L}\right|^{2}=A_{m}Le^{-BL}
\label{eq:overlaptheor}
\end{equation}
for large enough $L.$ Although in our simulations we cannot access such large volumes,
the theoretical curve fits very well the overlaps measured by TFSA (see Fig. \ref{fig:meson_overlap}). 

However, as can be seen from the table in Fig. \ref{fig:meson_overlap}, the critical volume $L_\text{crit}=1/B$ at which the one-particle 
overlap is expected to start decreasing is a few hundred times the correlation 
length. In such large volumes the finite size effects are already 
expected to be minuscule; eventually, as shown by the excellent 
agreement with the iTEBD lattice simulations discussed in Section \ref{sec:TFSA-versus-iTEBD}, 
already the volume $ML=40$ which is just about a hundred times 
the correlation length can be considered infinite in terms 
of field theoretic finite size effects.

This can be explained by observing that the smallness of the exponent 
$B$, which from Eq. (\ref{eq:theor_overlap}) is given by
\begin{equation}
B  =  \int_{0}^{\infty}\frac{dq}{2\pi}\log\left(1+|K(q,M,M_{0})|^{2}\right)
\end{equation}
which is small whenever the post-quench density of particles given by \cite{2012JSMTE..07..016C}
\begin{equation}
\rho = \int_{0}^{\infty}\frac{dq}{2\pi}\frac{|K(q,M,M_{0})|^{2}}{1+|K(q,M,M_{0})|^{2}}
\end{equation}
is small. For a small density of particles, the average distance between particles 
is much larger than the correlation length, so even though the infinite volume 
post-quench contains a finite density of particles, its dynamics is dominated 
by one-particle contributions ``semiclassically pasted together'' \cite{2011PhRvB..84p5117R,Evangelisti2013,2015arXiv150702708K}. This opens the way to measure the masses of the mesons through studying the post-quench time evolution of the magnetisation. This method was called ``quench spectroscopy'' in Ref. \cite{Gritsev2007}.

Fig. \ref{fig:ferro_Loschmidt} shows the Loschmidt echo
\begin{equation}
\mathcal{L}(t)=|\langle\Psi_{0}|\Psi(t)\rangle|^{2}
\end{equation}
obtained by the TFSA. It quickly relaxes from $1$ to some lower value
and then begins to oscillate with an amplitude that increases in time. Note that the extrapolation induces fast oscillations for
short times that are most likely numerical artefacts. These can be
removed by Gaussian smoothening, as plotted in blue dots in 
Fig. \ref{fig:ferro_Loschmidt}, which however spoils the steep
decreasing edge of the initial transient. In any case, the curves
seem to be reliable after $t\approx3$.

\section{Non-integrable quenches in the paramagnetic phase \label{sec:Non-integrable-quenches-in-PM}}

In this section we review the results obtained for quenches within
the paramagnetic phase. This phase does not feature confinement and consequently has a very different excitation spectrum containing magnons and no mesons. As a result, the time evolution
of the magnetisation and the Loschmidt echo significantly differs from those in the
ferromagnetic phase.

\subsection{Spectrum}

Similarly to the ferromagnetic case, we first show
the spectrum of the post-quench Hamiltonian in Fig. \ref{fig:Spectrum_para}. The first excited state
is the magnonic wave with a mass perturbatively corrected by the magnetic
field. It approaches its infinite value as fast as the meson states
in the ferromagnetic phase (cf. Fig. \ref{fig:Spectrum_ferro}). The
higher excited states shown in the plot correspond to two-particle states of the magnons and form
a continuum in the large volume limit. There are also states with
more than two magnons that are not shown in the plot.

\begin{figure}[t!]
\subfloat[$\bar{h}=-0.05$]{\includegraphics[width=0.5\textwidth]{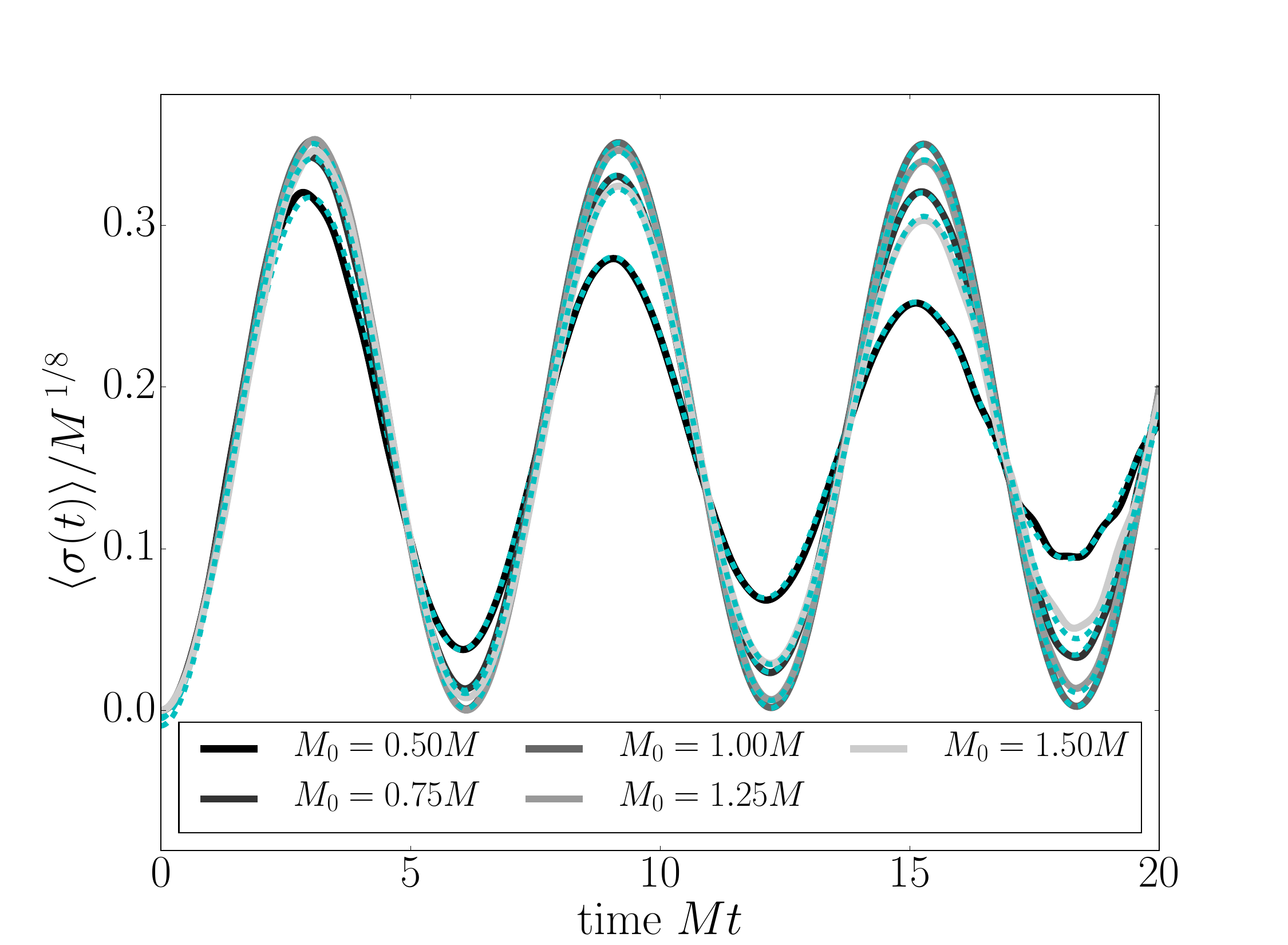}\label{fig:cut-off_extrapolation_smallh_para}}
\subfloat[$\bar{h}=-0.1$]{\includegraphics[width=0.5\textwidth]{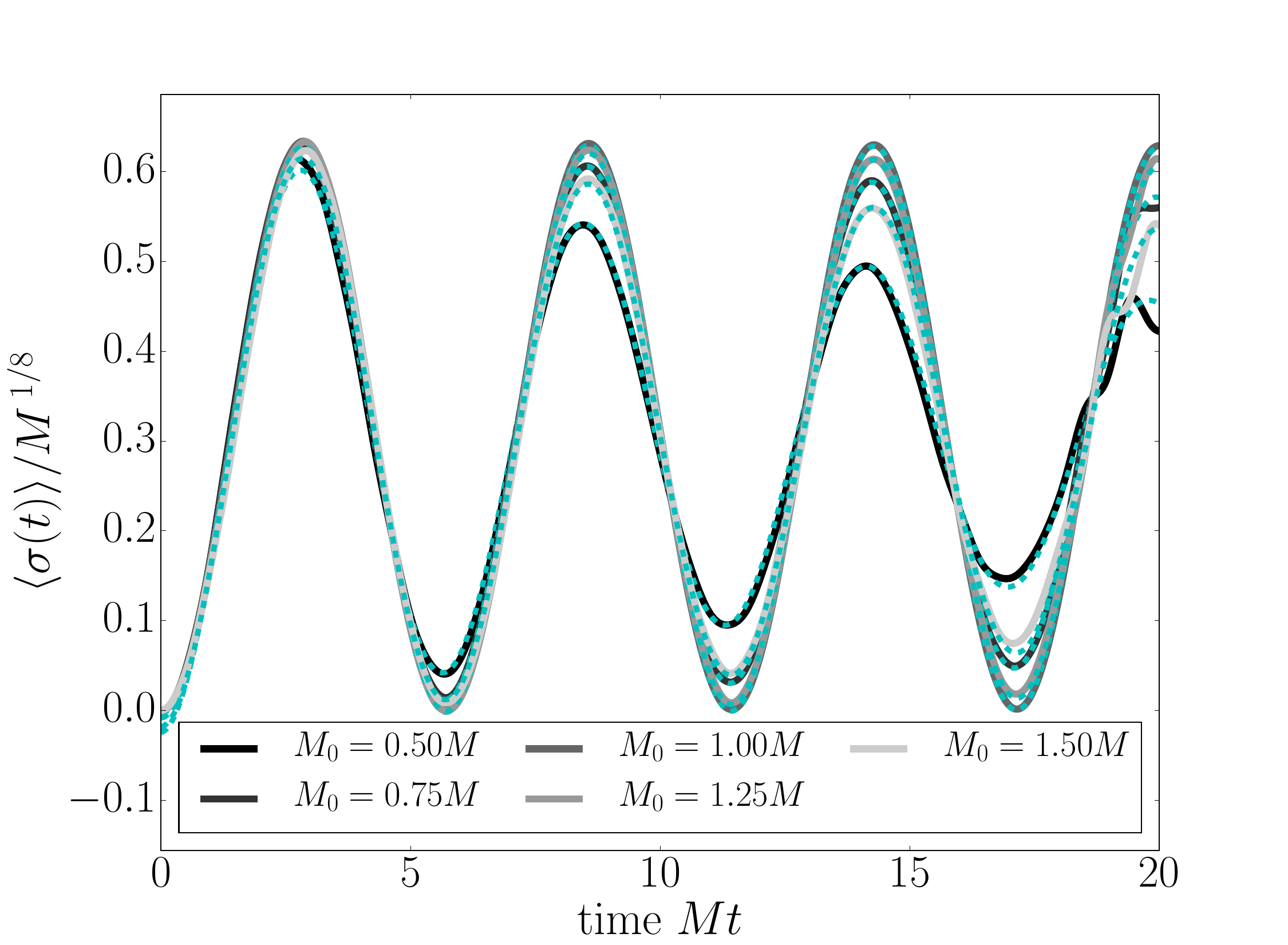}\label{fig:cut-off_extrapolation_largeh_para}}
\caption{  TFSA results (greyscale continuous lines) for $\langle\sigma(t)\rangle$ after cut-off extrapolation in the paramagnetic phase for quenches from different initial masses $M_{0}$ at system size
$ML=40$ and (a) $\bar{h}=-0.05$, (b) $\bar h = -0.1.$  The extrapolation details can be found in
Fig. \ref{fig:Cut-off-extrapolation-for-para-sigma-smallh} of Appendix
\ref{sub:Cut-off-extrapolation-schemes}. The result of fitting the function \eqref{eq:parafit} to the TFSA results is shown in green dashed lines.
%The blue dashed line is the (cut-off extrapolated) prediction for the asymptotic value obtained by evaluating the diagonal ensemble average using the overlaps numerically determined from TFSA.
}
\label{fig:sigma_para}
\end{figure}

\subsection{Time evolution of the order parameter}

To make contrast with the time evolution of $\langle\sigma(t)\rangle$ in the ferromagnetic
phase that was dominated by the presence of a few low energy meson
bound states arising due to the confinement of kink-like excitations,
we examine the dynamics of $\langle\sigma(t)\rangle$ in the paramagnetic case where no
such mesons are present. As shown in Fig. \ref{fig:sigma_para},  $\langle\sigma(t)\rangle$
shows a slightly damped sinusoidal oscillation.

In contrast to the ferromagnetic phase where the dynamics are dominated
by the non-perturbative effects of confinement, in
the paramagnetic phase a perturbative treatment of the post-quench
dynamics is expected to give a reasonable description of small quenches in $h$ (without changing the mass, i.e. $M_0=M$).
Such an approach was worked
out in Ref. \cite{2014JPhA...47N2001D}. We take the pre-quench Hamiltonian $H(M,0)$ to be the unperturbed Hamiltonian
and $\Delta H=h\int dx\,\sigma$
as the perturbation. Using the results of \cite{2014JPhA...47N2001D},
the leading perturbative contribution to the time evolution of
the magnetisation is given by the one-particle term

\begin{equation}
\langle\sigma(t)\rangle_\text{pert}=h\frac{2}{M^2}\left|F_{1,0}\right|^{2}\cos(Mt)+O(h^2)\,,
\label{Eq: Sigma(t) perturbative}
\end{equation}
where $F_{1,0}=\langle p=0|\sigma|0\rangle$ is the infinite volume form factor of $\sigma$
between the ground state $|0\rangle$ and a zero-momentum one-particle state
taken in the pre-quench basis (c.f.  \eqref{eq:sigmaFF}).
\begin{figure}[t!]
\centering{}\includegraphics[width=0.6\textwidth]{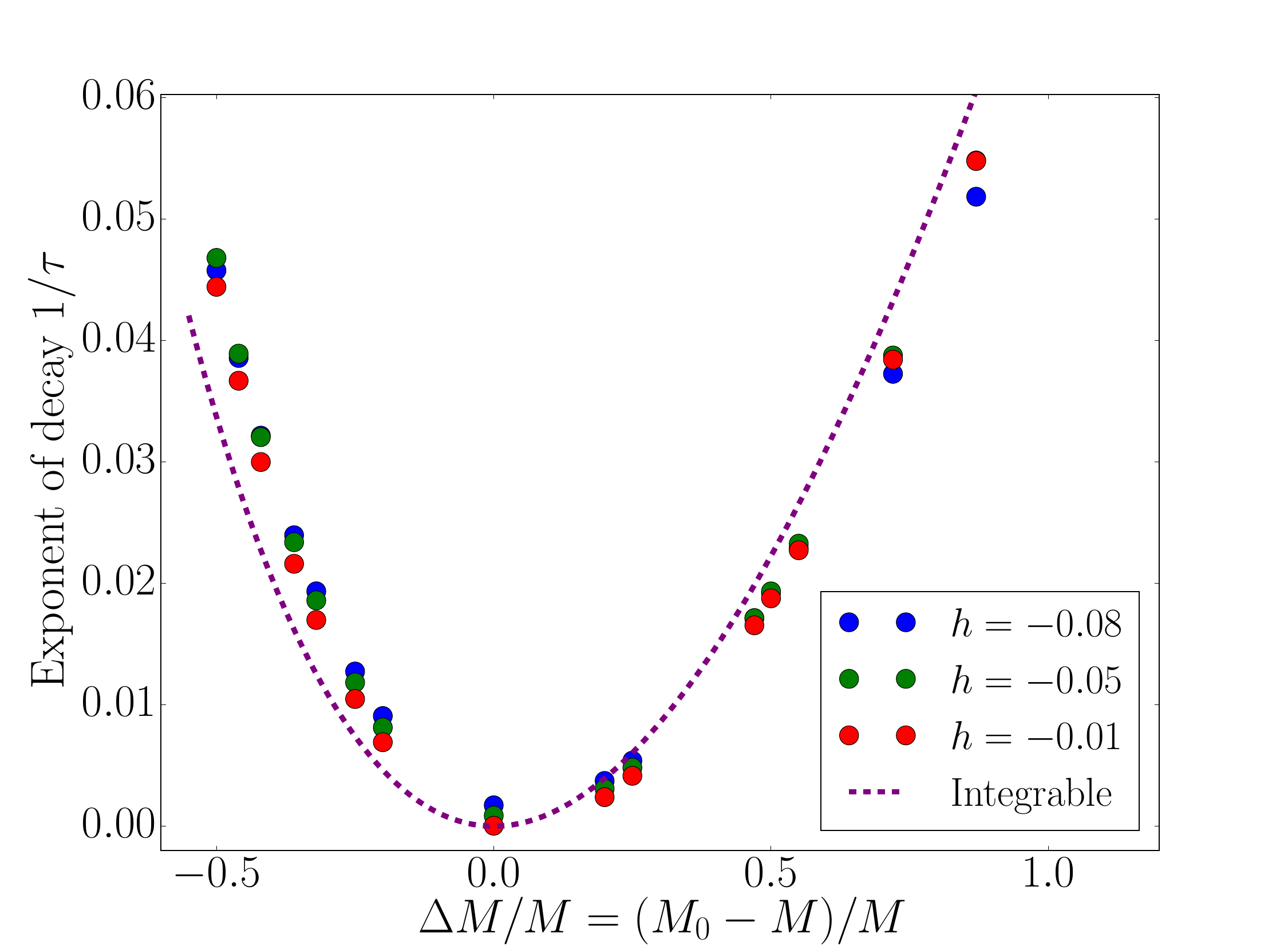}
\caption{Fitted value of the decay rate $\tau^{-1}$ in Eq. \eqref{eq:parafit} as a function of $\Delta M/M=(M-M_0)/M$ for different values of $h.$ The dashed purple line is the integrable decay rate given in Eq. \eqref{eq:tau_theor}.}
\label{fig:para_decay}
\end{figure}
\begin{table}
\centering
\begin{tabular}{|c|c|c|c|c|c|c|}
\hline
$h$  & $A_\text{pert.}$ & $A_\text{fit}$ & $m_\text{TFSA}$ & $\omega_\text{fit}$  & $\langle\sigma\rangle_\text{DE}$ & $C_\text{fit}$ \\ \hline
-0.01 & 0.0369 & 0.0368	& 1.0010		& 1.0011 & 0.035 & 0.037 \\ \hline
-0.02 & 0.0737 & 0.0732	& 1.0039		& 1.0045 & 0.069 & 0.073 \\ \hline
-0.03 & 0.1106 & 0.1087	& 1.0089 	& 1.0102 & 0.103 & 0.109 \\ \hline
-0.05 & 0.1843 & 0.1763	& 1.0244		& 1.0280 & 0.169 & 0.176 \\ \hline
-0.08 & 0.2950 & 0.2655	& 1.0607 	& 1.0683 & 0.245 & 0.264 \\ \hline
-0.10 & 0.3687 & 0.3172	& 1.0925 	& 1.0996 & 0.297 & 0.314 \\ \hline
\end{tabular}
\caption{Comparing fitted parameters of the $\langle\sigma(t)\rangle$ curves in the paramagnetic phase with other TFSA and perturbative data. $\langle\sigma\rangle_\text{DE}$  is the diagonal ensemble prediction for a mass quench $M_0 = 1.5 M$ at cut-off $\Lambda / M = 8$. $A$ and $f$ are fitted using the data for a quench with $M_0 = M$ while $C$ is fitted for a quench with $M_0 = 1.5 M$.}
\label{table:fits}
\end{table}
This formula does not account for the constant offset that is however expected due to the non-zero magnetic field. Moreover, for simultaneous quenches both in $h$ and $M$ the oscillation is seen to be damped (c.f. Fig. \ref{fig:sigma_para}). A natural function to fit the TFSA result is
\begin{equation}
\sigma(t) = A \cos(\omega t)e^{-t/\tau}+C\:.
\label{eq:parafit}
\end{equation}

For quenches in $h$ with $M_0=M$ there is no damping, so $\tau=0.$  In 
Table \ref{table:fits} we report the fitted values of the amplitude $A$ together 
with the perturbative prediction in \eqref{Eq: Sigma(t) perturbative}, the 
frequency $\omega$ with the $h$-dependent mass of the magnon, and the offset $C$ 
along with the expectation value of $\sigma$ in the diagonal ensemble (infinite 
time average). The agreement for all three parameters is excellent, showing that 
the offset is given by the infinite time average, and perturbation theory 
captures both the amplitude and the frequency of the oscillations although for 
the latter the post-quench value of the particle mass gives a better agreement. 

The mass correction can also be estimated in the framework of second order
form factor perturbation theory  \cite{2010NuPhB.825..466T} (including intermediate states 
with up to two particles) with the result
\begin{equation}
M(h)=M(1+\delta \bar h^2)\,,
\end{equation}
where 
\begin{align}
\delta & =  -\int_{0}^{\infty}\frac{d\theta}{2\pi}
\left(\frac{\left|M^{-1/8}F^{\sigma}_3(i\pi,\theta,-\theta)\right|^{2}}
{\cosh\theta(2\cosh\theta-1)}-
\frac{16\left|M^{-1/8} F^{\sigma}_1(0)\right|^{2}}{\sinh^{2}\theta\cosh\theta}\right) 
+  4\left|M^{-1/8}F_{1}^{\sigma}(0)\right|^{2}\nonumber \\
& =  10.7593\dots
\end{align}
which is quite close to the value extracted from fitting the mass gap 
data in Table \ref{table:fits} up to ${\bar h}=-0.05$: 
\begin{equation}
\delta=\begin{cases}
        9.77\dots&\quad \mathrm{from }\;M(h)\,,\\ 
        10.073\dots&\quad \mathrm{from }\;\omega_\text{fit}\,.
        \end{cases}
\end{equation}
This shows that the quench spectroscopy works also in the paramagnetic phase. Apart from the mass of the particle excitation, the quench dynamics also allows for the determination of the form factor appearing in the amplitude in 
\eqref{Eq: Sigma(t) perturbative}.

For combined quenches both in $h$ and $M,$ we found that the function 
\eqref{eq:parafit} with $\tau>0$ describes almost perfectly the evolution of the 
magnetisation, at least for moderate size quenches (see Fig. 
\ref{fig:sigma_para}). The frequency of the oscillation is found to be largely 
independent of $M_0,$ it changes around $1\%$ while $M_0$ is varied between 
$0.5M$ and $1.9M.$ The change in the offset $C$ is around $4-5\%,$  
while the amplitude is a bit more sensitive to the initial mass. 

The damping rate, $\tau^{-1},$ is shown in Fig. \ref{fig:para_decay} for various 
values of $h$ and $M_0.$ We found that it depends only very weakly on the value 
of $h$ for small $h.$ Moreover, it is close to the damping rate valid for 
integrable mass quenches at $h=0,$ given by Eq. \eqref{eq:tau_theor} and shown 
in dashed line in Fig. \ref{fig:para_decay}. For integrable quenches within the 
paramagnetic phase the magnetisation is zero for all times, but expression 
\eqref{eq:tau_theor} holds for quenches from the ferromagnetic phase to the 
paramagnetic phase \cite{2012JSMTE..07..016C}. Interestingly, we found that the same expression gives the 
decay rate for quenches within the paramagnetic phase when the post-quench 
magnetic field is non-zero. The reason for this is that the main source of 
damping is the finite density of magnons which is predominantly induced by the 
quench in the mass, described by Eq. \eqref{eq:tau_theor}. The small deviations 
are due to higher order effects both in $h$ and in $\Delta M.$

\subsection{Statistics of work and Loschmidt echo}

\begin{figure}
\subfloat[Statistics of work]{\includegraphics[width=0.515\textwidth]{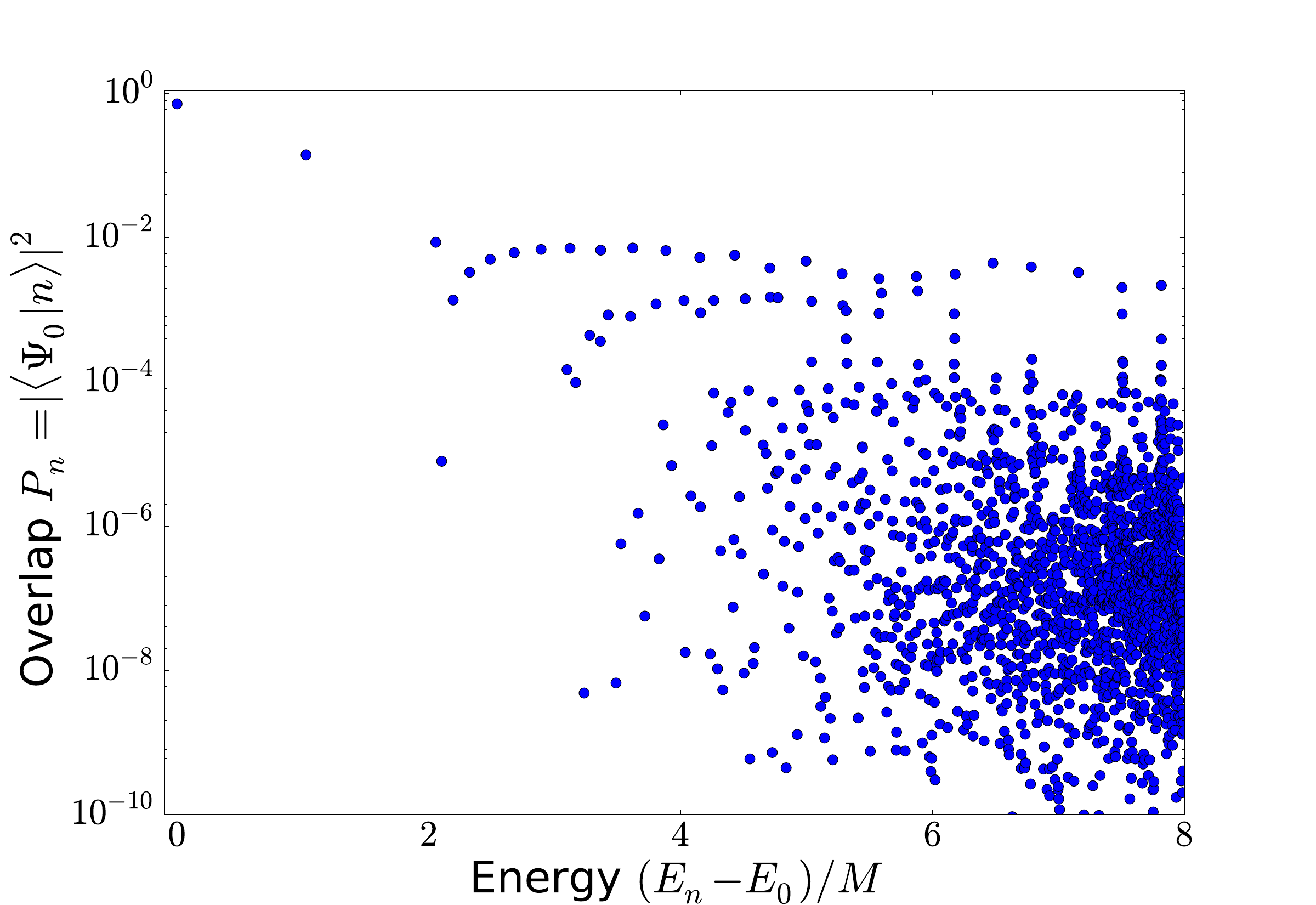}\label{fig:PW_para_nonint}}
\subfloat[Loschmidt echo]{\includegraphics[width=0.485\textwidth]{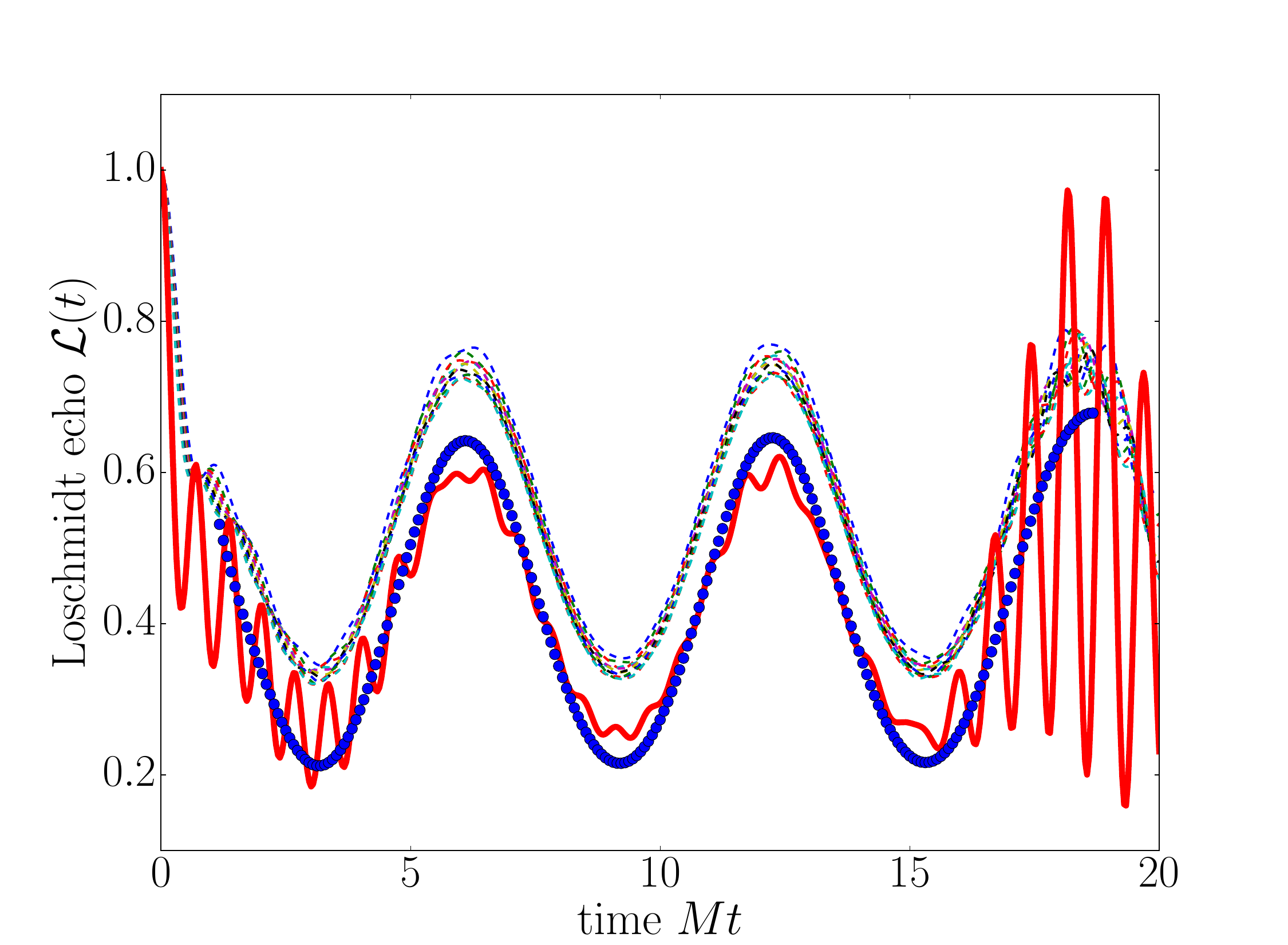}\label{fig:para_Loschmidt} }
\caption{ (a) Statistics of work $P(W)$ for a quench in the paramagnetic
phase with $M_{0}=1.5M,$ $\bar{h}=-0.05,$ $ML=40,$
and $\Lambda=8M.$ (b) Cut-off-extrapolation of the Loschmidt
echo for the same quench as in (a). The dashed lines are the raw data for different cut-offs, the solid red line shows the cut-off extrapolated curve, and the result of the extrapolation of the smoothened curves is shown in solid blue line.}
\end{figure}

As we saw above, the time evolution of $\sigma$ is dominated by a
single frequency in the paramagnetic case. To support this observation,
in Fig. \ref{fig:PW_para_nonint} we show the statistics of work in this phase that is similar to the
ferromagnetic case in that it is dominated by low lying states. The difference is that there is a single massive particle leading to a pure oscillation instead of multiple frequencies. In Fig. \ref{fig:para_Loschmidt} the Loschmidt
echo is plotted in the paramagnetic phase featuring a persistent oscillation with a single frequency. 

\section{Comparison with iTEBD lattice simulations \label{sec:TFSA-versus-iTEBD}}

As a final check of our TFSA method, in this section we compare our results for the time evolution of the magnetisation with simulations on the Ising spin chain with both transverse and longitudinal fields governed by the Hamiltonian \eqref{eq:QIM_Hamiltonian}. On the lattice we used the infinite volume Time Evolving Block Decimation (iTEBD) method which is free of finite size effects and thus allows for testing the finite size errors of our TFSA results.

\begin{figure}[t!]
\subfloat[Rescaled iTEBD results for $J=5,10,20$.]{\includegraphics[width=.48\textwidth]{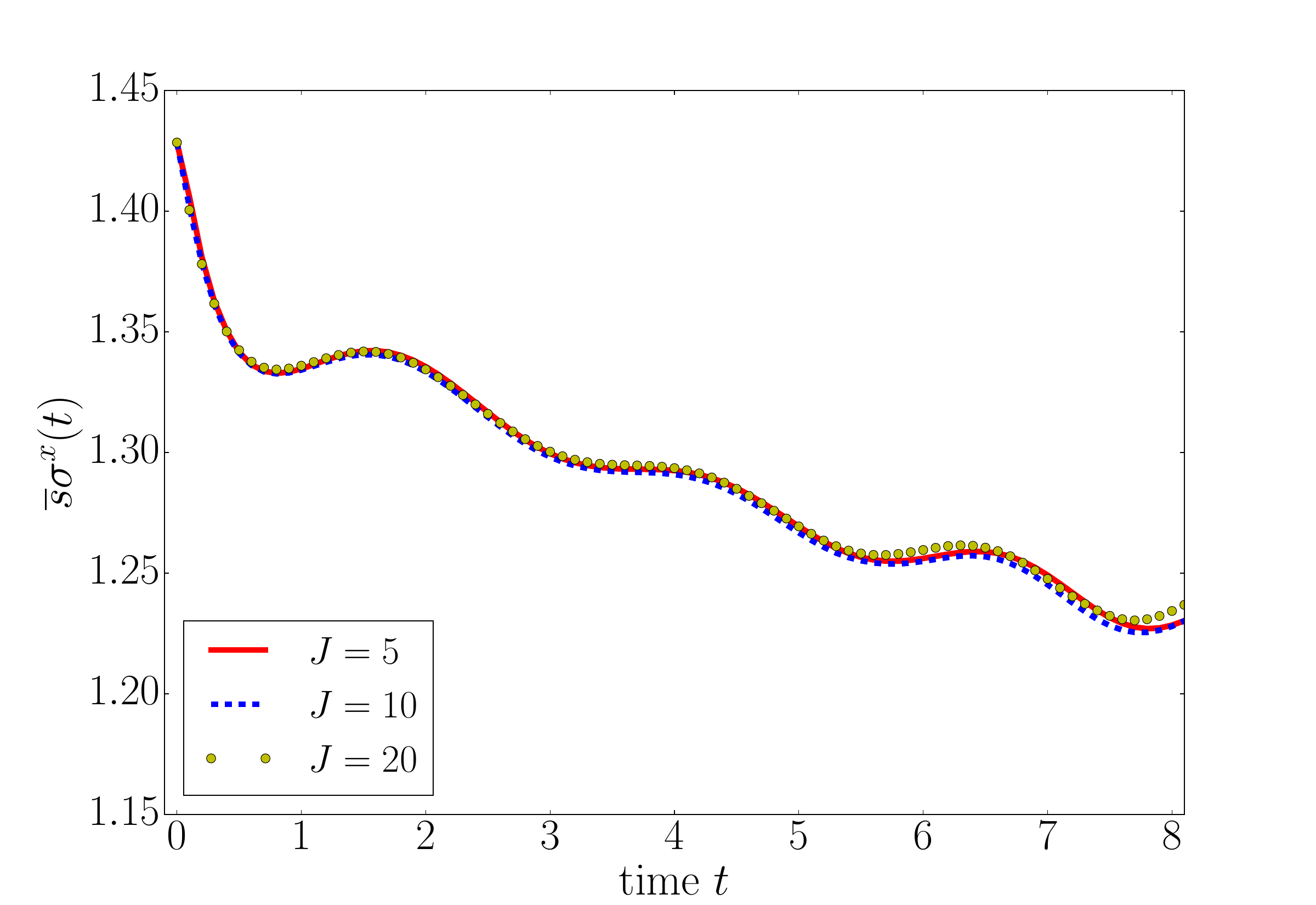}\label{fig:scalinglimit}}
\hfill
\subfloat[Quench from $H(1.5M,0)$ to $H(M,\bar h =-0.1)$ in the ferromagnetic phase.]{\includegraphics[width=.48\textwidth]{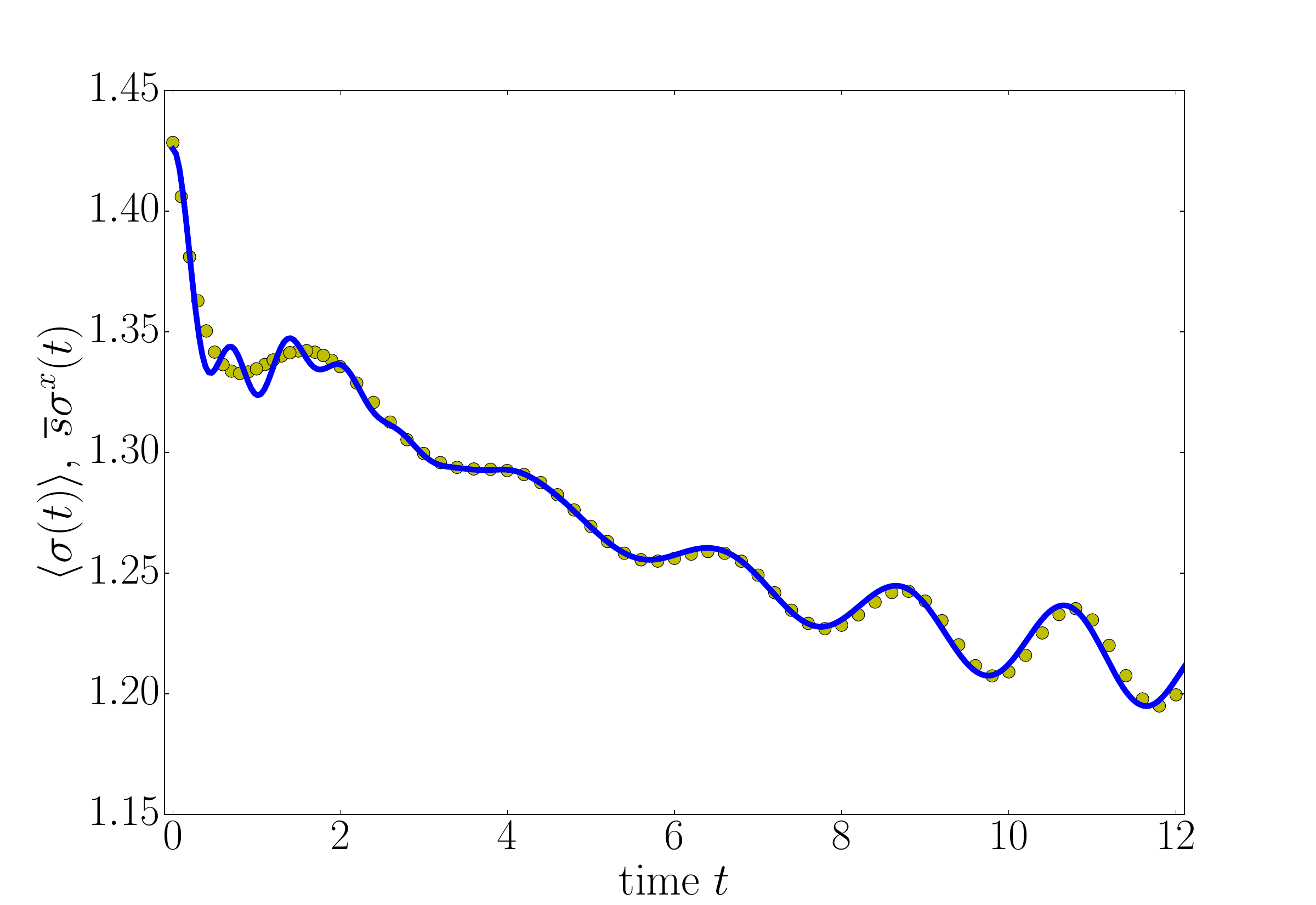}\label{fig:ferroJ5}}\\
\subfloat[Quench from $H(M,0)$ to $H(M,\bar h =-0.05)$ in the ferromagnetic phase.]{\includegraphics[width=.48\textwidth]{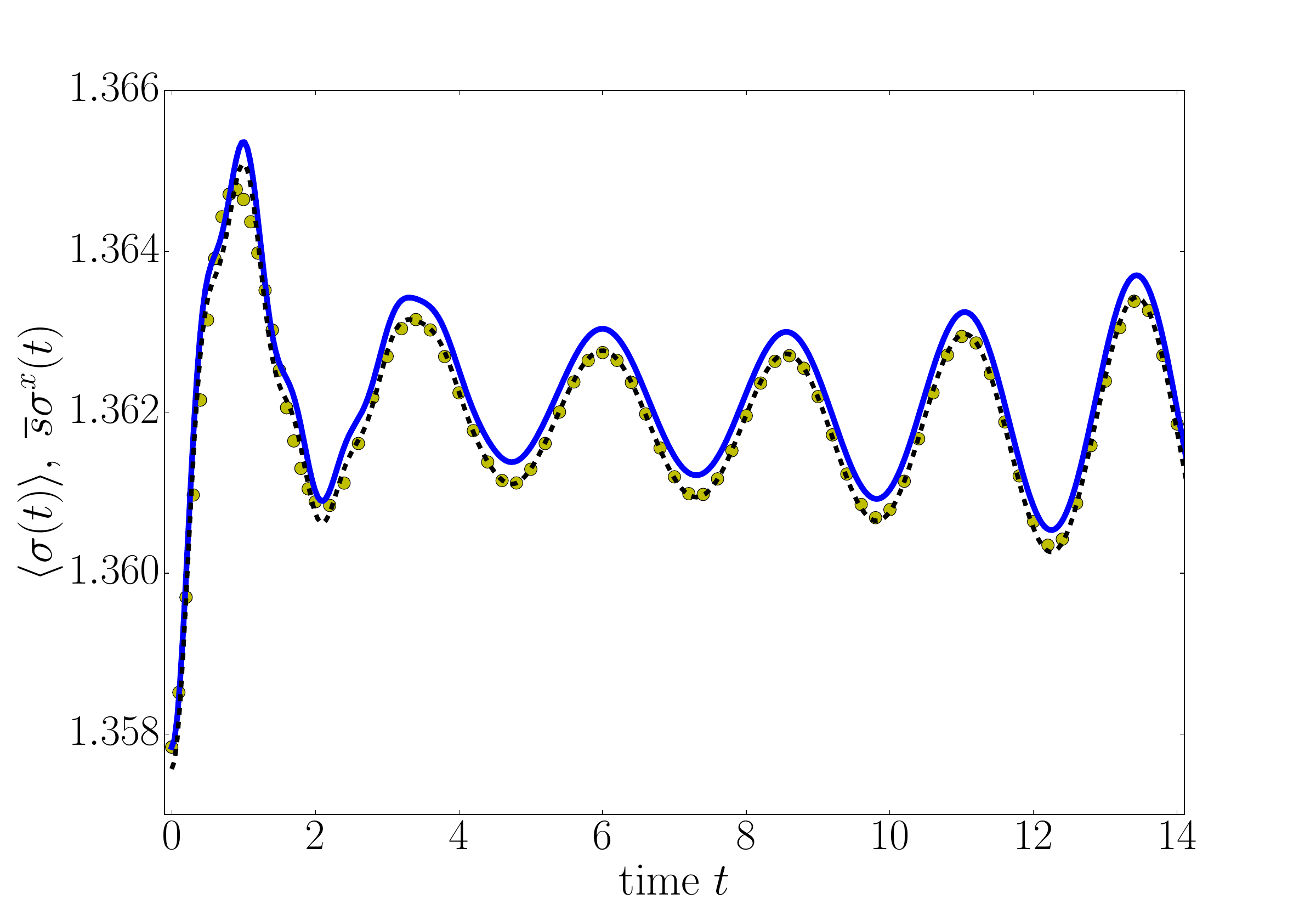}\label{fig:ferroJ5h005}}
\hfill
\subfloat[Quench from $H(1.5M,0)$ to $H(M,\bar h =-0.05)$ in the paramagnetic phase.]{\includegraphics[width=.48\textwidth]{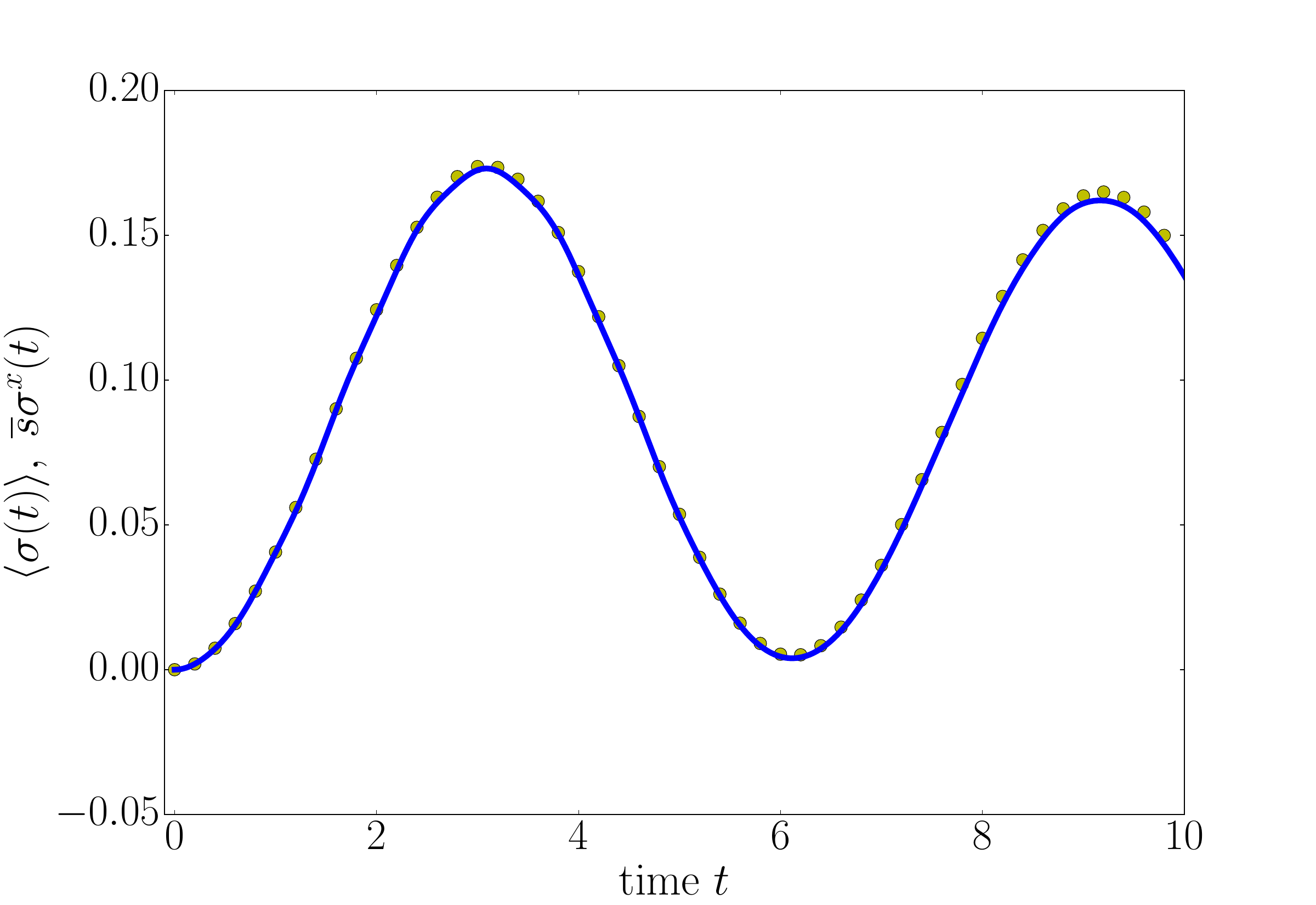}\label{fig:paraJ5}}
\caption{Comparison between iTEBD lattice simulations and the results of the TFSA. (a) Rescaled results of the iTEBD simulations at $J=5,10,20$ corresponding to the quench $H(1.5,0)\to H(1,-0.1)$ in the ferromagnatic phase. (b,c,d) Cut-off extrapolated TFSA results at system size $ML=40$ (solid blue line) and iTEBD results (green dots) for the quench (b) $H(1.5,0)\to H(1,-0.1)$ in the ferromegnatic phase, (c) $H(1,0)\to H(1,\bar h =-0.05)$ in the ferromagnetic phase, (d) $H(1.5,0)\to H(1,-0.05)$ in the paramagnetic phase. In (c) the dashed black line shows the TFSA result multiplied by 0.9998.}
\end{figure}

In order to carry out the comparison, the lattice simulation should approach the scaling limit, i.e. it must be performed at sufficiently large values of $J$ near the quantum critical point such that the mass $M=2J\left|1-h_{z}\right|$ is kept fixed. If the system is close enough to the critical point, the numerical results on the lattice and in the field theory should scale on top of each other indicating the validity of the field theoretical description. Using Eqs. \eqref{eq:M-hz}, \eqref{eq:sigmanorm}, \eqref{eq:hbardef} we can connect the various parameters and operators on the lattice and in the continuum. In particular, when $h_z$ is at a finite distance from $h_z^\text{crit.}=1$, relation \eqref{eq:sigmanorm} should be modified as
\begin{equation}
\label{eq:sigmanorm2}
\sigma(na) = \bar s J^{1/8}\left(\frac{1+h_z}2\right)^{-1/8} \sigma_n^x \equiv \tilde s\, \sigma_n^x\,.
\end{equation}

In Fig. \ref{fig:scalinglimit} the results for $\langle\sigma^x_n(t)\rangle$ of three iTEBD simulations are shown with $J=5,10,20$ rescaled according to Eq. \eqref{eq:sigmanorm2} such that they all correspond to the quench $H(M_0=1.5,\bar h = 0)\to H(M=1,\bar h=0.1)$ in the ferromagnetic phase. The three curves are almost indistinguishable from each other, so they are very close to the universal curve corresponding to the field theory. In Fig. \ref{fig:ferroJ5} we compare this curve with our TFSA result and find excellent agreement. We stress that this comparison has no free parameters. Similar comparisons are shown for a ferromagnetic quench in the $h$ direction without a mass quench in Fig. \ref{fig:ferroJ5h005} and for a combined quench both in $M$ and $h$ in the paramagnetic phase in Fig. \ref{fig:paraJ5}. For the quench only in $h$ shown in Fig. \ref{fig:ferroJ5h005} some deviations can be seen. However, the magnitude of these deviations is very small (note the scale on the vertical axis). Moreover, the discrepancy is largely eliminated by modifying relation \eqref{eq:sigmanorm2} by a factor $\approx 0.9998,$ the corresponding curve is shown in thin black line.

From the very good agreement between the lattice and the TFSA results we can draw the following conclusions. First, as the iTEBD does not suffer from finite size corrections, the finite size errors in the TFSA are very small, presumably exponentially suppressed. The finite size $L$ however restricts the simulation to times $t\le L/2.$ Secondly, since the iTEBD method truncates the Hilbert space based on entanglement entropy instead of energy and this truncation is very well controlled, we conclude that the extrapolation in the TFSA cut-off has a rather small error and eliminates the effect of the finite energy cut-off to a great extent. 
%Finally, the agreement also shows that the scaling limit of quantum quenches on the lattice agrees with quenches in the field theory, so beyond equilibrium phenomena the field theory also describes non-equilibrium dynamics of the spin chain.

Let us emphasise that the TFSA is much more efficient to study the dynamics of the field theory than the iTEBD on the lattice in the scaling limit. The reason is that approaching the quantum critical point, the lattice simulation takes more and more time due to the critical slowing down. In practice, this means that each simulation of a given quench can take days or even weeks depending on the distance from the critical point (i.e. the value of $J$) on an Intel 3.60 GHz processor. In contrast, in the TFSA the most time consuming step is the computation of the matrices   that can take days for Hilbert spaces of dimension $\sim 10^5.$ However, once the matrices are constructed they can be used to investigate many different quenches as the simulation of the time evolution only takes a few hours.

\section{Conclusions \label{sec:Conclusions}}

In this paper we studied the real time evolution following quantum quenches in 
the scaling Ising field theory using the Truncated Hilbert Space Approach. Our 
main results are twofold. 

On the one hand, we demonstrated that the TFSA method is suitable for studying 
the quench dynamics in one dimensional interacting field theories. Due to the 
microscopic nature of the method it has access to many quantities, including the 
individual overlaps, the work statistics, or the Loschmidt echo that are hard 
to access with other methods. As there is no universal numerical method 
available for studying the out of equilibrium dynamics in continuum systems, the 
present proof of principle study can have far reaching consequences. This 
is especially so since using already existing truncated Hamiltonian 
approaches, the method  can be easily generalised to other field theories. 
An obvious candidate is the sine--Gordon model for which the method was first developed
in \cite{1998PhLB..430..264F} and which is an important model providing the
 low energy effective description of various one dimensional systems, including coupled 1D quasi-condensates currently studied in matter wave interference experiments 
\cite{2015arXiv150503126S}. The method can also be generalised to arbitrary finite duration 
quench protocols or ramps.

On the other hand, we found interesting dynamics in the non-integrable Ising 
field theory. For quenches in the ferromagnetic phase we found that the weak 
confinement of the elementary excitations of the integrable model has drastic 
consequences for the quench dynamics, in accordance with a recent study of 
quenches in the lattice version of the model \cite{2016arXiv160403571K}. 
We also found that in both phases the quench dynamics is dominated by a 
few low energy states even for relatively long times. In the paramagnetic 
phase this can be explained borrowing a perturbative approach developed by 
Delfino in \cite{2014JPhA...47N2001D}, while in the ferromagnetic 
phase this turns out to be connected to low density during the quench, 
leading to the one-particle states dominating the system up to system 
sizes of hundreds of the correlation length. This means that measuring specific observables either in numerical simulations or in experiments allows one to extract important information about the spectrum and form factors of the post-quench Hamiltonian, realising a kind of quench spectroscopy.

Finally, the agreement between the lattice iTEBD and the continuum
TFSA also shows that the scaling limit of quantum quenches on the lattice
agrees with sudden quenches in the field theory. Therefore, the
universal description given by the scaling field theory
in the vicinity of the critical point is valid beyond the equilibrium and
captures the non-equilibrium dynamics of the spin chain.

\subsection*{Acknowledgements}

This research was supported by the Momentum grant LP2012-50 of the
Hungarian Academy of Sciences. M.K. was partially supported by the EU Marie Curie IIF Grant PIIF-GA-2012- 330076 and a J\'anos Bolyai Research Scholarship of the Hungarian Academy of Sciences. M.C. was partially supported by the ERC under Starting Grant 279391 EDEQS (MC). We are grateful to Gy\H{o}z\H{o} Egri who took part in the early stage of this work, and to Robert Konik and Neil Robinson 
for helpful discussions.

\providecommand{\href}[2]{#2}\begingroup\raggedright\endgroup

\newpage

\appendix

\section{Some details of the finite volume Ising field theory}

In this appendix we review some basic ingredients of the finite volume Ising field theory, such as the structure of the Hilbert space in the two phases, and the finite volume matrix elements of the magnetisation operator. We also provide some details on implementing the time evolution using the Chebyshev expansion.

\subsection{The ferro-/paramagnetic phases in field theory \label{sub:The-ferro/paramagnetic-phases}}

As discussed in the main text, the free fermion Hamiltonian \eqref{eq:HamiltonianMajorana}
is the same in both the paramagnetic and ferromagnetic phases. To
understand the eventual difference between the phases at the level
of the field theory, one can consider the Jordan--Wigner transformation
from the spin chain \eqref{eq:QIM_Hamiltonian} to the fermionic formulation.
This maps the $\mathbb{Z}_{2}$ symmetry of the Ising chain
\begin{equation}
\sigma^{x,y}\rightarrow-\sigma^{x,y}\qquad\sigma^{z}\rightarrow\sigma^{z}
\end{equation}
into the fermionic parity $e^{i\pi N}$, where $N$ is the fermion
number operator. The mapping to fermions specifies the boundary conditions
to be periodic in the odd $N$ sector and anti-periodic in the even
$N$ sector. Consequently, the set of possible wave numbers is given
by integer multiples of $2\pi/L$ for $N$ odd, and half-integer multiples
for $N$ even, corresponding to the Ramond (R) and Neveu-Schwarz (NS)
sectors:
\begin{align}
p_{n} & =\frac{2\pi n}{L}\quad n\in\mathbb{Z}\quad\text{Ramond,}\\
k_{n} & =\frac{2\pi n}{L}\quad n\in\mathbb{Z}+\frac{1}{2}\quad\text{Neveu-Schwarz}.
\end{align}
The excitation energy of the lowest lying excitation compared to
the Fock vacuum is proportional to $h_{z}-1$ which becomes negative
in the ferromagnetic phase, so the ground state must be redefined
by filling this negative energy level. As a consequence, the Ramond
sector in the ferromagnetic phase has states with even number of particles,
in contrast to the the paramagnetic phase. The number of particles
in the Neveu--Schwarz sector is even in both phases.

To summarise, the basis used in the ferromagnetic phase is given by
\begin{align}
|k_{1},\dots,k_{2m}\rangle_{NS} & =\prod_{j=1}^{2m}a_{k_{j}}^{\dagger}|0\rangle_\text{NS}\quad k_{j}\in \text{NS,}\\
|p_{1},\dots,p_{2m+1}\rangle_{R} & =\prod_{j=1}^{2m+1}a_{p_{j}}^{\dagger}|0\rangle_\text{R}\quad p_{j}\in \text{R,}
\end{align}
while the basis in the paramagnetic phase is 
\begin{align}
|k_{1},\dots,k_{2m}\rangle_{NS} & =\prod_{j=1}^{2m}a_{k_{j}}^{\dagger}|0\rangle_\text{NS}\quad k_{j}\in \text{NS,}\\
|p_{1},\dots,p_{2m}\rangle_{R} & =\prod_{j=1}^{2m}a_{p_{j}}^{\dagger}|0\rangle_\text{R}\quad p_{j}\in \text{R,}
\end{align}
where $|0\rangle_\text{R}$ and $|0\rangle_\text{NS}$ are the Ramond and Neveu--Schwarz
vacua, respectively, and $a_{p}^{\dagger}$ creates a fermion particle
with momentum $p$ and mass $M$.

\subsection{Matrix elements of $\sigma$ \label{sub:sigmaFF}}

In the Ising field theory, the order parameter field $\sigma$ only
has matrix elements between the even parity (Neveu--Schwarz) and odd
parity (Ramond) sectors. In the eigenbasis of $H(M,0)$ in Eq. \eqref{eq:HamiltonianMajorana},
the matrix elements have the following exact expression \cite{2001hep.th...12167F,Bugrij:2000is,2001hep.th....7117B}:
\begin{multline}
_{\text{NS}}\langle k_{1},k_{2},\dots,k_{m}|\sigma(0,0)|l_{1},l_{2},\dots,l_{n}\rangle_{\text{R}}\\
=S(ML)\prod_{j=1}^{m}\tilde{g}(\theta_{k_{j}})\prod_{i=1}^{n}g(\theta_{p_{i}})F_{m,n}(\theta_{k_{1}},\dots,\theta_{k_{m}}|\theta_{l_{1}},\dots,\theta_{l_{n}}),\label{Eq:Sigma matrix elements}
\end{multline}
where $\theta_{n}$ is the rapidity corresponding to momentum $p=2\pi/L$
as defined in \eqref{eq:finvol_rapidity}. $F_{m,n}$ is the infinite
volume matrix element (form factor) given by
\begin{multline}
F_{m,n}(\theta_{1},\dots,\theta_{m}|\theta'_{1},\dots,\theta'_{n})\\
=i^{\left[\frac{m+n}{2}\right]}\bar{\sigma}\prod_{0<i<j\leq m}\tanh\left(\frac{\theta_{i}-\theta_{j}}{2}\right)\prod_{0<p<q\leq n}\tanh\left(\frac{\theta'_{p}-\theta'_{q}}{2}\right)\prod_{\substack{0<s\leq m\\
0<t\leq n
}
}\coth\left(\frac{\theta_{s}-\theta'_{t}}{2}\right),\label{eq:sigmaFF}
\end{multline}
with $[x]$ being the integer part of $x$, and $\bar{\sigma}$ is
the infinite volume expectation value of the order parameter field
given in \eqref{eq:sigmabar}. 

The other ingredients in \eqref{Eq:Sigma matrix elements} account
for finite size effects. The overall factor $S(ML)$ is the ratio
between the vacuum expectation value in finite and infinite volumes
\begin{equation}
\bar{\sigma}S(L)=\begin{cases}
_{\text{NS}}\langle0|\sigma(0,0)|0\rangle_{\text{R}}\quad\mbox{ferromagnetic phase}\,,\\
_{\text{NS}}\langle0|\mu(0,0)|0\rangle_{\text{R}}\quad\text{paramagnetic phase}\,,
\end{cases}
\end{equation}
(where $\mu$ is the disorder field) and is given by the following
formula:
\begin{equation}
S(ML)=\exp\left\{ \frac{(ML)^{2}}{2}\int\int_{-\infty}^{\infty}\frac{\mathrm{d}\theta_{1}\mathrm{d}\theta_{2}}{(2\pi)^{2}}\frac{\sinh\theta_{1}\sinh\theta_{2}}{\sinh(ML\cosh\theta_{1})\sinh(ML\cosh\theta_{2})}\log\left|\coth\frac{\theta_{1}-\theta_{2}}{2}\right|\right\} \:.
\end{equation}
The momentum dependent leg factors $g$ and $\tilde{g}$ are defined
as
\begin{equation}
g(\theta)=\frac{e^{\kappa(\theta,ML)}}{\sqrt{ML\cosh{\theta}}}\,,\qquad\qquad\tilde{g}(\theta)=\frac{e^{-\kappa(\theta,ML)}}{\sqrt{ML\cosh{\theta}}}\,,
\end{equation}
where
\begin{equation}
\kappa(\theta,ML)=\int_{-\infty}^{\infty}\frac{\mathrm{d}\theta'}{2\pi}\frac{1}{\cosh(\theta-\theta')}\log\left(\frac{1-e^{-ML\cosh{\theta'}}}{1+e^{-ML\cosh{\theta'}}}\right)\,.
\end{equation}
It is important to note that the function $S(ML)$ rapidly approaches
the value $1$ for large $ML$: $S(ML=15)-1$ is already of the order
$\mathcal{O}(10^{-14}$). Similarly $\kappa(\theta,ML)\approx0$ to
very high precision for all $\theta$ when $ML\geq20$. For the volumes
used in most of our simulations ($ML\geq30$) we can therefore substitute
$\kappa=0$ and $S=1$.

\subsection{Implementing the quench time evolution\label{sub:Implementing-the-quench}}

The Chebyshev polynomials $T_{n}(x)$ are
defined by the recurrence relation
\begin{equation}
\label{eq:Trecursion}
T_{n+1}(x)  =2xT_{n}(x)-T_{n-1}(x)\,,\qquad
T_{0}(x)=1\,,\quad T_{1}(x)=x\,.
\end{equation}
They satisfy the orthogonality condition
\begin{equation}
\int_{-1}^{1}\frac{\mathrm{d}x}{\sqrt{1-x^{2}}}\,T_{n}(x)\,T_{m}(x)=\begin{cases}
\pi & \quad\text{if }n=m=0\\
\delta_{nm}\frac{\pi}{2} & \quad\text{otherwise}\:,
\end{cases}
\end{equation}
and form a complete basis for functions on the interval $[-1,1]$.
This can be used to evaluate functions of matrices whose eigenvalues
lie within the unit circle. In particular, the exponential function
appearing in the time evolution operator can be expressed as
\begin{equation}
e^{-iHt}=J_{0}(t)1\!\!1+2\sum_{n=1}^{\infty}(-i)^{n}J_{n}(t)T_{n}(H)\,,
\end{equation}
where $J_n(z)$ are the Bessel functions 
\begin{equation}
J_{n}(z)=\sum_{l=0}^{\infty}\frac{(-1)^{l}}{l!(k+l)!}\left(\frac{z}{2}\right)^{2l+k}\,.
\end{equation}
In order for the above expansion to be valid the Hamiltonian $H$
must be shifted and rescaled so that all its eigenvalues lie inside
the unit circle; the effect of this transformation on the time evolution
can be undone by an appropriate rescaling of the time $t$. In a practical
calculation the expansion is truncated at some $n$; for our purposes we
used a truncation at $n=1000$.

In practice, there is no need to evaluate the time evolution operator itself, only its action on the initial state, that is, the time evolved wave function. Thanks to the recursion relations \eqref{eq:Trecursion}, this only requires matrix-vector multiplication instead of matrix-matrix multiplications.

Once we have the time evolved wave function at a given time as an $N_\text{cut}$ dimensional vector, the calculation of expectation values simply amounts to multiplication between this vector and the numerical matrix of the operator.

\section{Cut-off extrapolation and finite size effects \label{sub:Cut-off-extrapolation-schemes}}

In this appendix we discuss the cut-off effects in the TFSA method and provide details on the extrapolation in the cut-off. 

\subsection{Renormalisation group improvement of TFSA \label{sub:Renormalization-group-improvemen}}

Since our interest is in the continuum field theory, it is desirable
to eliminate the cut-off dependence inherent in the TFSA. This is
achieved by introducing running parameters $m$ and $h$ that are
adjusted according to suitable renormalisation group equations; this
idea was recently introduced in various truncated Hamiltonian approaches
to field theories \cite{2006hep.th...12203F,2007PhRvL..98n7205K,2015PhRvD..91b5005H}.

The Ising field theory can be represented as a relevant perturbation
of its ultraviolet fixed point: 

\begin{equation}
H_\text{IFT}=H_{\text{CFT}}+\tau\int_{0}^{L}\text{d}x\,\varepsilon(x)+h\int_{0}^{L}\text{d}x\,\sigma(x).
\end{equation}
where the fixed point Hamiltonian 
\begin{equation}
H_\text{CFT}=\frac{1}{2\pi}\int_{-\infty}^{\infty}dx\,\frac{i}{2}\left(\psi(x)\partial_{x}\psi(x)-\bar{\psi}(x)\partial_{x}\bar{\psi}(x)\right)
\end{equation}
is the conformal field theory of the massless Majorana fermion, and
\begin{equation}
\tau=\frac{M}{2\pi}\:.
\end{equation}
The leading order renormalisation group equations for the couplings
$\tau$ and $h$ are \cite{2011arXiv1106.2448G}

\begin{align*}
\frac{\text{d}\tilde{h}}{\text{d}n} & =\frac{\tilde{h}\tilde{\tau}}{2\pi\,n^{2}}\,,\\
\frac{\text{d}\tilde{\tau}}{\text{d}n} & =\frac{\tilde{h}^{2}}{4\Gamma(-3/8)^{2}n^{15/4}}\,,
\end{align*}
where $\tilde{\tau}=L\tau$ and $\tilde{h}=L^{15/8}h/(2\pi)^{7/8}$
are the dimensionless versions of the couplings and $n$ is the dimensionless
conformal cut-off (specified as the highest level of conformal descendants
kept after truncation). 

To the leading order, the TFSA has the same running for the couplings since
the introduction of mass affects only low-lying states and is a sub-leading
effect, therefore we only need to transcribe the above RG equations
in terms of the parameters of the TFSA. The cut-off in the free fermion
theory can be written in terms of the conformal cut-off as $\Lambda=4\pi n/L$;
in order to have a dimensionless cut-off $\lambda$ and system length
$l$ we introduce some mass scale $\overline{m}$ (to be specified
later):

\begin{equation}
\Lambda=\lambda\,\overline{m}\,,\qquad L=l/\overline{m}\,,
\end{equation}
so that 
\begin{equation}
n=\frac{l\lambda}{4\pi}\,.
\end{equation}
Using these relations we can write the effective couplings at some
value of the free fermionic cut-off $\lambda$ in terms of $\tilde{\tau}(n)$
and $\tilde{h}(n)$ as

\begin{align}
M(\lambda) & =2\pi\tau\left(\frac{l\lambda}{4\pi}\right)=\overline{m}\frac{2\pi}{l}\tilde{\tau}\left(\frac{l\lambda}{4\pi}\right)\,,\nonumber \\
h(\lambda) & =\frac{(2\pi)^{7/8}}{L^{15/8}}\tilde{h}\left(\frac{l\lambda}{4\pi}\right)=\overline{m}^{15/8}\frac{(2\pi)^{7/8}}{l^{15/8}}\tilde{h}\left(\frac{l\lambda}{4\pi}\right)\,.\label{eq:ff_cut-off}
\end{align}
Ideally, it would be best to choose the mass scale $\overline{m}$
as the effective post-quench mass $M(\lambda)$ used in the numerical
simulations. However, in that case the first equation in \eqref{eq:ff_cut-off}
gives

\begin{equation}
\frac{l}{2\pi}=\tilde{\tau}\left(\frac{l\lambda}{4\pi}\right),\label{eq:l_dep}
\end{equation}
so in order to ensure $\overline{m}=M(\lambda)$ the value of $l$
must be adjusted as a function of the cut-off. This is very inconvenient
as it means that the basis for the TFSA must be regenerated each time
the cut-off is adjusted, which is quite time consuming.

On the other hand, choosing $l$ to be independent of $\lambda$,
the RG equations can be rewritten in terms of the dimensionless parameters 

\begin{equation}
\hat{m}=\frac{M}{\overline{m}}\,,\qquad\hat{h}=\frac{h}{\overline{m}^{15/8}}
\end{equation}
as follows:
\begin{equation}
\frac{\text{d}\hat{h}}{\text{d}\lambda}=\frac{\hat{h}\hat{m}}{\pi\lambda^{2}}\quad\frac{\text{d}\hat{m}}{\text{d}\lambda}=\frac{2^{3/4}(2\pi)^{2}}{\Gamma(-3/8)^{2}}\frac{\hat{h}^{2}}{\lambda^{15/4}}\,.
\end{equation}

The numerical solution of the above equations shows that the parameter
$h$ changes significantly while flowing from $\lambda=\infty$ to
the values of the cut-off used for the TFSA simulations ($\lambda<10$).
However, to a good approximation $M$ remains constant which means
that using $M=1$ without adjusting $l$ is a very good approximation.
This leads to the choice
\begin{equation}
\overline{m}=M\,,
\end{equation}
i.e. we choose units in which $M=1$.

In contrast, the value of $h$ must be adjusted according to the RG
flow; whenever we refer to the value of $h$ in the main text it means
the physical value at infinite cut-off ($\lambda=\infty$) while the
numerical simulations use the value corresponding to the cut-off implemented
in the TFSA.

\subsection{Cut-off dependence and extrapolation schemes}
\label{sec:smooth_extrapol}

\begin{table}
\centering
\subfloat[Ferromagnetic phase]{\begin{tabular}{|c|c|c|}
\hline
$L$  & $\Lambda_\text{max}$ & Dimension  \\ \hline
30 & 9.75             & 5746       \\ \hline
40 & 9.75             & 24262      \\ \hline
50 & 9.25             & 46498      \\ \hline
60 & 9                & 90161      \\ \hline
70 & 8.75 & 148149 \\ \hline
80 & 8.25 & 154307 \\ \hline
90 & 7.75 & 132582 \\ \hline
100 & 7.25 & 89494\\ \hline
\end{tabular}}
\hspace{2cm}
\subfloat[Paramagnetic phase]{\begin{tabular}{|c|c|c|}
\hline
$L$  & $\Lambda_\text{max}$ & Dimension  \\ \hline
30 & 9.75             & 5695      \\ \hline
40 & 9.75             & 23997      \\ \hline
50 & 9.25             & 45629      \\ \hline
60 & 9                & 88202      \\ \hline
70 & 8.5 & 102729 \\ \hline
80 & 7.75 & 68189 \\ \hline
90 & 7.5 & 80319 \\ \hline
100 & 7.25 & 86757 \\ \hline
\end{tabular}}
\caption{Maximal energy cut-offs and the corresponding Hilbert space dimensions for different volumes $L$ in the (a) ferromagnetic and (b) paramagnetic phase. Below $\Lambda_\text{max}$ the applied cut-offs decrease in steps of  $0.25.$}
\label{tab:cutoffs}
\end{table}

\begin{figure}[t]
\begin{centering}
\includegraphics[width=0.5\textwidth]{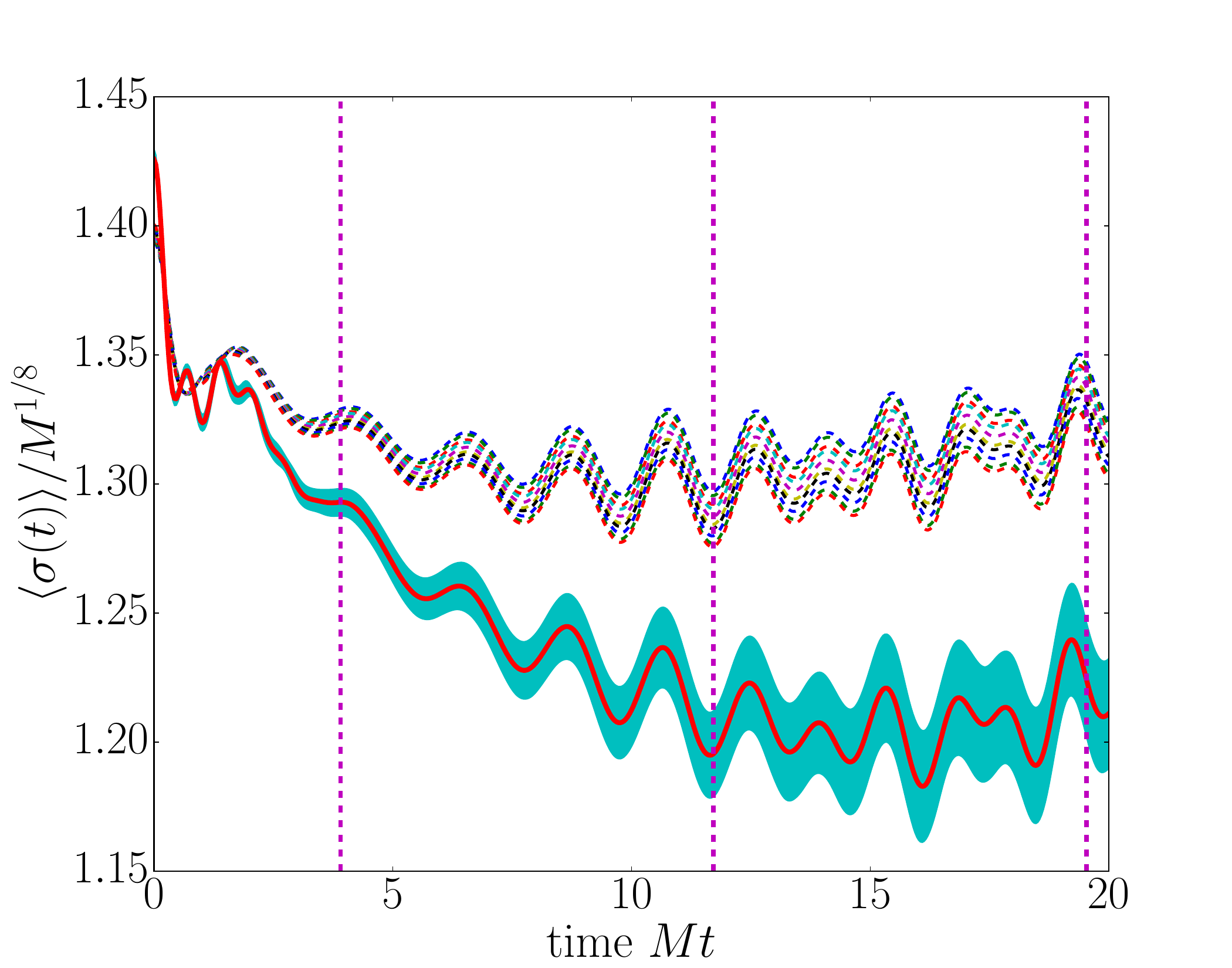}~\\
\includegraphics[width=0.3\textwidth]{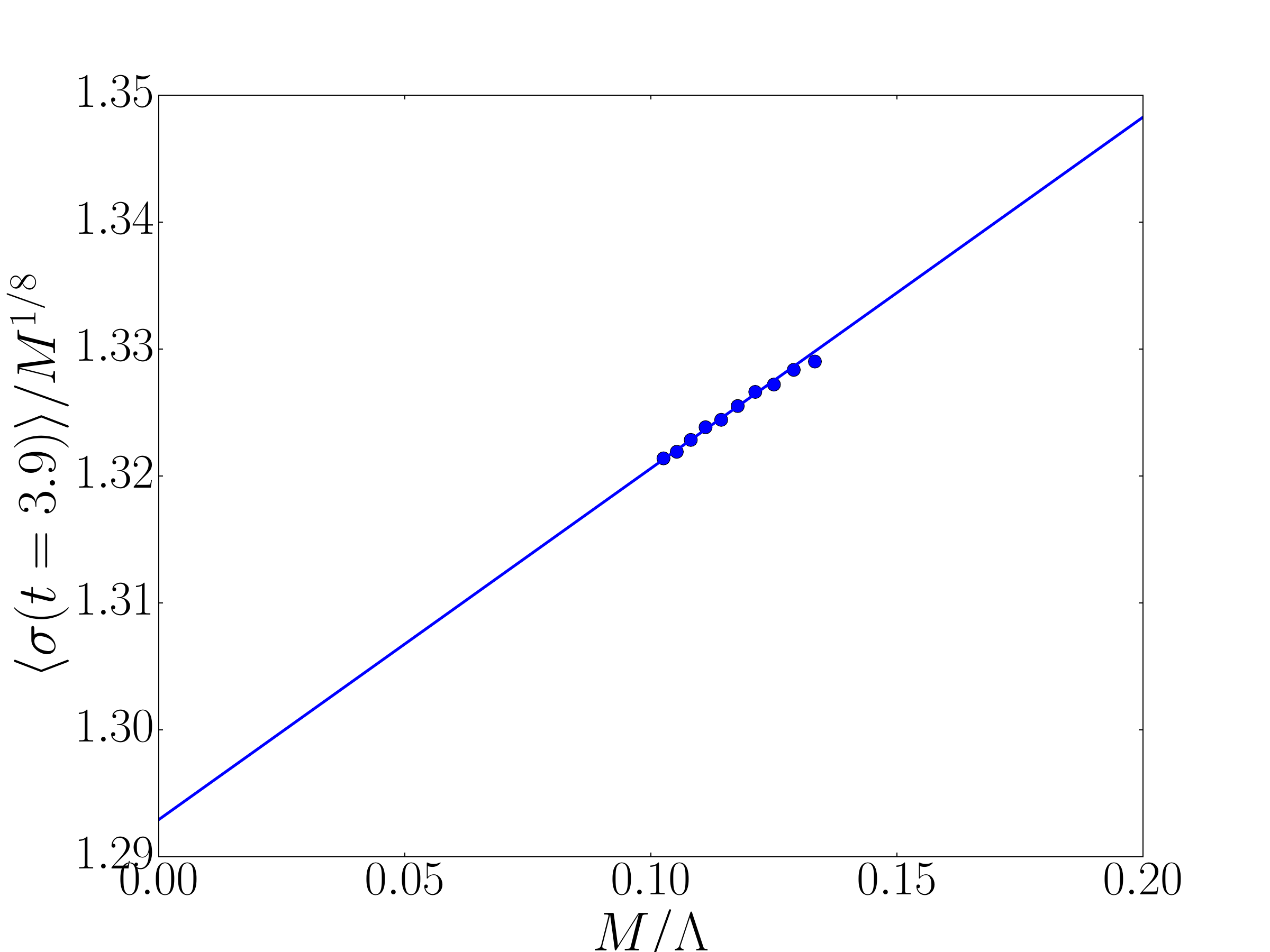}\includegraphics[width=0.3\textwidth]{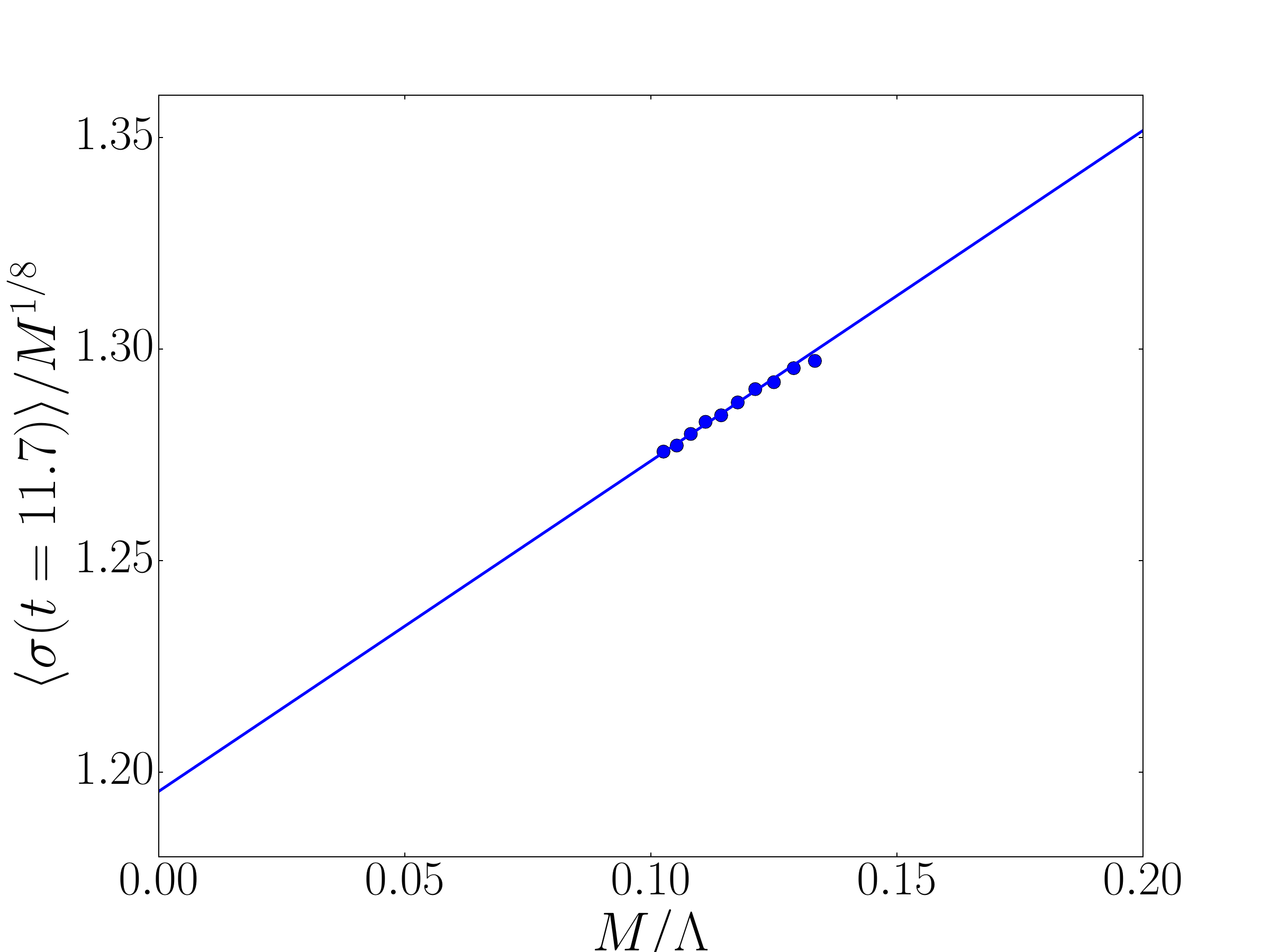}\includegraphics[width=0.3\textwidth]{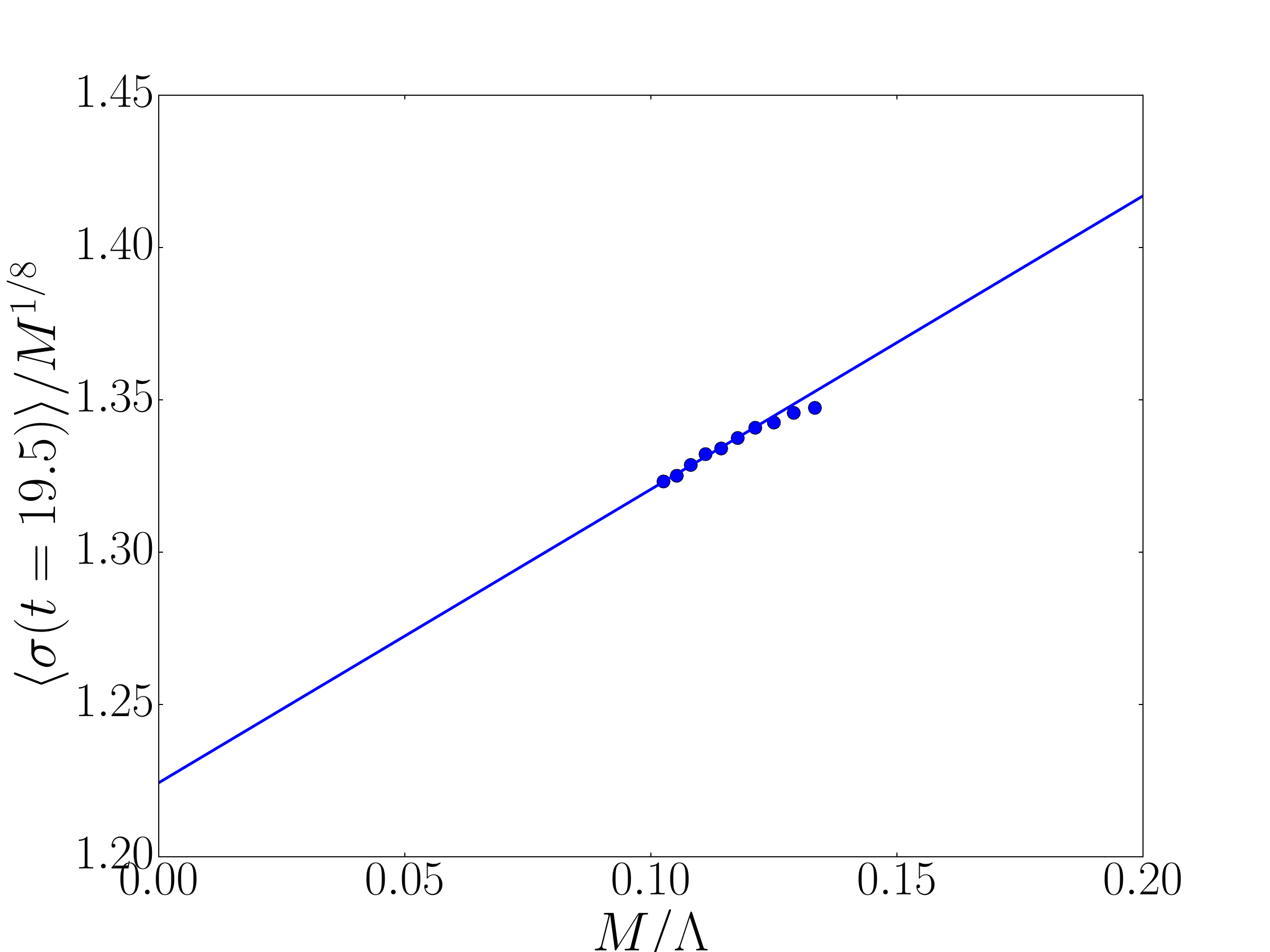}
\par\end{centering}
\caption{\label{fig:Cut-off-extrapolation-for-ferro-sigma-largeh} Cut-off extrapolation
for $\langle\sigma(t)\rangle$ for a quench in the ferromagnetic phase from $M_0=1.5M$ with $\bar h=-0.1$ (c.f. Fig. \ref{fig:sigma_extrapolation_largeh}).
The upper plot shows the extrapolation with a green band added
displaying the naive extrapolation errors from a least-squares fit error
estimate. The lower three plots show the extrapolation for the instances
of time indicated by the vertical dashed lines in the upper panel.}
\end{figure}

\begin{figure}[t]
\begin{centering}
\includegraphics[width=0.5\textwidth]{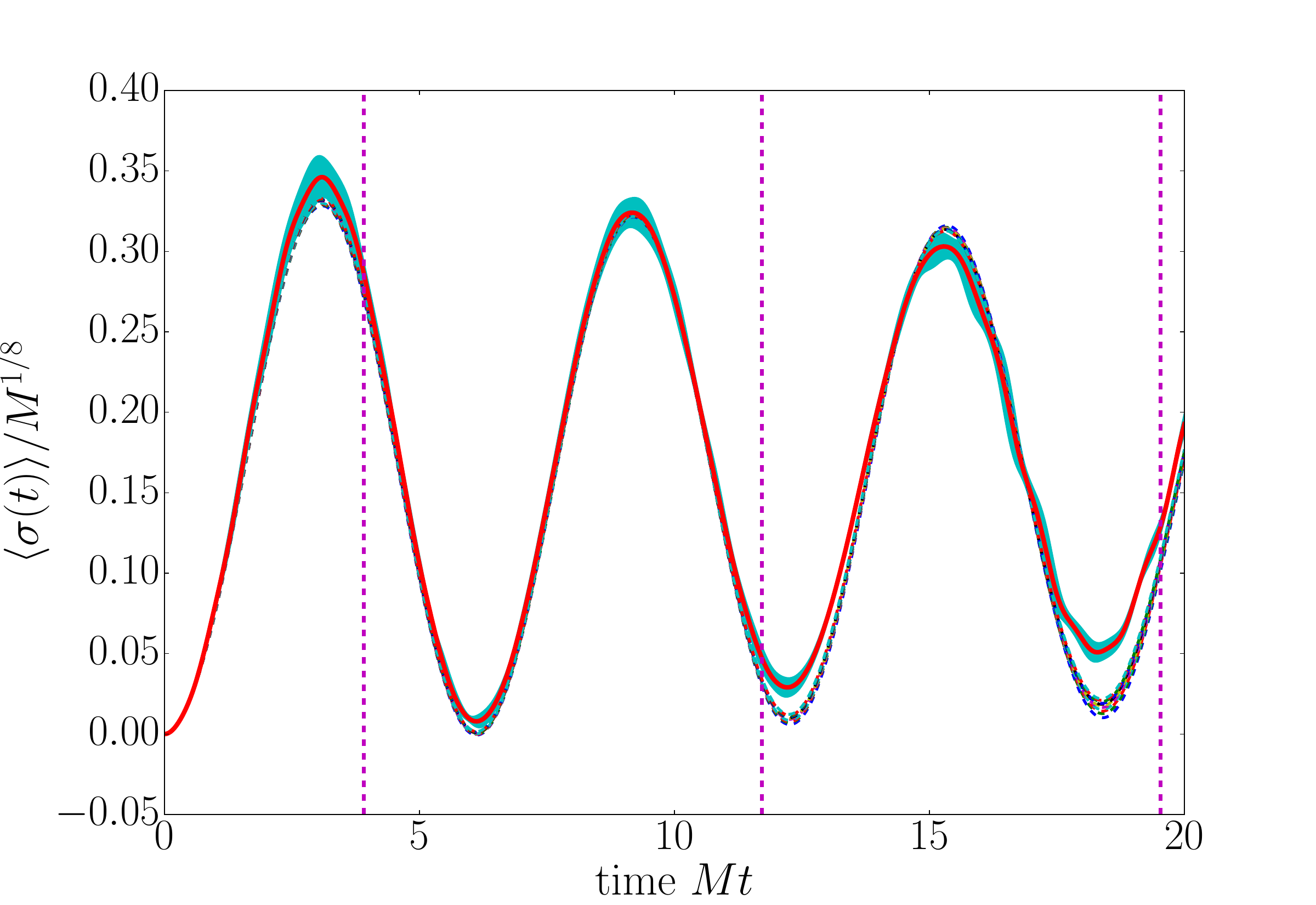}~\\
\includegraphics[width=0.3\textwidth]{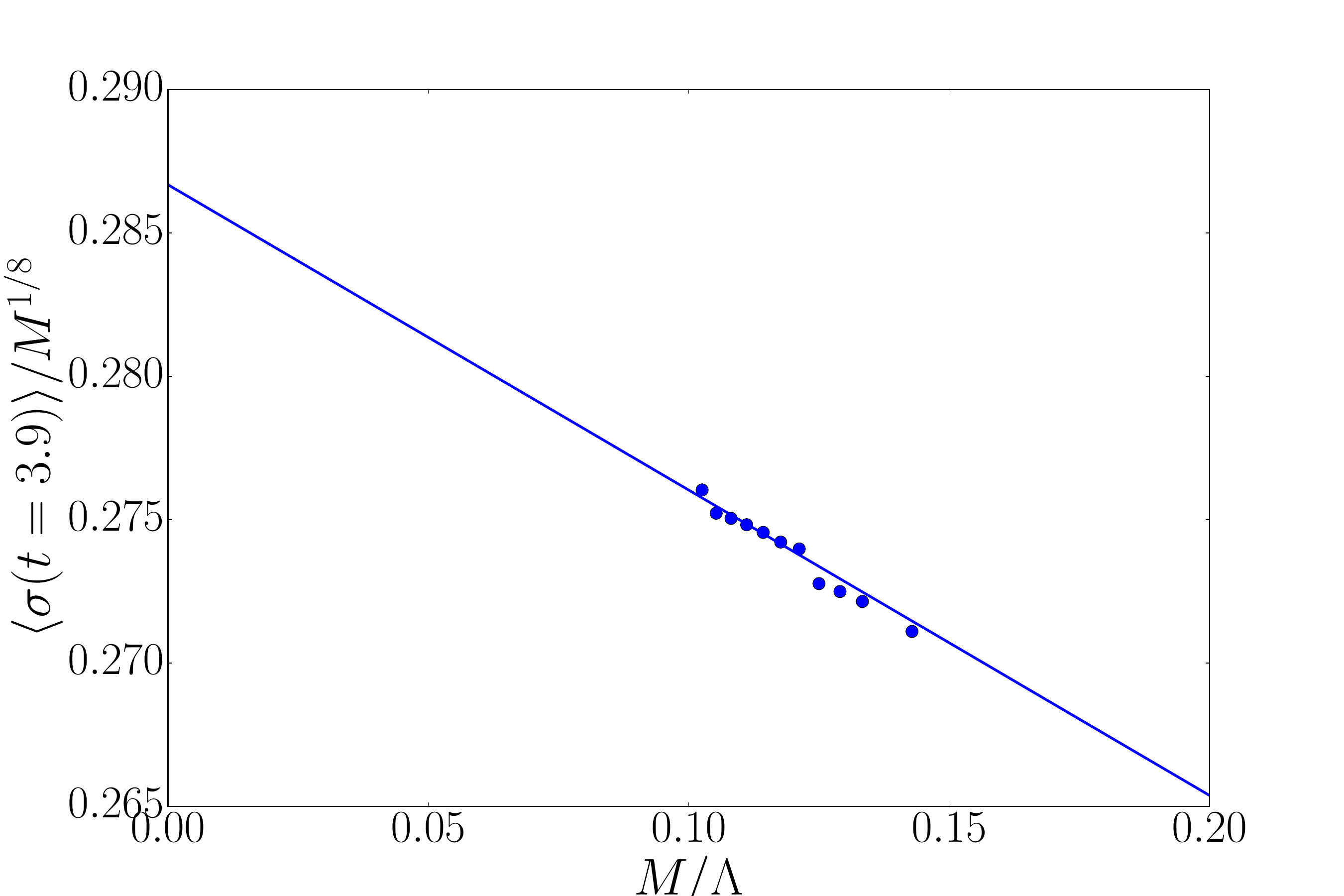}\includegraphics[width=0.3\textwidth]{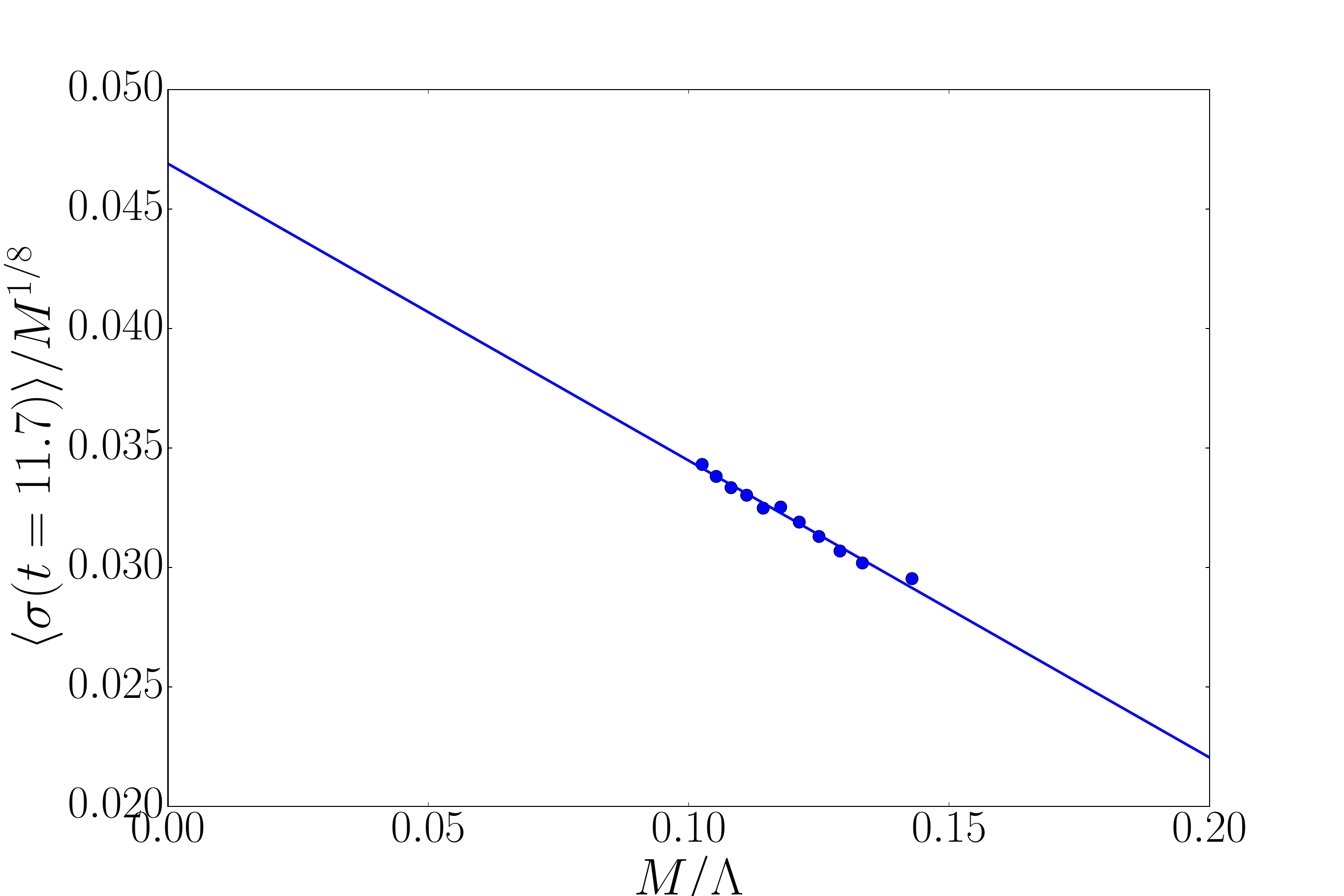}\includegraphics[width=0.3\textwidth]{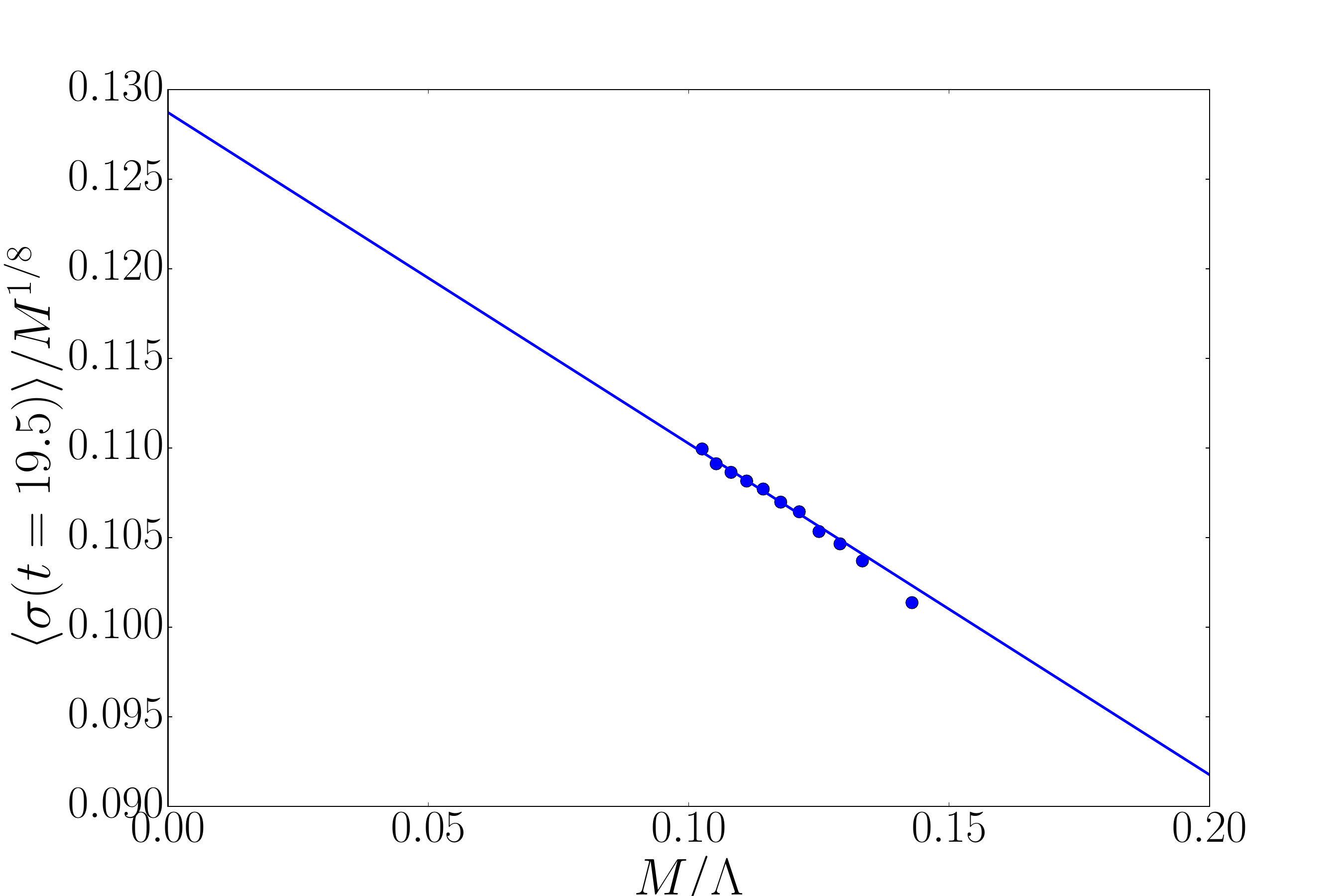}
\par\end{centering}
\caption{\label{fig:Cut-off-extrapolation-for-para-sigma-smallh} Cut-off extrapolation
for $\langle\sigma(t)\rangle$ for a quench in the paramagnetic phase from $M_0=1.5M$ with $\bar h=-0.1$ (c.f. Fig. \ref{fig:cut-off_extrapolation_smallh_para}).
The upper plot shows the extrapolation with a green band added
displaying the naive extrapolation errors from a least-squares fit error
estimate. The lower three plots show the extrapolation for the instances
of time indicated by the vertical dashed lines in the upper panel.}
\end{figure}

\begin{figure}[h!]
\begin{centering}
\includegraphics[width=0.4\textwidth]{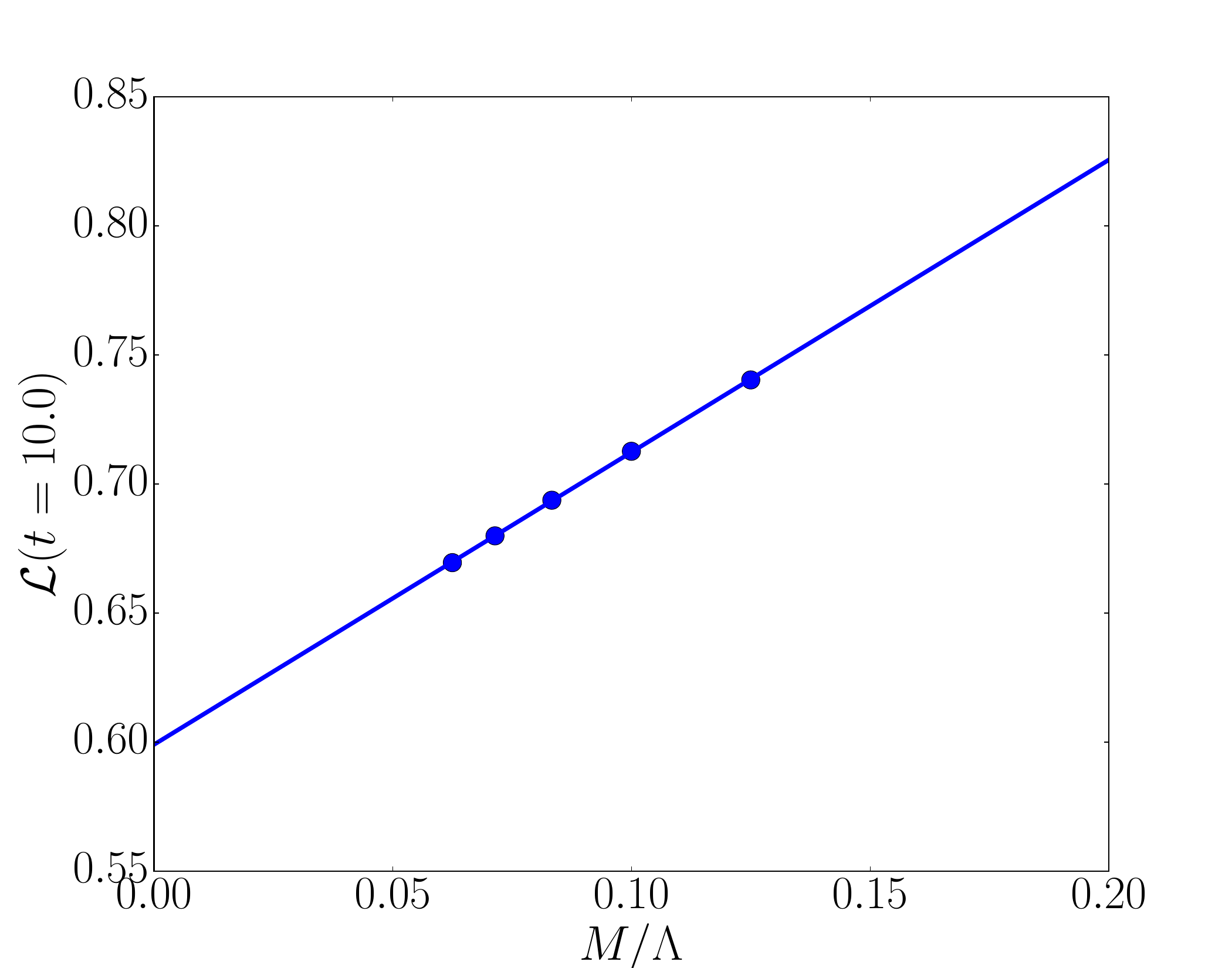}
\par\end{centering}
\caption{Cut-off extrapolation of the Loschmidt echo in Fig. \ref{fig:loschmidt_echo_m0_1,5} at time $Mt=10.$}
\label{fig:Loschmidt_extrap}
\end{figure}

Since the cut-off is imposed in the total energy of the states in
the TFSA space, the dimension of the Hilbert space increases quickly
with the volume $L$. As a consequence the maximum value of $\Lambda$
allowed by the available computing power is smaller for larger systems
and the actual maximum values of $\Lambda$ for different values of
$L$ are given in Table \ref{tab:cutoffs}. We made an exception for the overlaps with the initial
state, where for $L=30$ the maximum cut-off was extended to $\Lambda=11$
\eqref{fig:pw_ferro_nonint}.

To eliminate the cut-off dependence remaining after the introduction
of running couplings, we performed an extrapolation of certain quantities
in the cut-off. For this purpose we used a series of cut-offs up to
the maximum values indicated, with a spacing of $\Delta\Lambda=0.25$.
The extrapolation was always performed using simulations with the
$6$ largest available cut-offs, for example in Fig. \ref{fig:sigma_extrapolation_smallh}
the cut-offs $\Lambda=8.5,\,8.75,\,\dots,9.75$ were used. 

The extrapolation itself used a certain fitting function for the cut-off
dependence. For the evolution order parameter $\langle\sigma(t)\rangle$
there is a theoretical prediction available from the formalism introduced
in \cite{2013JHEP...08..094S}, which predicts that the leading cut-off
dependence is given by
\begin{equation}
\langle\sigma(t)\rangle_{\Lambda}=\langle\sigma(t)\rangle+A(t)\Lambda^{-1}+\dots\,,
\end{equation}
where the ellipses indicate sub-leading corrections. Cut-off extrapolation
for the results given in the main text are summarised in Figs. 
\ref{fig:Cut-off-extrapolation-for-ferro-sigma-largeh} and 
\ref{fig:Cut-off-extrapolation-for-para-sigma-smallh}. Moreover, for the Loschmidt echo $\mathcal{L}(t)$ we observed that 
\begin{equation}
\mathcal{L}(t)_{\Lambda}=\mathcal{L}(t)+B(t)\Lambda^{-1}+\dots
\end{equation}
gives a very accurate fit to the cut-off dependence, as illustrated
in Fig. \ref{fig:Loschmidt_extrap}. 

However, before extrapolation the oscillations related to the cut-off
must be eliminated, since their frequency depends on $\Lambda$ which 
leads to some unphysical features after extrapolation. We eliminate
them by convolving the curves $\langle\sigma(t)\rangle_{\Lambda},\mathcal{L}(t)_{\Lambda}$ with a Gaussian
\begin{equation}
\frac{1}{\sqrt{2\pi\Sigma^{2}}}e^{-\frac{t^{2}}{2\Sigma^{2}}}\:,
\label{eq:smoothening}
\end{equation}
where the parameter $\Sigma$ is adjusted according to the cut-off, $\Sigma=2\pi/\Lambda.$
Note that at the edges of the time window the Gaussian function would
involve a substantial contribution from nonexistent data points that
are outside the window; this is avoided by adjusting the width $\Sigma$
for time instances close to the edges of the window.

\section{Finite size behaviour of the overlaps\label{sec:overlaps}}

Here we present numerical data for the meson wave functions in order to 
show that two-quark states dominate the meson states. Based on this observation 
we present a theoretical calculation of the finite size dependence of the 
overlaps of the 1-meson states.

\subsection{The meson wave function}

The approximate wave functions of the mesons can be easily obtained
in the nonrelativistic two-quark approximation (where the role of
quarks if played by the domain wall excitations). The meson is modeled
as a bound state of two particles of mass $M$ in a linear potential
\begin{equation}
V(x)=2h\left\langle \sigma\right\rangle |x|\,,
\end{equation}
where $x$ is the relative coordinates of the two quarks. Standard
quantum mechanics then yields the normalised wave function
\begin{align}
\Psi(x) =  \begin{cases}
\frac{(M\lambda)^{1/6}\mbox{Ai}\left((M\lambda)^{1/3}(x-\mathcal{E}/\lambda)\right)}{\sqrt{2}\left|\mbox{Ai}'(-z)\right|}  \quad &x>0\,,\\
-\Psi(-x)  \quad &x<0\,,
\end{cases}
\end{align}
where $\mbox{Ai}$ is the Airy function, 
\begin{align}
\lambda & =  2h\left\langle \sigma\right\rangle \,,\\
\mathcal{E} & =  \left(\frac{\lambda^{2}}{M}\right)^{1/3}z
\end{align}
with $z>0$ corresponding to a zero of the Airy function 
\begin{equation}
\mbox{Ai}(-z)=0\,.
\end{equation}
There is an infinite number of zeros which can be ordered as $z_{1}<z_{2}<\dots$,
with $z_{n}$ giving the $n$th meson state. The variable $\mathcal{E}$
gives the mass $m$ of the meson via
\begin{equation}
m=2M+\mathcal{E} \,.
\label{eq:airymass}\end{equation}
Introducing the momentum space wave function (with a phase redefinition
to make the result purely real)
\begin{equation}
\Psi(q)  =  -i\int_{-\infty}^{\infty}dxe^{iqx}\Psi(x)\,,
\label{eq:momentumspace_wavefun_theor}\end{equation}
the properly normalised finite volume wave function can be written as 
\begin{equation}
|\Psi\rangle_{L}=\frac{1}{\sqrt{L}}\sum_{n=1/2}^{\infty}\Psi(q_{n})|q_{n},-q_{n}\rangle_{L}\,.
\end{equation}
Note that the index here runs over both integers and half-integers,
corresponding to the Ramond and Neveu--Schwarz components of the state,
respectively.

\begin{figure}
\subfloat[First meson]{\includegraphics[width=0.33\textwidth]{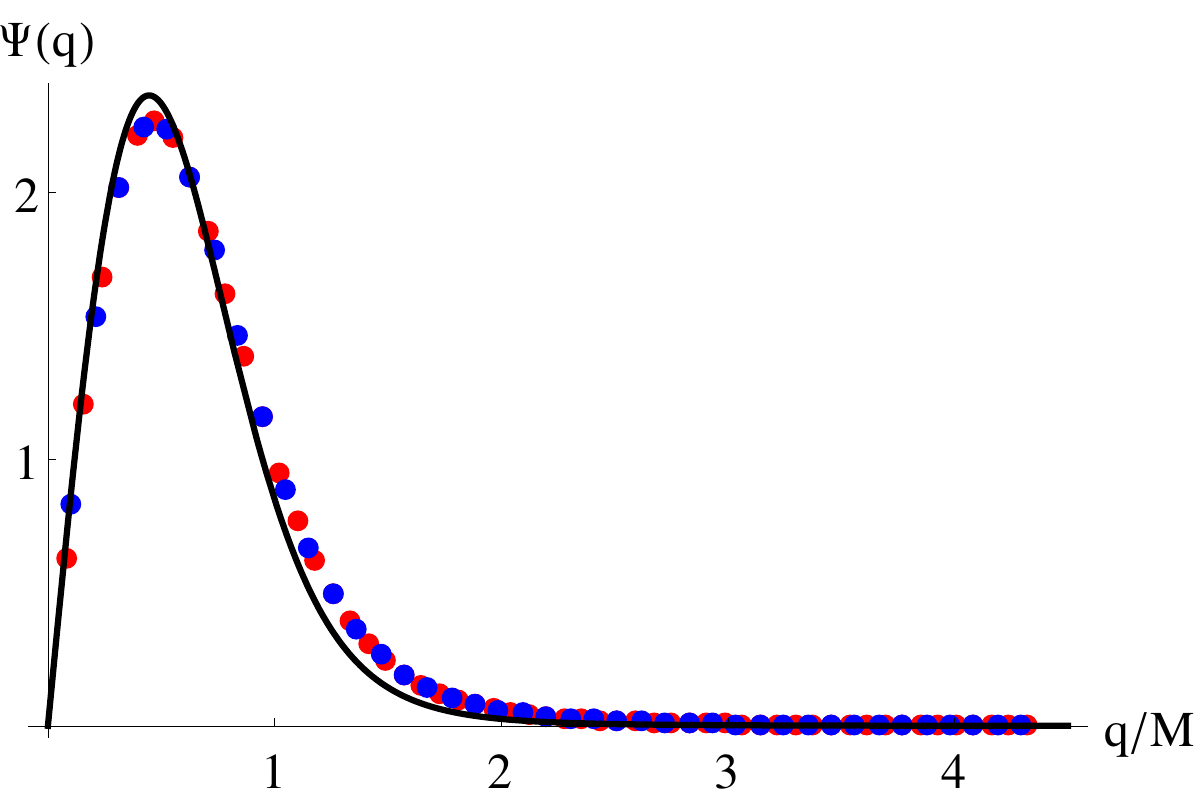}}
\subfloat[Second meson]{\includegraphics[width=0.33\textwidth]{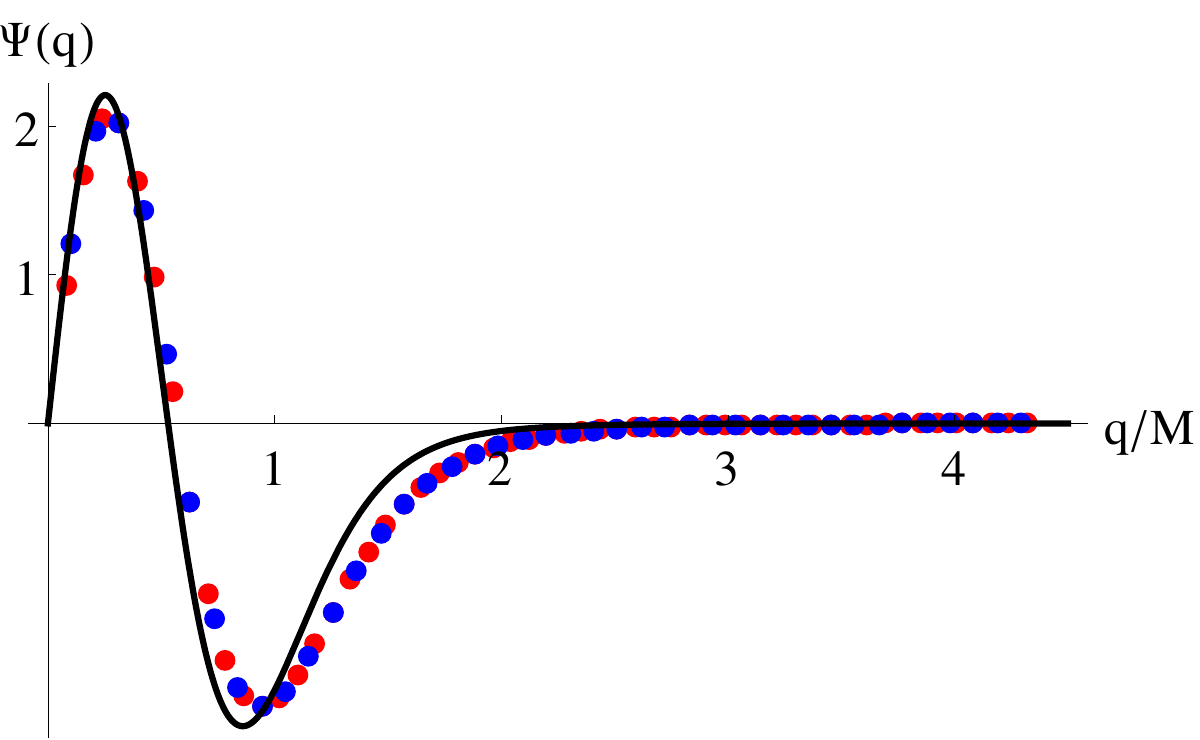}}
\subfloat[Third meson]{\includegraphics[width=0.33\textwidth]{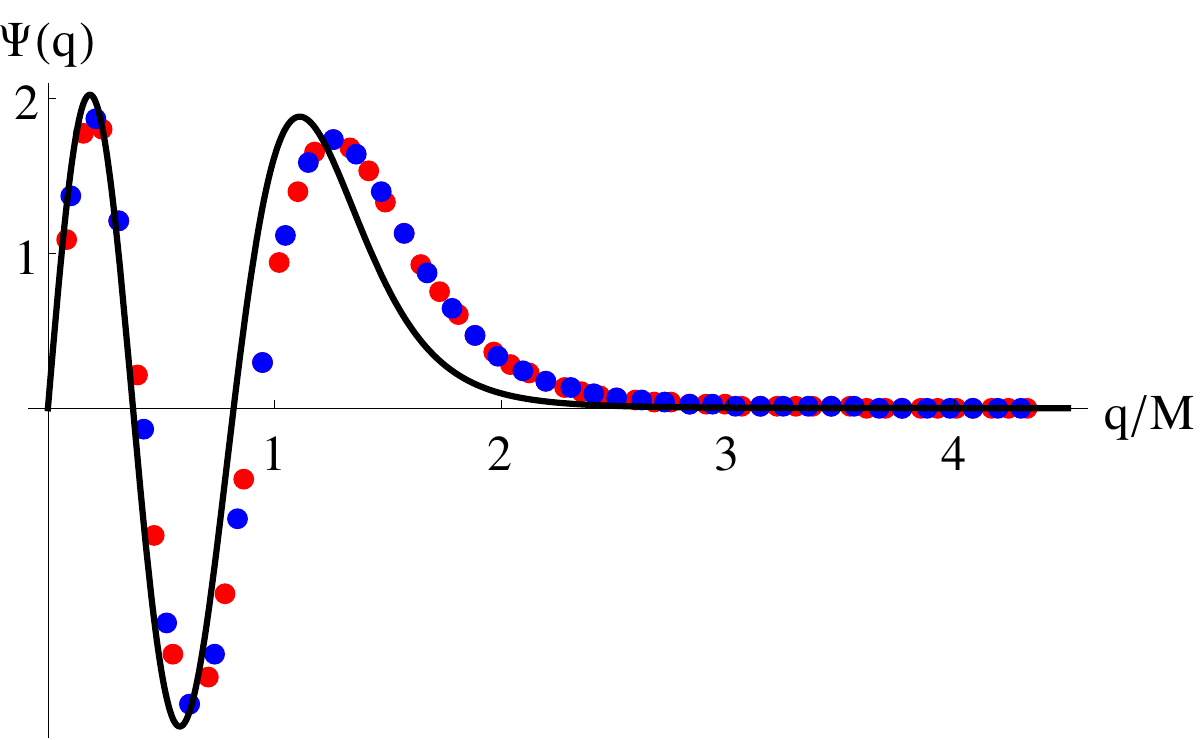}}
\caption{\label{fig:Meson-wave-functions}Meson wave functions obtained from 
(\ref{eq:momentumspace_wavefun_theor}) for $\bar{h}=-0.1$ (continuous line),
compared to TFSA results with $ML=30$ (blue dots) and $ML=40$ (red dots) at $\Lambda=9.0M$
which corresponds to a momentum cut-off $q/M<\sqrt{77}/2\approx 4.387$.}
\end{figure}

Using the TFSA, the meson wave function can be explicitly evaluated
in the fermionic basis. We have performed this calculation for values
of longitudinal magnetic field $\bar{h}=-0.05,-0.1,$ and volumes $ML=30,40.$
The first important observation is that in each case, the two-quark components
accounted for more than $99\%$ of the squared norm, so the two-quark
approximation is very accurate.

The finite volume wave function
can be directly compared to the numerical components of the meson
state vectors computed in the TFSA basis via the relation
\begin{equation}
 \Psi(q)=\sqrt{L}\Psi_L(q)
\end{equation}

As illustrated in Fig. \ref{fig:Meson-wave-functions},
we found good agreement which gets worse for the higher mesons
(and also with increasing $\bar{h}$), primarily because the non-relativistic
approximation is less accurate. Note also that the finite volume wave 
functions corresponding to different volumes scale on top of each other, and 
the momentum range where the wave-functions have their support lies well below 
the cut-off, which excludes the presence of substantial finite size or volume
dependence remaining. Therefore the deviations of the TFSA wave function from 
the prediction Eq. (\ref{eq:momentumspace_wavefun_theor}) observed 
in Fig.  \ref{fig:Meson-wave-functions} are really due to the 
non-relativistic two-body approximation.

\subsection{Volume dependence of meson overlaps}

The initial state in finite volume can be expanded as (c.f. Eq. \eqref{eq:initialStateExpansion})
\begin{align}
|\Psi_{0}\rangle_{L} & =  \mathcal{N}\left(|0\rangle_{L}+i\sum_{n}K(q_{n})|q_{n},-q_{n}\rangle_{L}+\dots\right)\,,\nonumber \\
   \mathcal{N}&=\prod_{n}\left(1+|K(q_{n},M,M_{0})|^{2}\right)^{-1/2}\,.
\end{align}
The normalisation factor $\mathcal{N}$ can be written as
\begin{align}
\log\mathcal{N} & =  -\frac{1}{2}\sum_{n}\log\left(1+|K(q_{n},M,M_{0})|^{2}\right)\nonumber \\
 & =  -\frac{1}{2}L\int_{0}^{\infty}\frac{dq}{\pi}\log\left(1+|K(q,M,M_{0})|^{2}\right)\,,
\end{align}
where care must be taken since $n$ runs in steps of $1/2.$ The overlap
with the $m$th meson state is 
\begin{align}
\langle\Psi_{0}|\Psi_{m}\rangle_{L} & =  \mathcal{N}\sqrt{\frac{1}{L}}\sum_{n}iK(q_{n})\Psi_{m}(q_{n})\nonumber \\
 & = \mathcal{N}\sqrt{L}\int_{0}^{\infty}\frac{dq}{\pi}iK(q)\Psi_{m}(q)\,.
\end{align}
Substituting (\ref{eq:InitialStateFerroPara}), the predicted volume dependence is 
\begin{equation}
\left|\langle\Psi_{0}|\Psi_{m}\rangle_{L}\right|^{2}=A_{m}Le^{-BL}\label{eq:overlaptheor_app}
\end{equation}
with 
\begin{align}
A_{m} & =  \frac{1}{2}\left|\int_{0}^{\infty}\frac{dq}{\pi}K(q)\Psi_{m}(q)\right|^{2}\,,\nonumber \\
B & =  \int_{0}^{\infty}\frac{dq}{2\pi}\log\left(1+|K(q,M,M_{0})|^{2}\right)\,.
\label{eq:theor_overlap}\end{align}
Indeed, the volume dependence of overlaps measure by the TFSA can be
fitted very precisely with functions of the predicted form as illustrated
in Fig. \ref{fig:meson_overlap}. Due to the presence of the cut-off and 
the inaccuracy inherited from the non-relativistic approximation inherent 
in the wave function \ref{eq:momentumspace_wavefun_theor},
however, the eventual values obtained for $A_{m}$ and $B$ only agree
with the theoretical predictions within order of magnitude.

Note that Eq. (\ref{eq:overlaptheor_app}) eventually predicts that the
increase of the meson overlaps is reversed for large enough volumes
\[
L>L_\text{crit}=\frac{1}{B}\,.
\]
From the table in Fig. \ref{fig:meson_overlap} one can see that this value
is 
\[
ML_\text{crit}\apprge300
\]
which is much larger than the volumes that can be reasonably reached
using TFSA, and is also too large to have any finite size dependence
for field theoretic correlators. The consequences of this observation
are discussed in the main text.

\section{Meson masses in the ferromagnetic phase \label{sec:Meson-masses-in}}

The simplest mass estimation was obtained by McCoy and Wu \cite{1978PhRvD..18.1259M}
and is given by Eq. (\ref{eq:airymass}).

Another approximation is provided by the WKB method. The WKB mass spectrum can be obtained by solving the quantisation
condition \cite{2005PhRvL..95y0601R,2006hep.th...12304F} 
\begin{align}
\frac{\sinh\left(2\vartheta_{n}\right)-2\vartheta_{n}}{\lambda} & =  2\pi\left(n-1/4\right)\,,\nonumber \\
m_{n}^{WKB} & =  2M\cosh\left(\vartheta_{n}\right)\,.\label{eq:wkb_ising}
\end{align}
It can be improved further by adding higher corrections in $\lambda$:
\begin{align}
\sinh\left(2\vartheta_{n}\right)-2\vartheta_{n} & =  2\pi\left(n-1/4\right)\lambda+\sum_{k=1}^{\infty}\lambda^{k+1}S_{k}\left(\vartheta_{n}\right)\,,\nonumber \\
m_{n}^{iWKB} & =  2M\cosh\left(\vartheta_{n}\right)\,.\label{eq:iwkb_ising}
\end{align}
The first term in this expansion is given by \cite{2006hep.th...12304F}
\begin{equation}
S_{1}\left(\vartheta\right)=\frac{1}{\sinh\left(2\vartheta\right)}\left(-\frac{1}{6}\sinh^{2}\left(\vartheta\right)+\frac{5}{24\sinh^{2}\left(\vartheta\right)}+\frac{1}{4\cosh^{2}\left(\vartheta\right)}-\frac{1}{12}\right)\,.
\end{equation}
Another approach to compute the meson mass is by solving a Bethe--Salpeter
equation \cite{2005PhRvL..95y0601R,2006hep.th...12304F,2009JPhA...42D4025R},
but it is only accurate for weak magnetic field $h$, while the semiclassical
results give a very good approximation for a wide range of parameters
as demonstrated in \cite{2015JHEP...09..146L}.

\section{Some details of the iTEBD simulations}

In order to obtain an external confirmation of our calculations 
with independent methodology we employed iTEBD (infinite Time Evolving Block Decimation) 
numerical simulations to follow the entire relaxation process toward the equilibrium state. 
The algorithm is based on the infinite Matrix Product State (iMPS) description 
of one-dimensional translational invariant lattice models in the thermodynamic limit which is free of any finite size effect. The canonical iMPS representation of a generic 
many-body state is therefore
\begin{equation}
\ket{\Psi} = 
\sum_{\{s\}} 
{\rm Tr}[\cdots {\bf\Gamma}_{o}^{s_{j}} {\bf\Lambda}_{o} 
{\bf \Gamma}_{e}^{s_{j+1}} {\bf\Lambda}_{e} \!\cdots ]
\ket{\cdots s_{j}s_{j+1}\!\cdots},
\end{equation}
where ${\bf \Gamma}_{o/e}^{s_{j}}$ are $\chi\times\chi$ matrices associated with odd/even lattice sites,
with $s_{j}$ spanning the Hilbert space of the $j^{\rm th}$ site with canonical basis 
$\{\ket{\uparrow}, \ket{\downarrow}\}$; similarly, ${\bf \Lambda}_{o/e}$ are diagonal matrices 
with entries being the singular values associated with Schmidt decomposition for the bipartition 
of the system on the odd/even bonds. Notice that the iMPS representation is properly 
defined thanks to the right/left orthonormalisation conditions
\begin{equation}
\sum_{s} ({\bf\Gamma}_{o/e}^{s} {\bf\Lambda}_{o/e}) ({\bf\Gamma}_{o/e}^{s} {\bf\Lambda}_{o/e})^{\dag} = {\bf I}\,,\quad
\sum_{s}  ({\bf\Lambda}_{e/o}{\bf\Gamma}_{o/e}^{s} )^{\dag} ({\bf\Lambda}_{e/o}{\bf\Gamma}_{o/e}^{s}) = {\bf I}\,,
\end{equation}
which essentially state that  the leading eigenvalue of the right/left transfer matrices is unitary and the corresponding right/left eigenvector is the identity  matrix (reshuffled as a vector). 
These conditions guarantee the correct normalisation of the vector $|\Psi\rangle$
as well as the possibility to perform well-defined operations in the  MPS representation.

Although both the initial state and the post-quench Hamiltonian are one-site shift invariant, 
we need to partially break translational symmetry in order to simulate the action 
of the $2^\text{nd}$-order Suzuki--Trotter approximation of the evolution operator, namely
\begin{equation}
e^{-i H dt } \simeq \bigotimes_{j\,odd}e^{-i \mathfrak{h}  dt/2 }
\bigotimes_{j\, even}e^{-i \mathfrak{h} dt }
\bigotimes_{j\,odd}e^{-i  \mathfrak{h}  dt/2}\,,
\end{equation}
where $\mathfrak{h}  = - J [\sigma^{x}\otimes\sigma^{x} 
+ h_{z}(\sigma^{z}\otimes\mathbb{I}+\mathbb{I}\otimes\sigma^{z})/2
+ h_{x}(\sigma^{x}\otimes\mathbb{I}+\mathbb{I}\otimes\sigma^{x})/2]$ 
represents the local interaction between nearest neighbour spins. 

Since the Hamiltonian density $\mathfrak{h}$ keeps an overall energy 
scale factor $J$ which tends to infinity in the scaling limit, 
we need to rescale the Trotter time step as $dt = 0.05/J$ in order
to keep small the related Trotter error. With this protocol, time in the iTEBD 
simulations is naturally measured in units of $J.$ As a consequence,
to reach a given iTEBD time, we need to perform a number of Trotter steps
which increases linearly with $J$, so approaching
the scaling limit more and more Trotter steps are necessary. In practice, we focused on $J=5,\, 10,\, 20$; therefore,
in order to reach a $t_{max} \simeq 10J$, the algorithm performed $\sim1000,\, \sim2000,$ and $\sim4000$ Trotter iterations, respectively.

Approaching the field theory limit, the quench in the transverse $h_z$ direction
becomes smaller and smaller; however, $J$ becomes larger and larger in such a way that
the mass $M=2J|1-h_{z}|$ is kept fixed. Therefore the entanglement entropy increases with time, 
and the auxiliary dimension $\chi$ has to be dynamically updated in order 
to optimally control the truncation error. 
At each local step, all the Schmidt vectors corresponding to singular values 
larger than $\lambda_\text{min} = 10^{-12}$ are retained. 
For practical reasons, to prevent the algorithm from getiting stuck in neverending computations,
this condition is relaxed when $\chi$ reaches the maximal 
value $\chi_\text{max} = 1024.$ Because of the upper bound $\chi_\text{max},$
the truncation procedure is the main source of error of the algorithm.

Let us note that for the above reasons it is extremely CPU-time consuming to reach the continuum limit
of any iTEBD lattice calculation.
Each simulation was executed on a \texttt{GenuineIntel Core i7-4790 3.60GHz} processor
with \texttt{32GB} of RAM.  After reaching the maximum value of the auxiliary dimension
$\chi_\text{max}$, the CPU-time needed for a Singular Value Decomposition (SVD)
of a matrix with dimensions $2048\times 2048$ is $\sim 200$sec.
Therefore, the total CPU-time needed to reach $t_\text{max} \simeq 10J$
is between $\sim5$ days and $18$ days, depending on the value of $J.$


\begin{thebibliography}{10}

\bibitem{2006PhRvL..96m6801C}
P.~{Calabrese} and J.~{Cardy}, ``{Time Dependence of Correlation Functions
  Following a Quantum Quench},''
  \href{http://dx.doi.org/10.1103/PhysRevLett.96.136801}{{\em Physical Review
  Letters} {\bfseries 96} (2006) 136801},
  \href{http://arxiv.org/abs/cond-mat/0601225}{{\ttfamily
  arXiv:cond-mat/0601225}}.

\bibitem{2007JSMTE..06....8C}
P.~{Calabrese} and J.~{Cardy}, ``{Quantum quenches in extended systems},''
  \href{http://dx.doi.org/10.1088/1742-5468/2007/06/P06008}{{\em Journal of
  Statistical Mechanics: Theory and Experiment} {\bfseries 6} (2007) 06008},
  \href{http://arxiv.org/abs/0704.1880}{{\ttfamily arXiv:0704.1880}}.

\bibitem{2006Natur.440..900K}
T.~{Kinoshita}, T.~{Wenger}, and D.~S. {Weiss}, ``{A quantum Newton's
  cradle},'' \href{http://dx.doi.org/10.1038/nature04693}{{\em Nature}
  {\bfseries 440} (2006) 900--903}.

\bibitem{2007Natur.449..324H}
S.~{Hofferberth}, I.~{Lesanovsky}, B.~{Fischer}, T.~{Schumm}, and
  J.~{Schmiedmayer}, ``{Non-equilibrium coherence dynamics in one-dimensional
  Bose gases},'' \href{http://dx.doi.org/10.1038/nature06149}{{\em Nature}
  {\bfseries 449} (2007) 324--327},
  \href{http://arxiv.org/abs/0706.2259}{{\ttfamily arXiv:0706.2259}}.

\bibitem{2012NatPh...8..325T}
S.~{Trotzky}, Y.-A. {Chen}, A.~{Flesch}, I.~P. {McCulloch},
  U.~{Schollw{\"o}ck}, J.~{Eisert}, and I.~{Bloch}, ``{Probing the relaxation
  towards equilibrium in an isolated strongly correlated one-dimensional Bose
  gas},'' \href{http://dx.doi.org/10.1038/nphys2232}{{\em Nature Physics}
  {\bfseries 8} (2012) 325--330},
  \href{http://arxiv.org/abs/1101.2659}{{\ttfamily arXiv:1101.2659}}.

\bibitem{2012Sci...337.1318G}
M.~{Gring}, M.~{Kuhnert}, T.~{Langen}, T.~{Kitagawa}, B.~{Rauer},
  M.~{Schreitl}, I.~{Mazets}, D.~A. {Smith}, E.~{Demler}, and
  J.~{Schmiedmayer}, ``{Relaxation and Prethermalization in an Isolated Quantum
  System},'' \href{http://dx.doi.org/10.1126/science.1224953}{{\em Science}
  {\bfseries 337} (2012) 1318},
  \href{http://arxiv.org/abs/1112.0013}{{\ttfamily arXiv:1112.0013}}.

\bibitem{2013NatPh...9..640L}
T.~{Langen}, R.~{Geiger}, M.~{Kuhnert}, B.~{Rauer}, and J.~{Schmiedmayer},
  ``{Local emergence of thermal correlations in an isolated quantum many-body
  system},'' \href{http://dx.doi.org/10.1038/nphys2739}{{\em Nature Physics}
  {\bfseries 9} (2013) 640--643},
  \href{http://arxiv.org/abs/1305.3708}{{\ttfamily arXiv:1305.3708}}.

\bibitem{2013PhRvL.111e3003M}
F.~{Meinert}, M.~J. {Mark}, E.~{Kirilov}, K.~{Lauber}, P.~{Weinmann}, A.~J.
  {Daley}, and H.-C. {N{\"a}gerl}, ``{Quantum Quench in an Atomic
  One-Dimensional Ising Chain},''
  \href{http://dx.doi.org/10.1103/PhysRevLett.111.053003}{{\em Physical Review
  Letters} {\bfseries 111} (2013) 053003},
  \href{http://arxiv.org/abs/1304.2628}{{\ttfamily arXiv:1304.2628}}.

\bibitem{2013Natur.502...76F}
T.~{Fukuhara}, P.~{Schau{\ss}}, M.~{Endres}, S.~{Hild}, M.~{Cheneau},
  I.~{Bloch}, and C.~{Gross}, ``{Microscopic observation of magnon bound states
  and their dynamics},'' \href{http://dx.doi.org/10.1038/nature12541}{{\em
  Nature} {\bfseries 502} (2013) 76--79},
  \href{http://arxiv.org/abs/1305.6598}{{\ttfamily arXiv:1305.6598}}.

\bibitem{2015Sci...348..207L}
T.~{Langen}, S.~{Erne}, R.~{Geiger}, B.~{Rauer}, T.~{Schweigler}, M.~{Kuhnert},
  W.~{Rohringer}, I.~E. {Mazets}, T.~{Gasenzer}, and J.~{Schmiedmayer},
  ``{Experimental observation of a generalized Gibbs ensemble},''
  \href{http://dx.doi.org/10.1126/science.1257026}{{\em Science} {\bfseries
  348} (2015) 207--211}, \href{http://arxiv.org/abs/1411.7185}{{\ttfamily
  arXiv:1411.7185}}.

\bibitem{2016arXiv160304409K}
A.~M. {Kaufman}, M.~E. {Tai}, A.~{Lukin}, M.~{Rispoli}, R.~{Schittko}, P.~M.
  {Preiss}, and M.~{Greiner}, ``{Quantum thermalization through entanglement in
  an isolated many-body system},'' {\em ArXiv e-prints} (2016) ,
  \href{http://arxiv.org/abs/1603.04409}{{\ttfamily arXiv:1603.04409}}.

\bibitem{2007PhRvL..98e0405R}
M.~{Rigol}, V.~{Dunjko}, V.~{Yurovsky}, and M.~{Olshanii}, ``{Relaxation in a
  Completely Integrable Many-Body Quantum System: An AbInitio Study of the
  Dynamics of the Highly Excited States of 1D Lattice Hard-Core Bosons},''
  \href{http://dx.doi.org/10.1103/PhysRevLett.98.050405}{{\em Physical Review
  Letters} {\bfseries 98} (2007) 050405},
  \href{http://arxiv.org/abs/cond-mat/0604476}{{\ttfamily cond-mat/0604476}}.

\bibitem{2014PhRvL.113k7202W}
B.~{Wouters}, J.~{De Nardis}, M.~{Brockmann}, D.~{Fioretto}, M.~{Rigol}, and
  J.-S. {Caux}, ``{Quenching the Anisotropic Heisenberg Chain: Exact Solution
  and Generalized Gibbs Ensemble Predictions},''
  \href{http://dx.doi.org/10.1103/PhysRevLett.113.117202}{{\em Physical Review
  Letters} {\bfseries 113} (2014) 117202},
  \href{http://arxiv.org/abs/1405.0172}{{\ttfamily arXiv:1405.0172}}.

\bibitem{2014PhRvL.113k7203P}
B.~{Pozsgay}, M.~{Mesty{\'a}n}, M.~A. {Werner}, M.~{Kormos}, G.~{Zar{\'a}nd},
  and G.~{Tak{\'a}cs}, ``{Correlations after Quantum Quenches in the XXZ Spin
  Chain: Failure of the Generalized Gibbs Ensemble},''
  \href{http://dx.doi.org/10.1103/PhysRevLett.113.117203}{{\em Physical Review
  Letters} {\bfseries 113} (2014) 117203},
  \href{http://arxiv.org/abs/1405.2843}{{\ttfamily arXiv:1405.2843}}.

\bibitem{2015PhRvL.115o7201I}
E.~{Ilievski}, J.~{De Nardis}, B.~{Wouters}, J.-S. {Caux}, F.~H.~L. {Essler},
  and T.~{Prosen}, ``{Complete Generalized Gibbs Ensembles in an Interacting
  Theory},'' \href{http://dx.doi.org/10.1103/PhysRevLett.115.157201}{{\em
  Physical Review Letters} {\bfseries 115} (2015) 157201},
  \href{http://arxiv.org/abs/1507.02993}{{\ttfamily arXiv:1507.02993}}.

\bibitem{2016arXiv160300440I}
E.~{Ilievski}, M.~{Medenjak}, T.~{Prosen}, and L.~{Zadnik}, ``{Quasilocal
  charges in integrable lattice systems},'' {\em ArXiv e-prints} (2016) ,
  \href{http://arxiv.org/abs/1603.00440}{{\ttfamily arXiv:1603.00440}}.

\bibitem{2014PhRvA..90d3625G}
G.~{Goldstein} and N.~{Andrei}, ``{Failure of the local generalized Gibbs
  ensemble for integrable models with bound states},''
  \href{http://dx.doi.org/10.1103/PhysRevA.90.043625}{{\em Physical Review A}
  {\bfseries 90} (2014) 043625},
  \href{http://arxiv.org/abs/1405.4224}{{\ttfamily arXiv:1405.4224}}.

\bibitem{2014JSMTE..09..026P}
B.~{Pozsgay}, ``{Failure of the generalized eigenstate thermalization
  hypothesis in integrable models with multiple particle species},''
  \href{http://dx.doi.org/10.1088/1742-5468/2014/09/P09026}{{\em Journal of
  Statistical Mechanics: Theory and Experiment} {\bfseries 9} (2014) 09026},
  \href{http://arxiv.org/abs/1406.4613}{{\ttfamily arXiv:1406.4613}}.

\bibitem{1991PhRvA..43.2046D}
J.~M. {Deutsch}, ``{Quantum statistical mechanics in a closed system},''
  \href{http://dx.doi.org/10.1103/PhysRevA.43.2046}{{\em Physical Review A}
  {\bfseries 43} (1991) 2046--2049}.

\bibitem{1994PhRvE..50..888S}
M.~{Srednicki}, ``{Chaos and quantum thermalization},''
  \href{http://dx.doi.org/10.1103/PhysRevE.50.888}{{\em Physical Review E}
  {\bfseries 50} (1994) 888--901},
  \href{http://arxiv.org/abs/cond-mat/9403051}{{\ttfamily cond-mat/9403051}}.

\bibitem{2004PhRvL..93n2002B}
J.~{Berges}, S.~{Bors{\'a}nyi}, and C.~{Wetterich}, ``{Prethermalization},''
  \href{http://dx.doi.org/10.1103/PhysRevLett.93.142002}{{\em Physical Review
  Letters} {\bfseries 93} (2004) 142002},
  \href{http://arxiv.org/abs/hep-ph/0403234}{{\ttfamily hep-ph/0403234}}.

\bibitem{2011PhRvB..84e4304K}
M.~{Kollar}, F.~A. {Wolf}, and M.~{Eckstein}, ``{Generalized Gibbs ensemble
  prediction of prethermalization plateaus and their relation to nonthermal
  steady states in integrable systems},''
  \href{http://dx.doi.org/10.1103/PhysRevB.84.054304}{{\em Physical Review B}
  {\bfseries 84} (2011) 054304},
  \href{http://arxiv.org/abs/1102.2117}{{\ttfamily arXiv:1102.2117}}.

\bibitem{2015PhRvL.115r0601B}
B.~{Bertini}, F.~H.~L. {Essler}, S.~{Groha}, and N.~J. {Robinson},
  ``{Prethermalization and Thermalization in Models with Weak Integrability
  Breaking},'' \href{http://dx.doi.org/10.1103/PhysRevLett.115.180601}{{\em
  Physical Review Letters} {\bfseries 115} (2015) 180601},
  \href{http://arxiv.org/abs/1506.02994}{{\ttfamily arXiv:1506.02994}}.

\bibitem{2015PhRvX...5d1043B}
G.~P. {Brandino}, J.-S. {Caux}, and R.~M. {Konik}, ``{Glimmers of a Quantum KAM
  Theorem: Insights from Quantum Quenches in One-Dimensional Bose Gases},''
  \href{http://dx.doi.org/10.1103/PhysRevX.5.041043}{{\em Physical Review X}
  {\bfseries 5} (2015) 041043},
  \href{http://arxiv.org/abs/1407.7167}{{\ttfamily arXiv:1407.7167}}.

\bibitem{2016arXiv160403571K}
M.~{Kormos}, M.~{Collura}, G.~{Tak{\'a}cs}, and P.~{Calabrese}, ``{Real time
  confinement following a quantum quench to a non-integrable model},'' {\em
  ArXiv e-prints} (2016) , \href{http://arxiv.org/abs/1604.03571}{{\ttfamily
  arXiv:1604.03571}}.

\bibitem{2004AIPC..739....3B}
J.~{Berges}, \href{http://dx.doi.org/10.1063/1.1843591}{``{Introduction to
  Nonequilibrium Quantum Field Theory},''} in {\em American Institute of
  Physics Conference Series}, M.~E. {Bracco}, M.~{Chiapparini}, E.~{Ferreira},
  and T.~{Kodama}, eds., vol.~739 of {\em American Institute of Physics
  Conference Series}, pp.~3--62.
\newblock 2004.
\newblock \href{http://arxiv.org/abs/hep-ph/0409233}{{\ttfamily
  arXiv:hep-ph/0409233}}.

\bibitem{2010NJPh...12e5015F}
D.~{Fioretto} and G.~{Mussardo}, ``{Quantum quenches in integrable field
  theories},'' \href{http://dx.doi.org/10.1088/1367-2630/12/5/055015}{{\em New
  Journal of Physics} {\bfseries 12} (2010) 055015},
  \href{http://arxiv.org/abs/0911.3345}{{\ttfamily arXiv:0911.3345}}.

\bibitem{2011PhRvL.106s1601B}
V.~{Balasubramanian}, A.~{Bernamonti}, J.~{de Boer}, N.~{Copland}, B.~{Craps},
  E.~{Keski-Vakkuri}, B.~{M{\"u}ller}, A.~{Sch{\"a}fer}, M.~{Shigemori}, and
  W.~{Staessens}, ``{Thermalization of Strongly Coupled Field Theories},''
  \href{http://dx.doi.org/10.1103/PhysRevLett.106.191601}{{\em Physical Review
  Letters} {\bfseries 106} (2011) 191601},
  \href{http://arxiv.org/abs/1012.4753}{{\ttfamily arXiv:1012.4753}}.

\bibitem{2012JSMTE..02..017S}
S.~{Sotiriadis}, D.~{Fioretto}, and G.~{Mussardo}, ``{Zamolodchikov-Faddeev
  algebra and quantum quenches in integrable field theories},''
  \href{http://dx.doi.org/10.1088/1742-5468/2012/02/P02017}{{\em Journal of
  Statistical Mechanics: Theory and Experiment} {\bfseries 2} (2012) 02017},
  \href{http://arxiv.org/abs/1112.2963}{{\ttfamily arXiv:1112.2963}}.

\bibitem{2012JSMTE..04..017S}
D.~{Schuricht} and F.~H.~L. {Essler}, ``{Dynamics in the Ising field theory
  after a quantum quench},''
  \href{http://dx.doi.org/10.1088/1742-5468/2012/04/P04017}{{\em Journal of
  Statistical Mechanics: Theory and Experiment} {\bfseries 4} (2012) 04017},
  \href{http://arxiv.org/abs/1203.5080}{{\ttfamily arXiv:1203.5080}}.

\bibitem{mussardoPRL111}
G.~Mussardo, ``{Infinite-time Average of Local Fields in an Integrable Quantum
  Field Theory after a Quantum Quench},''
  \href{http://dx.doi.org/10.1103/PhysRevLett.111.100401}{{\em Physical Review
  Letters} {\bfseries 111} (2013) 100401},
  \href{http://arxiv.org/abs/1308.4551}{{\ttfamily arXiv:1308.4551}}.

\bibitem{2014PhLB..734...52S}
S.~{Sotiriadis}, G.~{Takacs}, and G.~{Mussardo}, ``{Boundary state in an
  integrable quantum field theory out of equilibrium},''
  \href{http://dx.doi.org/10.1016/j.physletb.2014.04.058}{{\em Physics Letters
  B} {\bfseries 734} (2014) 52--57},
  \href{http://arxiv.org/abs/1311.4418}{{\ttfamily arXiv:1311.4418}}.

\bibitem{2014JSMTE..10..035B}
B.~{Bertini}, D.~{Schuricht}, and F.~H.~L. {Essler}, ``{Quantum quench in the
  sine-Gordon model},''
  \href{http://dx.doi.org/10.1088/1742-5468/2014/10/P10035}{{\em Journal of
  Statistical Mechanics: Theory and Experiment} {\bfseries 10} (2014) 10035},
  \href{http://arxiv.org/abs/1405.4813}{{\ttfamily arXiv:1405.4813}}.

\bibitem{2014JPhA...47N2001D}
G.~{Delfino}, ``{Quantum quenches with integrable pre-quench dynamics},''
  \href{http://dx.doi.org/10.1088/1751-8113/47/40/402001}{{\em Journal of
  Physics A Mathematical General} {\bfseries 47} (2014) 402001},
  \href{http://arxiv.org/abs/1405.6553}{{\ttfamily arXiv:1405.6553}}.

\bibitem{2015PhRvA..91e1602E}
F.~H.~L. {Essler}, G.~{Mussardo}, and M.~{Panfil}, ``{Generalized Gibbs
  ensembles for quantum field theories},''
  \href{http://dx.doi.org/10.1103/PhysRevA.91.051602}{{\em Physical Review A}
  {\bfseries 91} (2015) 051602},
  \href{http://arxiv.org/abs/1411.5352}{{\ttfamily arXiv:1411.5352}}.

\bibitem{2015JSMTE..11..004S}
D.~{Schuricht}, ``{Quantum quenches in integrable systems: constraints from
  factorisation},''
  \href{http://dx.doi.org/10.1088/1742-5468/2015/11/P11004}{{\em Journal of
  Statistical Mechanics: Theory and Experiment} {\bfseries 11} (2015) 11004},
  \href{http://arxiv.org/abs/1509.00435}{{\ttfamily arXiv:1509.00435}}.

\bibitem{2016NuPhB.902..508H}
D.~X. {Horv{\'a}th}, S.~{Sotiriadis}, and G.~{Tak{\'a}cs}, ``{Initial states in
  integrable quantum field theory quenches from an integral equation
  hierarchy},'' \href{http://dx.doi.org/10.1016/j.nuclphysb.2015.11.025}{{\em
  Nuclear Physics B} {\bfseries 902} (2016) 508--547},
  \href{http://arxiv.org/abs/1510.01735}{{\ttfamily arXiv:1510.01735}}.

\bibitem{2016JSMTE..03.3115C}
A.~{Cort{\'e}s Cubero}, G.~{Mussardo}, and M.~{Panfil}, ``{Quench dynamics in
  two-dimensional integrable SUSY models},''
  \href{http://dx.doi.org/10.1088/1742-5468/2016/03/033115}{{\em Journal of
  Statistical Mechanics: Theory and Experiment} {\bfseries 3} (2016) 033115},
  \href{http://arxiv.org/abs/1511.02712}{{\ttfamily arXiv:1511.02712}}.

\bibitem{2016JSMTE..06.3102B}
B.~{Bertini}, L.~{Piroli}, and P.~{Calabrese}, ``{Quantum quenches in the
  sinh-Gordon model: steady state and one-point correlation functions},''
  \href{http://dx.doi.org/10.1088/1742-5468/2016/06/063102}{{\em Journal of
  Statistical Mechanics: Theory and Experiment} {\bfseries 6} (2016) 063102},
  \href{http://arxiv.org/abs/1602.08269}{{\ttfamily arXiv:1602.08269
  [cond-mat.stat-mech]}}.

\bibitem{2016arXiv160403879C}
A.~{Cort{\'e}s Cubero}, ``{Planar quantum quenches: Computation of exact
  time-dependent correlation functions at large $N$},'' {\em ArXiv e-prints}
  (2016) , \href{http://arxiv.org/abs/1604.03879}{{\ttfamily
  arXiv:1604.03879}}.

\bibitem{1991IJMPA...6.4557Y}
V.~P. {Yurov} and A.~B. {Zamolodchikov}, ``{Truncated-Fermionic Approach to the
  Critical 2d Ising Model with Magnetic Field},''
  \href{http://dx.doi.org/10.1142/S0217751X91002161}{{\em International Journal
  of Modern Physics A} {\bfseries 6} (1991) 4557--4578}.

\bibitem{2001hep.th...12167F}
P.~{Fonseca} and A.~{Zamolodchikov}, ``{Ising field theory in a magnetic field:
  analytic properties of the free energy},''
  \href{http://dx.doi.org/10.1023/A:1022147532606}{{\em Journal of Statistical
  Physics} {\bfseries 111} (2003) 527},
  \href{http://arxiv.org/abs/hep-th/0112167}{{\ttfamily arXiv:hep-th/0112167}}.

\bibitem{1990IJMPA...5.3221Y}
V.~P. {Yurov} and A.~B. {Zamolodchikov}, ``{Truncated Comformal Space Approach
  to Scaling Lee-Yang Model},''
  \href{http://dx.doi.org/10.1142/S0217751X9000218X}{{\em International Journal
  of Modern Physics A} {\bfseries 5} (1990) 3221--3245}.

\bibitem{1991NuPhB.348..591L}
M.~{L{\"a}ssig}, G.~{Mussardo}, and J.~L. {Cardy}, ``{The scaling region of the
  tricritical Ising model in two dimensions},''
  \href{http://dx.doi.org/10.1016/0550-3213(91)90206-D}{{\em Nuclear Physics B}
  {\bfseries 348} (1991) 591--618}.

\bibitem{1998PhLB..430..264F}
G.~{Feverati}, F.~{Ravanini}, and G.~{Tak{\'a}cs}, ``{Truncated conformal space
  at c=1, nonlinear integral equation and quantization rules for multi-soliton
  states},'' \href{http://dx.doi.org/10.1016/S0370-2693(98)00543-7}{{\em
  Physics Letters B} {\bfseries 430} (1998) 264--273},
  \href{http://arxiv.org/abs/hep-th/9803104}{{\ttfamily arXiv:hep-th/9803104}}.

\bibitem{2014JSMTE..12..010C}
A.~{Coser}, M.~{Beria}, G.~P. {Brandino}, R.~M. {Konik}, and G.~{Mussardo},
  ``{Truncated conformal space approach for 2D Landau-Ginzburg theories},''
  \href{http://dx.doi.org/10.1088/1742-5468/2014/12/P12010}{{\em Journal of
  Statistical Mechanics: Theory and Experiment} {\bfseries 12} (2014) 12010},
  \href{http://arxiv.org/abs/1409.1494}{{\ttfamily arXiv:1409.1494}}.

\bibitem{2015arXiv151206901B}
Z.~{Bajnok} and M.~{Lajer}, ``{Truncated Hilbert space approach to the 2d
  $\phi^{4}$ theory},'' {\em ArXiv e-prints} (2015) ,
  \href{http://arxiv.org/abs/1512.06901}{{\ttfamily arXiv:1512.06901}}.

\bibitem{2015PhRvD..91h5011R}
S.~{Rychkov} and L.~G. {Vitale}, ``{Hamiltonian truncation study of the
  $\phi^{4}$ theory in two dimensions},''
  \href{http://dx.doi.org/10.1103/PhysRevD.91.085011}{{\em Physical Review D}
  {\bfseries 91} (2015) 085011},
  \href{http://arxiv.org/abs/1412.3460}{{\ttfamily arXiv:1412.3460}}.

\bibitem{2016PhRvD..93f5014R}
S.~{Rychkov} and L.~G. {Vitale}, ``{Hamiltonian truncation study of the
  $\phi^{4}$ theory in two dimensions. II. The Z$_{2}$ -broken phase and the
  Chang duality},'' \href{http://dx.doi.org/10.1103/PhysRevD.93.065014}{{\em
  Physical Review D} {\bfseries 93} (2016) 065014},
  \href{http://arxiv.org/abs/1512.00493}{{\ttfamily arXiv:1512.00493}}.

\bibitem{2013NuPhB.877..457B}
M.~{Beria}, G.~P. {Brandino}, L.~{Lepori}, R.~M. {Konik}, and G.~{Sierra},
  ``{Truncated conformal space approach for perturbed Wess-Zumino-Witten
  SU(2$_{}$ models},''
  \href{http://dx.doi.org/10.1016/j.nuclphysb.2013.10.005}{{\em Nuclear Physics
  B} {\bfseries 877} (2013) 457--483},
  \href{http://arxiv.org/abs/1301.0084}{{\ttfamily arXiv:1301.0084}}.

\bibitem{2015NuPhB.899..547K}
R.~M. {Konik}, T.~{P{\'a}lmai}, G.~{Tak{\'a}cs}, and A.~M. {Tsvelik},
  ``{Studying the perturbed Wess-Zumino-Novikov-Witten $SU(2)_k$ theory using
  the truncated conformal spectrum approach},''
  \href{http://dx.doi.org/10.1016/j.nuclphysb.2015.08.016}{{\em Nuclear Physics
  B} {\bfseries 899} (2015) 547--569},
  \href{http://arxiv.org/abs/1505.03860}{{\ttfamily arXiv:1505.03860}}.

\bibitem{2016arXiv160102979A}
P.~{Azaria}, R.~M. {Konik}, P.~{Lecheminant}, T.~{Palmai}, G.~{Takacs}, and
  A.~M. {Tsvelik}, ``{Particle Formation and Ordering in Strongly Correlated
  Fermionic Systems: Solving a Model of Quantum Chromodynamics},'' {\em ArXiv
  e-prints} (2016) , \href{http://arxiv.org/abs/1601.02979}{{\ttfamily
  arXiv:1601.02979}}.

\bibitem{2015PhRvD..91b5005H}
M.~{Hogervorst}, S.~{Rychkov}, and B.~C. {van Rees}, ``{Truncated conformal
  space approach in d dimensions: A cheap alternative to lattice field
  theory?},'' \href{http://dx.doi.org/10.1103/PhysRevD.91.025005}{{\em Physical
  Review D} {\bfseries 91} (2015) 025005},
  \href{http://arxiv.org/abs/1409.1581}{{\ttfamily arXiv:1409.1581}}.

\bibitem{2015PhRvB..92p1111J}
A.~J.~A. {James} and R.~M. {Konik}, ``{Quantum quenches in two spatial
  dimensions using chain array matrix product states},''
  \href{http://dx.doi.org/10.1103/PhysRevB.92.161111}{{\em Physical Review B}
  {\bfseries 92} (2015) 161111},
  \href{http://arxiv.org/abs/1504.00237}{{\ttfamily arXiv:1504.00237}}.

\bibitem{1961AnPhy..16..407L}
E.~{Lieb}, T.~{Schultz}, and D.~{Mattis}, ``{Two soluble models of an
  antiferromagnetic chain},''
  \href{http://dx.doi.org/10.1016/0003-4916(61)90115-4}{{\em Annals of Physics}
  {\bfseries 16} (1961) 407--466}.

\bibitem{1970AnPhy..57...79P}
P.~{Pfeuty}, ``{The one-dimensional Ising model with a transverse field},''
  \href{http://dx.doi.org/10.1016/0003-4916(70)90270-8}{{\em Annals of Physics}
  {\bfseries 57} (1970) 79--90}.

\bibitem{1978PhRvD..18.1259M}
B.~M. {McCoy} and T.~T. {Wu}, ``{Two-dimensional Ising field theory in a
  magnetic field: Breakup of the cut in the two-point function},''
  \href{http://dx.doi.org/10.1103/PhysRevD.18.1259}{{\em Physical Review D}
  {\bfseries 18} (1978) 1259--1267}.

\bibitem{2014JPhA...47q5002B}
L.~{Bucciantini}, M.~{Kormos}, and P.~{Calabrese}, ``{Quantum quenches from
  excited states in the Ising chain},''
  \href{http://dx.doi.org/10.1088/1751-8113/47/17/175002}{{\em Journal of
  Physics A Mathematical General} {\bfseries 47} (2014) 175002},
  \href{http://arxiv.org/abs/1401.7250}{{\ttfamily arXiv:1401.7250}}.

\bibitem{2009EL.....8720002S}
S.~{Sotiriadis}, P.~{Calabrese}, and J.~{Cardy}, ``{Quantum quench from a
  thermal initial state},''
  \href{http://dx.doi.org/10.1209/0295-5075/87/20002}{{\em EPL (Europhysics
  Letters)} {\bfseries 87} (2009) 20002},
  \href{http://arxiv.org/abs/0903.0895}{{\ttfamily arXiv:0903.0895}}.

\bibitem{2011PhRvL.106v7203C}
P.~{Calabrese}, F.~H.~L. {Essler}, and M.~{Fagotti}, ``{Quantum Quench in the
  Transverse-Field Ising Chain},''
  \href{http://dx.doi.org/10.1103/PhysRevLett.106.227203}{{\em Physical Review
  Letters} {\bfseries 106} (2011) 227203},
  \href{http://arxiv.org/abs/1104.0154}{{\ttfamily arXiv:1104.0154}}.

\bibitem{2012JSMTE..07..016C}
P.~{Calabrese}, F.~H.~L. {Essler}, and M.~{Fagotti}, ``{Quantum quench in the
  transverse field Ising chain: I. Time evolution of order parameter
  correlators},''
  \href{http://dx.doi.org/10.1088/1742-5468/2012/07/P07016}{{\em Journal of
  Statistical Mechanics: Theory and Experiment} {\bfseries 7} (2012) 07016},
  \href{http://arxiv.org/abs/1204.3911}{{\ttfamily arXiv:1204.3911}}.

\bibitem{2012JSMTE..07..022C}
P.~{Calabrese}, F.~H.~L. {Essler}, and M.~{Fagotti}, ``{Quantum quenches in the
  transverse field Ising chain: II. Stationary state properties},''
  \href{http://dx.doi.org/10.1088/1742-5468/2012/07/P07022}{{\em Journal of
  Statistical Mechanics: Theory and Experiment} {\bfseries 7} (2012) 07022},
  \href{http://arxiv.org/abs/1205.2211}{{\ttfamily arXiv:1205.2211}}.

\bibitem{2008Natur.452..854R}
M.~{Rigol}, V.~{Dunjko}, and M.~{Olshanii}, ``{Thermalization and its mechanism
  for generic isolated quantum systems},''
  \href{http://dx.doi.org/10.1038/nature06838}{{\em Nature} {\bfseries 452}
  (2008) 854--858}, \href{http://arxiv.org/abs/0708.1324}{{\ttfamily
  arXiv:0708.1324}}.

\bibitem{2014JSMTE..12..009B}
M.~{Brockmann}, B.~{Wouters}, D.~{Fioretto}, J.~{De Nardis}, R.~{Vlijm}, and
  J.-S. {Caux}, ``{Quench action approach for releasing the N{\'e}el state into
  the spin-1/2 XXZ chain},''
  \href{http://dx.doi.org/10.1088/1742-5468/2014/12/P12009}{{\em Journal of
  Statistical Mechanics: Theory and Experiment} {\bfseries 12} (2014) 12009},
  \href{http://arxiv.org/abs/1408.5075}{{\ttfamily arXiv:1408.5075}}.

\bibitem{2015JSMTE..04..001M}
M.~{Mesty{\'a}n}, B.~{Pozsgay}, G.~{Tak{\'a}cs}, and M.~A. {Werner},
  ``{Quenching the XXZ spin chain: quench action approach versus generalized
  Gibbs ensemble},''
  \href{http://dx.doi.org/10.1088/1742-5468/2015/04/P04001}{{\em Journal of
  Statistical Mechanics: Theory and Experiment} {\bfseries 4} (2015) 04001},
  \href{http://arxiv.org/abs/1412.4787}{{\ttfamily arXiv:1412.4787}}.

\bibitem{2006hep.th...12203F}
G.~{Feverati}, K.~{Graham}, P.~A. {Pearce}, G.~Z. {Toth}, and G.~{Watts}, ``{A
  Renormalisation group for TCSA},'' {\em ArXiv e-prints} (2006) ,
  \href{http://arxiv.org/abs/hep-th/0612203}{{\ttfamily hep-th/0612203}}.

\bibitem{2007PhRvL..98n7205K}
R.~M. {Konik} and Y.~{Adamov}, ``{Numerical Renormalization Group for Continuum
  One-Dimensional Systems},''
  \href{http://dx.doi.org/10.1103/PhysRevLett.98.147205}{{\em Physical Review
  Letters} {\bfseries 98} (2007) 147205},
  \href{http://arxiv.org/abs/cond-mat/0701605}{{\ttfamily
  arXiv:cond-mat/0701605}}.

\bibitem{2008PhRvL.101l0603S}
A.~{Silva}, ``{Statistics of the Work Done on a Quantum Critical System by
  Quenching a Control Parameter},''
  \href{http://dx.doi.org/10.1103/PhysRevLett.101.120603}{{\em Physical Review
  Letters} {\bfseries 101} (2008) 120603},
  \href{http://arxiv.org/abs/0806.4301}{{\ttfamily arXiv:0806.4301}}.

\bibitem{2009PhRvL.102l7204R}
D.~{Rossini}, A.~{Silva}, G.~{Mussardo}, and G.~E. {Santoro}, ``{Effective
  Thermal Dynamics Following a Quantum Quench in a Spin Chain},''
  \href{http://dx.doi.org/10.1103/PhysRevLett.102.127204}{{\em Physical Review
  Letters} {\bfseries 102} (2009) 127204},
  \href{http://arxiv.org/abs/0810.5508}{{\ttfamily arXiv:0810.5508}}.

\bibitem{2010PhRvB..82n4302R}
D.~{Rossini}, S.~{Suzuki}, G.~{Mussardo}, G.~E. {Santoro}, and A.~{Silva},
  ``{Long time dynamics following a quench in an integrable quantum spin chain:
  Local versus nonlocal operators and effective thermal behavior},''
  \href{http://dx.doi.org/10.1103/PhysRevB.82.144302}{{\em Phys. Rev. B}
  {\bfseries 82} (2010) 144302},
  \href{http://arxiv.org/abs/1002.2842}{{\ttfamily arXiv:1002.2842}}.

\bibitem{2011PhRvB..84p5117R}
H.~{Rieger} and F.~{Igl{\'o}i}, ``{Semiclassical theory for quantum quenches in
  finite transverse Ising chains},''
  \href{http://dx.doi.org/10.1103/PhysRevB.84.165117}{{\em Phys. Rev. B}
  {\bfseries 84} (2011) 165117},
  \href{http://arxiv.org/abs/1106.5248}{{\ttfamily arXiv:1106.5248}}.

\bibitem{2015arXiv150702708K}
M.~{Kormos} and G.~{Zar{\'a}nd}, ``{Quantum quenches in the sine--Gordon model:
  a semiclassical approach},''
  \href{http://dx.doi.org/10.1103/PhysRevE.93.062101}{{\em Physical Review E}
  {\bfseries 93} (2015) 062101},
  \href{http://arxiv.org/abs/1507.02708}{{\ttfamily arXiv:1507.02708}}.

\bibitem{Evangelisti2013}
S.~Evangelisti, ``{Semi-classical theory for quantum quenches in the O(3)
  non-linear sigma model},''
  \href{http://dx.doi.org/10.1088/1742-5468/2013/04/P04003}{{\em Journal of
  Statistical Mechanics: Theory and Experiment} (2013) P04003},
  \href{http://arxiv.org/abs/1210.4028}{{\ttfamily arXiv:1210.4028}}.

\bibitem{Gritsev2007}
V.~Gritsev, E.~Demler, M.~Lukin, and A.~Polkovnikov, ``{Spectroscopy of
  collective excitations in interacting low-dimensional many-body systems using
  quench dynamics},''
  \href{http://dx.doi.org/10.1103/PhysRevLett.99.200404}{{\em Physical Review
  Letters} {\bfseries 99} (2007) 200404},
  \href{http://arxiv.org/abs/cond-mat/0702343}{{\ttfamily
  arXiv:cond-mat/0702343}}.

\bibitem{2010NuPhB.825..466T}
G.~{Tak{\'a}cs}, ``{Form factor perturbation theory from finite volume},''
  \href{http://dx.doi.org/10.1016/j.nuclphysb.2009.10.001}{{\em Nuclear Physics
  B} {\bfseries 825} (2010) 466--481},
  \href{http://arxiv.org/abs/0907.2109}{{\ttfamily arXiv:0907.2109}}.

\bibitem{2015arXiv150503126S}
T.~{Schweigler}, V.~{Kasper}, S.~{Erne}, B.~{Rauer}, T.~{Langen},
  T.~{Gasenzer}, J.~{Berges}, and J.~{Schmiedmayer}, ``{On solving the quantum
  many-body problem},'' {\em ArXiv e-prints} (2015) ,
  \href{http://arxiv.org/abs/1505.03126}{{\ttfamily arXiv:1505.03126}}.

\bibitem{Bugrij:2000is}
A.~I. Bugrij, ``{The Correlation function in two-dimensional Ising model on the
  finite size lattice. 1.},''
  \href{http://dx.doi.org/10.1023/A:1010320126700}{{\em Theor. Math. Phys.}
  {\bfseries 127} (2001) 528--548},
  \href{http://arxiv.org/abs/hep-th/0011104}{{\ttfamily arXiv:hep-th/0011104}}.

\bibitem{2001hep.th....7117B}
A.~I. {Bugrij}, ``{Form factor representation of the correlation function of
  the two dimensional Ising model on a cylinder},'' in {\em Integrable
  Structures of Exactly Solvable Two-Dimensional Models of Quantum Field
  Theory}, S.~{Pakuliak} and G.~{von Gehlen}, eds., vol.~35 of {\em NATO
  Science Series}.
\newblock 2001.
\newblock \href{http://arxiv.org/abs/hep-th/0107117}{{\ttfamily
  arXiv:hep-th/0107117}}.

\bibitem{2011arXiv1106.2448G}
P.~{Giokas} and G.~{Watts}, ``{The renormalisation group for the truncated
  conformal space approach on the cylinder},'' {\em ArXiv e-prints} (2011) ,
  \href{http://arxiv.org/abs/1106.2448}{{\ttfamily arXiv:1106.2448}}.

\bibitem{2013JHEP...08..094S}
I.~M. {Sz{\'e}cs{\'e}nyi}, G.~{Tak{\'a}cs}, and G.~M.~T. {Watts}, ``{One-point
  functions in finite volume/temperature: a case study},''
  \href{http://dx.doi.org/10.1007/JHEP08(2013)094}{{\em Journal of High Energy
  Physics} {\bfseries 8} (2013) 94},
  \href{http://arxiv.org/abs/1304.3275}{{\ttfamily arXiv:1304.3275}}.

\bibitem{2005PhRvL..95y0601R}
S.~B. {Rutkevich}, ``{Large-n Excitations in the Ferromagnetic Ising Field
  Theory in a Weak Magnetic Field: Mass Spectrum and Decay Widths},''
  \href{http://dx.doi.org/10.1103/PhysRevLett.95.250601}{{\em Physical Review
  Letters} {\bfseries 95} (2005) 250601},
  \href{http://arxiv.org/abs/hep-th/0509149}{{\ttfamily arXiv:hep-th/0509149}}.

\bibitem{2006hep.th...12304F}
P.~{Fonseca} and A.~{Zamolodchikov}, ``{Ising Spectroscopy I: Mesons at
  $T<T_c$},'' {\em ArXiv e-prints} (2006) ,
  \href{http://arxiv.org/abs/hep-th/0612304}{{\ttfamily arXiv:hep-th/0612304}}.

\bibitem{2009JPhA...42D4025R}
S.~B. {Rutkevich}, ``{Formfactor perturbation expansions and confinement in the
  Ising field theory},''
  \href{http://dx.doi.org/10.1088/1751-8113/42/30/304025}{{\em Journal of
  Physics A Mathematical General} {\bfseries 42} (2009) 304025},
  \href{http://arxiv.org/abs/0901.1571}{{\ttfamily arXiv:0901.1571}}.

\bibitem{2015JHEP...09..146L}
M.~{Lencs{\'e}s} and G.~{Tak{\'a}cs}, ``{Confinement in the q-state Potts
  model: an RG-TCSA study},''
  \href{http://dx.doi.org/10.1007/JHEP09(2015)146}{{\em Journal of High Energy
  Physics} {\bfseries 9} (2015) 146},
  \href{http://arxiv.org/abs/1506.06477}{{\ttfamily arXiv:1506.06477}}.

\end{thebibliography}
\end{document}